\documentclass[aps,prd,reprint,superscriptaddress]{revtex4-2}

\usepackage{hyperref}
\usepackage{bm}
\usepackage{physics}
\usepackage{graphicx}
\usepackage{comment}
\usepackage[dvipsnames]{xcolor}
\usepackage{xspace}

\newcommand{\GION}{\texttt{GION}\xspace}
\newcommand{\hkpc}{h^{-1} \, \mathrm{kpc}}
\newcommand{\hMpc}{h^{-1} \, \mathrm{Mpc}}
\newcommand{\hMpcinv}{h \, \mathrm{Mpc}^{-1}}
\newcommand{\hGpc}{h^{-1} \, \mathrm{Gpc}}
\newcommand{\hMsun}{h^{-1} \, \mathrm{M}_\odot}

\begin{document}


\title{Introducing the GION Lightcone Simulation Suite for Multi-Wavelength Large-Scale Structure Cosmology}

\author{Ken Osato}
\email[]{ken.osato@chiba-u.jp}
\affiliation{Center for Frontier Science, Chiba University, Chiba 263-8522, Japan}
\affiliation{Department of Physics, Graduate School of Science, Chiba University, Chiba 263-8522, Japan}
\affiliation{RIKEN Center for Advanced Intelligence Project, Tokyo 103-0027, Japan}
\affiliation{Kavli Institute for the Physics and Mathematics of the Universe, The University of Tokyo Institutes for Advanced Study, Chiba, 277-8583, Japan}

\author{Satoshi Tanaka}
\affiliation{Graduate School of Social Data Science, Hitotsubashi University, Tokyo 186-8601, Japan}
\affiliation{Center for Information and Communication Technology, Hitotsubashi University, Tokyo 186-8601, Japan}

\author{Tomoaki Ishiyama}
\affiliation{Digital Transformation Enhancement Council, Chiba University, Chiba 263-8522, Japan}

\author{Hironao Miyatake}
\affiliation{Kobayashi-Maskawa Institute for the Origin of Particles and the
Universe (KMI), Nagoya University, Nagoya 464-8602, Japan}
\affiliation{Institute for Advanced Research, Nagoya University, Nagoya
464-8601, Japan}
\affiliation{Kavli Institute for the Physics and Mathematics of the Universe, The University of Tokyo Institutes for Advanced Study, Chiba, 277-8583, Japan}

\author{Takahiro Nishimichi}
\affiliation{Department of Astrophysics and Atmospheric Sciences, Faculty of Science, Kyoto Sangyo University, Kyoto 603-8555, Japan}
\affiliation{Center for Gravitational Physics and Quantum Information, Yukawa Institute for Theoretical Physics, Kyoto University, Kyoto 606-8502, Japan}
\affiliation{RIKEN Center for Advanced Intelligence Project, Tokyo 103-0027, Japan}
\affiliation{Kavli Institute for the Physics and Mathematics of the Universe, The University of Tokyo Institutes for Advanced Study, Chiba, 277-8583, Japan}

\author{Masato Shirasaki}
\affiliation{National Astronomical Observatory of Japan, National Institutes of Natural Sciences, Tokyo 181-8588, Japan}
\affiliation{The Institute of Statistical Mathematics, Tokyo 190-8562, Japan}
\affiliation{RIKEN Center for Advanced Intelligence Project, Tokyo 103-0027, Japan}

\date{\today}

\begin{abstract}
We present the \GION (Ginkaku Lightcone Multi-wavelength Simulation Project) simulation suite: 10 full-sky high-resolution cosmological lightcone simulations and their associated products for multi-wavelength large-scale structure cosmology.
We use the TreePM $N$-body code \texttt{GINKAKU}, which is highly optimized in the massively parallel environment, and integrate the on-the-fly lightcone construction technique.
Each lightcone simulation is run with $6144^3$ particles in a volume of $(1.5 \, \hGpc)^3$, which corresponds to a particle mass of $1.28 \times 10^9 \, \hMsun$,
and extends to $z = 5$.
The fine mass resolution and the depth of the lightcone simulations allow us to resolve dark matter halos down to $10^{11} \, \hMsun$ in the high-redshift Universe, which is crucial for applications to the Stage-IV cosmological surveys.
The fundamental statistics of the simulations, including the matter power spectrum, halo power spectrum, and halo mass function, are validated against theoretical predictions.
We also generate various large-scale structure observables from the lightcone outputs: the dark matter halo catalogs, mock galaxy samples, weak lensing convergence maps, and thermal and kinematic Sunyaev--Zel'dovich (tSZ and kSZ) maps.
The angular power spectra of these observables, including the cross-correlation between weak lensing and the tSZ effect, are in good agreement with theoretical predictions, demonstrating that the suite provides mutually consistent multi-wavelength observables on the same lightcone.
The \GION simulation suite will be publicly available and can be used for various cosmological analyses to address the observational systematic effects and validate theoretical models.
\end{abstract}


\maketitle

\section{Introduction}
\label{sec:intro}
The growth and geometry of the large-scale structure (LSS) of the Universe serve as a unique probe into its dark constituents, dark matter and dark energy \cite{Weinberg2013, Huterer2018}.
To unveil the nature of these dark components, wide and deep cosmological surveys aim to measure cosmological statistics with unprecedented precision.
The ongoing and forthcoming surveys, collectively referred to as Stage-IV surveys \cite{DETF2006}, include Dark Energy Spectroscopic Instrument (DESI) \cite{DESI2022}, \textit{Euclid} \cite{EuclidI}, Subaru Prime Focus Spectrograph (PFS) \cite{Takada2014}, Rubin Observatory Legacy Survey of Space and Time (LSST) \cite{LSST2019}, and \textit{Nancy Grace Roman Space Telescope} \cite{Roman2015}.
Their primary science drivers are the measurement of weak gravitational lensing \cite{Bartelmann2001, Mandelbaum2018} and galaxy clustering statistics \cite{Bassett2010, Aubourg2015} and the resulting constraints on cosmological models.

Complementary to LSS observations, which probe the late-time Universe, the cosmic microwave background (CMB) provides independent information on the early Universe \cite{Hu2002, Durrer2015}.
In the last decade, \textit{WMAP} \cite{Bennett2013} and \textit{Planck} \cite{Planck2018I} satellites measured the temperature and polarization anisotropies on large scales, and the ground-based telescopes such as Atacama Cosmology Telescope \cite{Swetz2011} and South Pole Telescope \cite{Carlstrom2011} enable precise measurements on small scales with the large aperture.
In the Stage-IV survey era, CMB experiments with upgraded instruments are ongoing and planned, which include the Simons Observatory \cite{SO2019} and LiteBIRD \cite{LiteBIRD2023}.
In addition to the primary anisotropies, these experiments will map the secondary anisotropies \cite{Aghanim2008} imprinted by the intervening LSS with high sensitivity: the integrated Sachs--Wolfe effect \cite{Sachs1967}, gravitational lensing of CMB \cite{Lewis2006}, and the thermal and kinematic Sunyaev--Zel'dovich (tSZ and kSZ) effects \cite{SZ1972, SZ1980}.
Because these signals and the LSS observables discussed above are sourced by the same underlying matter distribution, they are strongly correlated, and their cross-correlations carry information that is inaccessible to the auto-correlations alone.
The information from cross-correlations can break parameter degeneracies and deproject the projected observable with a tomographic approach.
The various cross-correlation pairs of LSS and CMB observables have already been measured and employed to constrain cosmological and astrophysical models, e.g., CMB lensing and galaxies \cite{Singh2017, Chen2022} and the tSZ effect and weak lensing \cite{Hojjati2017, Osato2018, Osato2020, Troester2022, Gatti2022, Pandey2022}.
Joint multi-wavelength analyses are therefore becoming standard practice.

To exploit the plethora of data from Stage-IV surveys, numerical simulations play a central role in data analysis \cite{Angulo2022}.
High-fidelity simulations act as a controlled synthetic realization of the observable sky: the same analysis pipeline as used for the observations can be applied to the simulated data, so that the framework can be validated in its entirety.
Simulations also give direct access to the statistical variation of the observables, i.e., the cosmic variance, provided that a sufficient number of independent realizations is available.
In addition, observational systematics such as survey masks and selection effects can be forward modeled.
The challenge specific to the Stage-IV era is that a single simulation must simultaneously cover the survey volume and resolve the small-scale structures probed by state-of-the-art telescopes, while delivering multiple observables that are mutually consistent on the same lightcone.

Several suites have addressed parts of this problem: full-sky lightcone simulations dedicated to weak lensing \cite{Fosalba2015, Takahashi2017}, survey-oriented mock catalogs \cite{CosmoDC2, DeRose2019, Ishiyama2021a, EuclidV}, and multi-wavelength products combining LSS and CMB secondaries \cite{Sehgal2010, Websky, Agora, HalfDome}.
However, simultaneously achieving the mass resolution needed to resolve the halos that host the faint, high-redshift galaxies targeted by Stage-IV spectroscopic surveys, full-sky coverage, and multiple independent realizations remains challenging.

In this paper, we introduce the new multi-wavelength lightcone simulation suite \GION (\textbf{GI}nkaku lightc\textbf{ON}e Multi-wavelength Simulation Project).
The suite consists of $10$ independent realizations with full-sky coverage extending to $z = 5$.
Each realization evolves $6144^3$ particles in a box of $(1.5 \, \hGpc)^3$, corresponding to a particle mass of $m_\mathrm{part} = 1.28 \times 10^9 \, \hMsun$, so that halos hosting emission-line galaxies ($\gtrsim 10^{11} \, \hMsun$) are resolved out to high redshift.
This combination of resolution and volume is made possible by the TreePM code \texttt{GINKAKU} \cite{Nishimichi2026}, which is highly optimized in the massively parallelized environment for the architecture of the supercomputer Fugaku.
We have implemented an on-the-fly lightcone output algorithm in \texttt{GINKAKU}, which was pioneered by the Hubble Volume Simulation \cite{Evrard2002} and integrated into modern $N$-body simulation codes \cite{Springel2021, Schaller2024}.
Because the algorithm captures the evolution of structures along the line-of-sight direction, it enables a more consistent comparison with real data than a stack of fixed-redshift snapshots.

\begin{figure*}
\includegraphics[width=\textwidth]{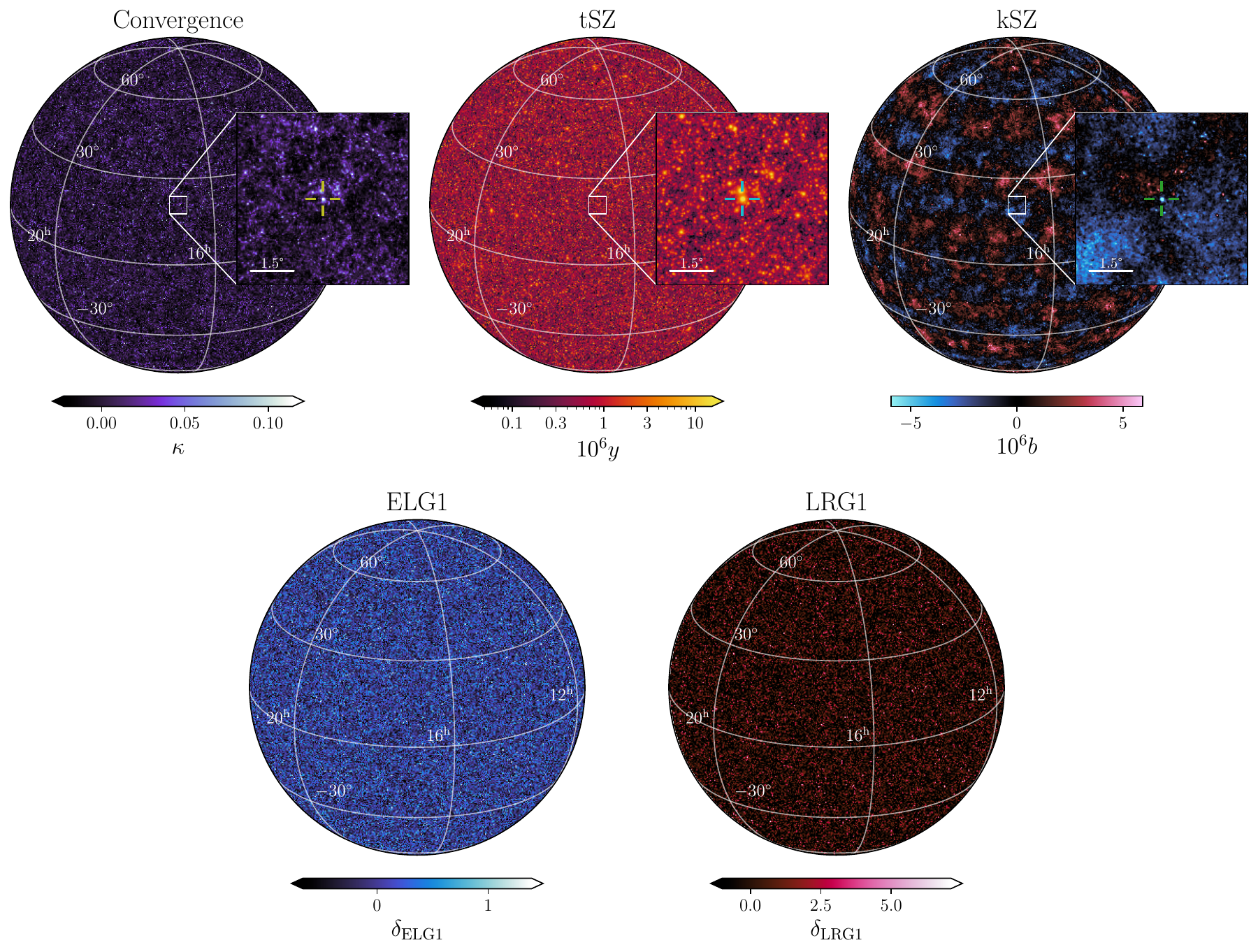}
\caption{The simulated maps from one of ten full-sky lightcone simulations in the \GION simulation suite; the lensing convergence map with the source redshift $z_\mathrm{s} = 1$ (top left), the tSZ Compton-$y$ map (top middle), the kSZ Doppler-$b$ map (top right), and the projected halo number density contrast map of halos with $10^{11.5} \, \hMsun < M_\mathrm{vir} < 10^{12.5} \, \hMsun$ and $0.8 < z < 1.1$ (ELG1, bottom left) and with $10^{12.5} \, \hMsun < M_\mathrm{vir} < 10^{13.5} \, \hMsun$ and $0.4 < z < 0.6$ (LRG1, bottom right).
For visualization purposes, the maps are smoothed with a Gaussian kernel with full-width at half-maximum of $2 \, \mathrm{arcmin}$.
For the convergence, tSZ, and kSZ maps, the zoomed-in regions centered at one of the convergence peaks are also shown.}
\label{fig:maps}
\end{figure*}

The \GION simulations follow only dark matter, and their raw outputs are not directly comparable to real observations.
We therefore also describe the post-processing algorithms used to construct mock observations from the lightcone outputs.
In this paper, we consider halo distributions, weak lensing, and the thermal and kinematic SZ effects, which are primary science targets of the Stage-IV surveys and the associated CMB experiments.
Figure~\ref{fig:maps} illustrates the simulated maps from the \GION simulation suite.
In subsequent papers, we will extend the dataset to more observables including CMB lensing and X-ray emissions, where the post-processing algorithms are already established.
All of these data products will be made publicly available at the \GION website \footnote{\url{https://github.com/GION-simulations/GION_Data_Release}}.

This paper is structured as follows.
Section~\ref{sec:numerical_methods} describes the numerical methods and data structures of the \GION simulation suite.
Section~\ref{sec:validation} validates the simulations by comparing their fundamental statistics with theoretical expectations at fixed redshifts.
Section~\ref{sec:lightcone_outputs} presents the main lightcone data products, and Section~\ref{sec:future_plans} outlines future extensions of the suite.
Finally, we make concluding remarks in Section~\ref{sec:conclusions}.
Appendix~\ref{sec:data_release} summarizes the data products and their public release, and
Appendix~\ref{sec:convergence_tests} summarizes convergence tests for the numerical parameters of the simulations.
Appendix~\ref{sec:exact_model} presents the effects of the Limber approximation and the flat-sky projection in power spectrum computations.

Throughout this paper, we assume the flat $\Lambda$ cold dark matter ($\Lambda$CDM) cosmology with the best-fit parameters of \textit{Planck} 2018 results \cite{Planck2018VI}: matter density $\Omega_\mathrm{m} = 0.3158$, baryon density $\Omega_\mathrm{b} = 0.04939$, dark energy density $\Omega_\Lambda = 1 - \Omega_\mathrm{m} = 0.6842$, scaled Hubble constant $h = 0.67321$, the amplitude of primordial perturbation $10^9 A_\mathrm{s} = 2.1005$, and its slope $n_\mathrm{s} = 0.96605$ at the pivot scale $k_\mathrm{piv} = 0.05 \, \mathrm{Mpc}^{-1}$.
The neutrino sector has one massive species with the mass $m_\nu = 0.06 \, \mathrm{eV}$ and two massless species.

As a halo mass definition, we mainly adopt the spherical overdensity mass $M_\mathrm{200b}$, which is defined as the mass enclosed within a sphere of radius $R_\mathrm{200b}$, where the mean density is $200$ times the mean matter density of the Universe.
Unless noted otherwise, we use this halo mass definition throughout this paper, i.e., $M \equiv M_\mathrm{200b}$.
We also use the alternative halo mass definition of the virial mass $M_\mathrm{vir}$, where the mean density is $\Delta_\mathrm{vir}(z)$ times the mean matter density of the Universe, and $M_\mathrm{500c}$, where the mean density is $500$ times the critical density of the Universe.
The overdensity parameter $\Delta_\mathrm{vir}(z)$ is derived in Ref.~\cite{Bryan1998}.
The radii $R_\mathrm{vir}$ and $R_\mathrm{500c}$ are defined in the same manner.

\section{Numerical methods}
\label{sec:numerical_methods}
\begin{figure}
\includegraphics[width=\columnwidth]{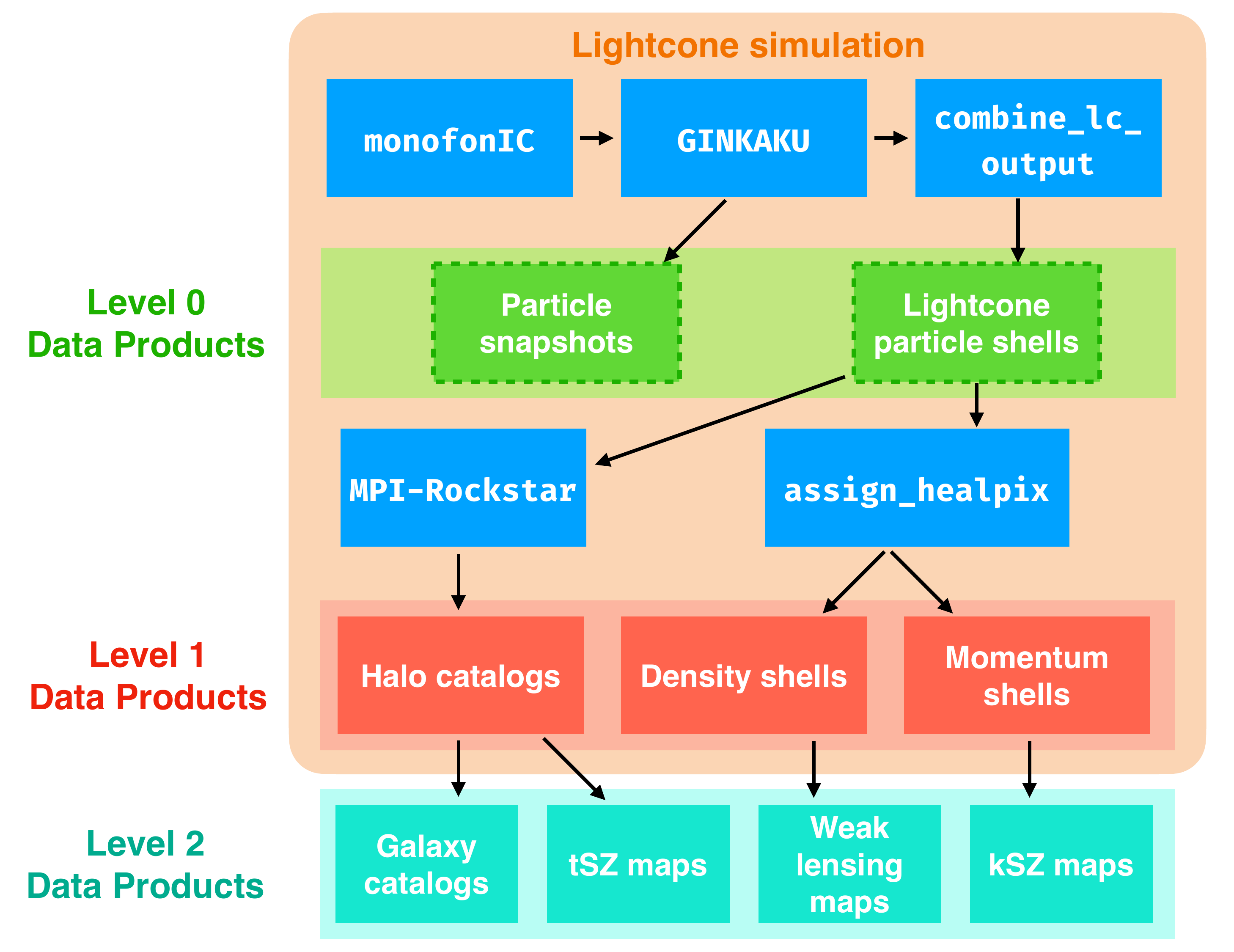}
\caption{The workflow and data products of the \GION simulation suite.}
\label{fig:GION_flowchart}
\end{figure}

In this section, we describe the numerical methods used to construct the \GION simulation suite.
The workflow of the simulation suite is illustrated in Figure~\ref{fig:GION_flowchart}.
There are three steps in the workflow: the $N$-body simulation with the on-the-fly lightcone output and the subsequent two post-processing steps.

We first generate the initial conditions with \texttt{monofonIC} \cite{Michaux2021} \footnote{\url{https://bitbucket.org/ohahn/monofonic}} and then run the cosmological $N$-body simulations with the TreePM code \texttt{GINKAKU} \cite{Nishimichi2026}.
In this work, the code is augmented with the on-the-fly lightcone construction technique to generate continuous particle outputs.
Every time the particle outputs reach the predefined threshold size, the output is passed to the \texttt{combine\_lc\_output} code to combine the output with the buffer region produced at the previous step so that halos near the boundary are not cropped.
In the second step, we process the lightcone particle outputs to produce the dark matter halo catalogs with \texttt{MPI-Rockstar} \cite{Tokuue2026} and the projected density and momentum shells with the auxiliary code \texttt{assign\_healpix}.
After this second step finishes, the lightcone particle outputs are deleted to reduce the storage requirements.
We retain the particle snapshots at multiple redshifts for one realization for validation and at $z = 0$ for all realizations.
The main ``lightcone simulation'' corresponds to these two steps and occupies most of the computing time.
We label the data products directly generated during $N$-body simulations, e.g., lightcone particle output, as ``Level 0 Data Products (DPs)'' and the post-processed data, e.g., halo catalogs, as ``Level 1 DPs''.
Level 0 products are temporary outputs for subsequent processing and are used for validation.
After the Level 1 DPs are generated, the Level 0 DPs are deleted to save storage space.
The Level 1 DPs are one of the primary deliverables of the \GION simulation suite and are also used for the subsequent post-processing to generate the large-scale structure observables.

The last step is to generate the LSS observables by further post-processing the Level 1 DPs.
There are a wide variety of possible applications but in this work, for demonstration purposes, we focus on the galaxy catalog, weak lensing, and thermal and kinematic Sunyaev--Zel'dovich effects.
The resultant data products are labeled as ``Level 2 DPs''.
All the Level 1 and Level 2 DPs will be publicly available.
The details of the data products and their public release are summarized in Appendix~\ref{sec:data_release}.

\subsection{Lightcone simulations}
We first summarize the baseline $N$-body simulations.
The initial condition is generated with \texttt{monofonIC} \cite{Michaux2021} based on third-order Lagrangian perturbation theory at the redshift $z_\mathrm{start} = 24$.
Though massive neutrinos are taken into account in the linear theory computation, in dark-matter-only simulations, the $N$-body code treats baryons and massive neutrinos as CDM for gravitational evolution.
To be consistent, the linear power spectrum is first computed at $z = 0$ and then back-scaled to $z_\mathrm{start}$ with the linear growth factor without massive neutrino effects.
We use \texttt{CLASS} Boltzmann code \cite{Blas2011} to compute the linear power spectrum.
As the $N$-body code, we employ the TreePM code \texttt{GINKAKU} \cite{Nishimichi2026}, which is highly optimized for the massively parallelized environment.
The simulation is run in a periodic cubic box with the length on a side $L = 1.5 \, \hGpc$ and the number of particles is $6144^3$, which corresponds to the particle mass $m_\mathrm{part} = 1.28 \times 10^9 \, \hMsun$.
The Plummer softening length is $1.28\%$ of the mean particle separation, which corresponds to $3.13 \, \hkpc$.
The simulation parameters are summarized in Table~\ref{tab:simulation_parameters}.
In addition to the main lightcone simulations, we also run smaller simulations to address the effects of the number of particle-mesh grids, the softening length, and the starting redshift.
The results are discussed in Appendix~\ref{sec:convergence_tests}.
We run $10$ simulations by varying the random seed to generate different initial conditions with the same cosmological parameters.
In addition to lightcone outputs, which will be discussed later, we also save the particle snapshots at $z = 5, 4, 3, 2.5, 2, 1.5, 1, 0.5, 0.3, 0$ for one of $10$ realizations and $z = 0$ for all realizations for sanity checks.
The validation with the snapshots will be presented in Section~\ref{sec:validation}.

\begin{table*}[htbp]
    \caption{Simulation parameters.}
    \label{tab:simulation_parameters}
    \begin{ruledtabular}
        \begin{tabular}{ccl}
            Parameter & Value & Description \\
            \hline
            \multicolumn{3}{c}{\texttt{monofonIC}} \\
            \hline
            \texttt{GridRes} & 6144 & Number of particles per dimension \\
            \texttt{BoxLength} & 1500.0 & Box length in $\hMpc$ \\
            \texttt{zstart} & 24.0 & Starting redshift \\
            \texttt{LPTorder} & 3 & Order of Lagrangian perturbation theory \\
            \hline
            \multicolumn{3}{c}{\texttt{GINKAKU}} \\
            \hline
            \texttt{pm\_factor}          & 1.0 & Particle-mesh grid factor \\
            \texttt{pm\_dt\_factor}      & 0.125 & Particle-mesh time step factor \\
            \texttt{tree\_dt\_factor}    & 0.25 & Tree time step factor \\
            \texttt{maximum\_dt}         & 0.03 & Maximum time step \\
            \texttt{global\_dt\_factor}  & 1.0 & Global time step factor \\
            \texttt{pp\_softening\_ratio}& 0.03 & Softening length parameter \\
            \texttt{cutoff\_radius}      & 3.0 & Particle-mesh cutoff radius \\
            \texttt{tree\_theta}         & 0.5 & Tree opening angle \\
            \texttt{output\_list} & [5, 4, 3, 2.5, 2, 1.5, 1, 0.5, 0.3, 0] or [0] & Redshifts for particle snapshot outputs \footnotemark[1] \\
            \hline
            \multicolumn{3}{c}{\texttt{combine\_lc\_output}} \\
            \hline
            \texttt{lc\_redshift\_max\_part} & 5.15 & Maximum redshift for lightcone particle output \\
            \texttt{lc\_redshift\_min\_part} & 0.0 & Minimum redshift for lightcone particle output \\
            \texttt{lc\_redshift\_max\_halo} & 5.0 & Maximum redshift for lightcone halo output \\
            \texttt{lc\_redshift\_min\_halo} & 0.0 & Minimum redshift for lightcone halo output\\
            \texttt{pad\_size} & 10.0 & Padding size in $\hMpc$ \\
            \hline
            \multicolumn{3}{c}{\texttt{assign\_healpix}} \\
            \hline
            \texttt{hpix\_cd\_max}   & 5400 & Maximum comoving distance in $\hMpc$\\
            \texttt{hpix\_cd\_min}   & 0 & Minimum comoving distance in $\hMpc$ \\
            \texttt{hpix\_cd\_delta} & 25 & Comoving distance interval in $\hMpc$ \\
            \texttt{hpix\_nside}     & 8192 & \texttt{HEALPix} resolution parameter \\
            \hline
            \multicolumn{3}{c}{\texttt{MPI-Rockstar}} \\
            \hline
            \texttt{FOF\_LINKING\_LENGTH} & 0.28 & Linking length in terms of the mean particle separation \\
            \texttt{MIN\_HALO\_PARTICLES} & 10 & Minimum number of particles in a halo \\
            \texttt{MIN\_HALO\_OUTPUT\_SIZE} & 50 & Minimum number of particles of halos to be output \\
            \texttt{SHAPE\_ITERATIONS} & 10 & Maximum number of iterations for shape calculation \\
        \end{tabular}
    \end{ruledtabular}
    \footnotetext[1]{Only one of 10 realizations has snapshots at multiple redshifts and the rest of the realizations have only $z = 0$ snapshot output.}
\end{table*}

Next, we describe the associated lightcone modules, which process the direct particle outputs from \texttt{GINKAKU} and produce Level 1 DPs, i.e., lightcone halo catalogs and density and momentum shells.
When the ``lightcone'' flag is switched on, \texttt{GINKAKU} continuously outputs particles when they pass the lightcone surface.
To cover the entire volume encompassed by the lightcone, the simulation box is replicated multiple times under the periodic boundary condition.
Due to the replications, the same structures appear multiple times in the lightcone outputs, though they appear at different redshifts and angular positions.
The replication effect has an appreciable impact on the covariance of the statistics but the impact on the mean statistics is sub-percent level \cite{Tokiwa2026}.
The output particles are distributed in a shell shape, and thus we call the output a ``lightcone particle shell''.
When the stored particle size or the thickness of the output shell exceeds the threshold, \texttt{GINKAKU} is paused and \texttt{combine\_lc\_output} launches.
The size limit and the maximum thickness are set to $1 \, \mathrm{TB}$ and $200 \, \hMpc$.
The \texttt{combine\_lc\_output} code combines the lightcone particle shell at the current step with the buffer shell from the previous step to keep the continuity for halo finding.
The thickness of the buffer region must be larger than the maximum radius of halos, and in our case, we set it to $10 \, \hMpc$, which is larger than the radius of the possible largest halos at $z = 0$.
The lightcone particle shells and particle snapshots constitute Level 0 DPs.
These DPs are temporary data used only for subsequent post-processing and testing and will be deleted once these processes are completed.
Therefore, Level 0 DPs are not regarded as the main data products of the \GION simulation suite.

\begin{figure*}
\includegraphics[width=\textwidth]{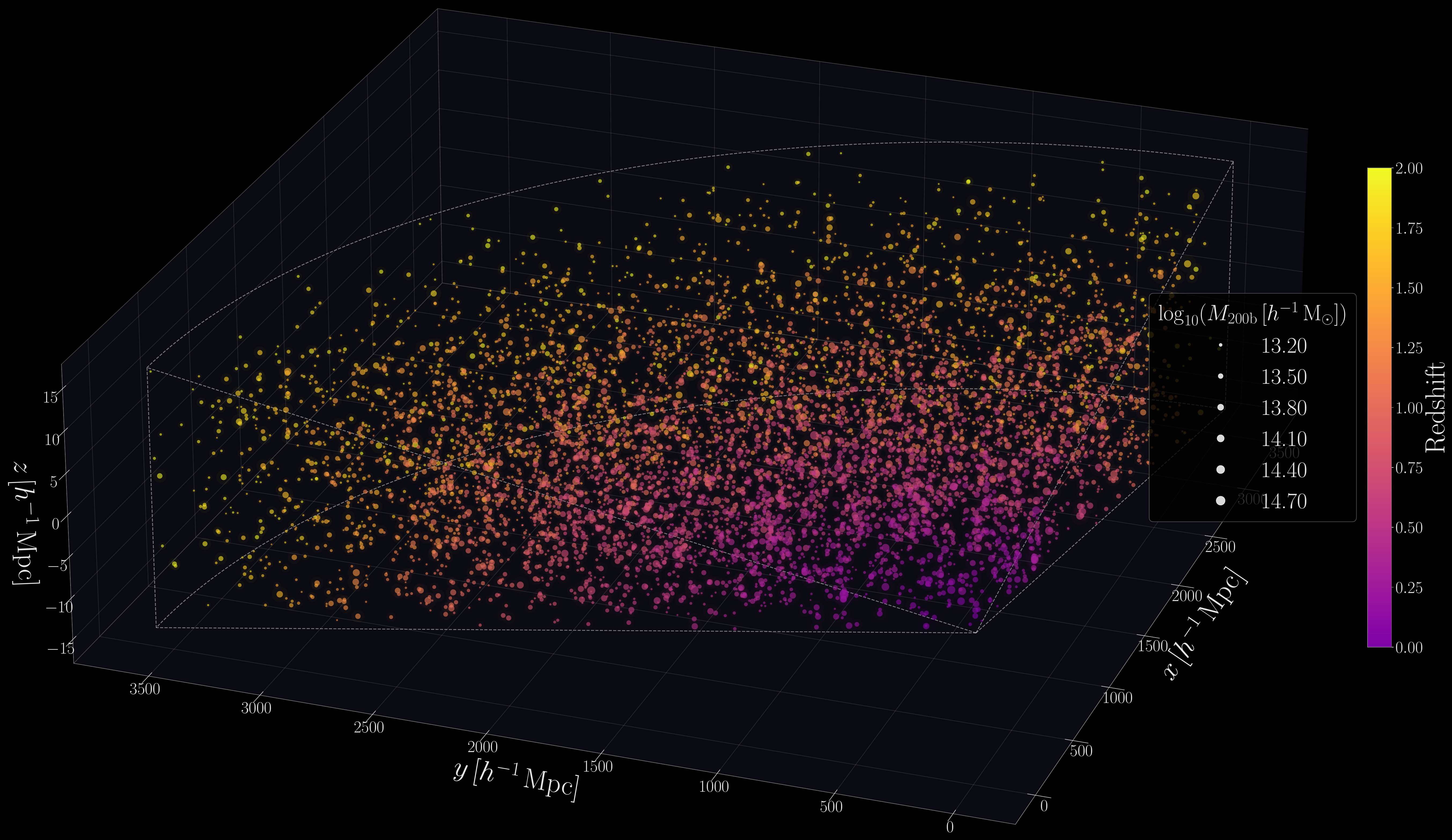}
\caption{The 3D lightcone halo distribution in the \GION simulation suite.
The halos are selected with $M_\mathrm{200b} > 10^{13} \, \hMsun$ in the range of $89^\circ < \theta < 91^\circ$, $0^\circ < \phi < 90^\circ$, and $0 < z < 2$.}
\label{fig:lc_halo_3d}
\end{figure*}

\begin{figure}
\includegraphics[width=\columnwidth]{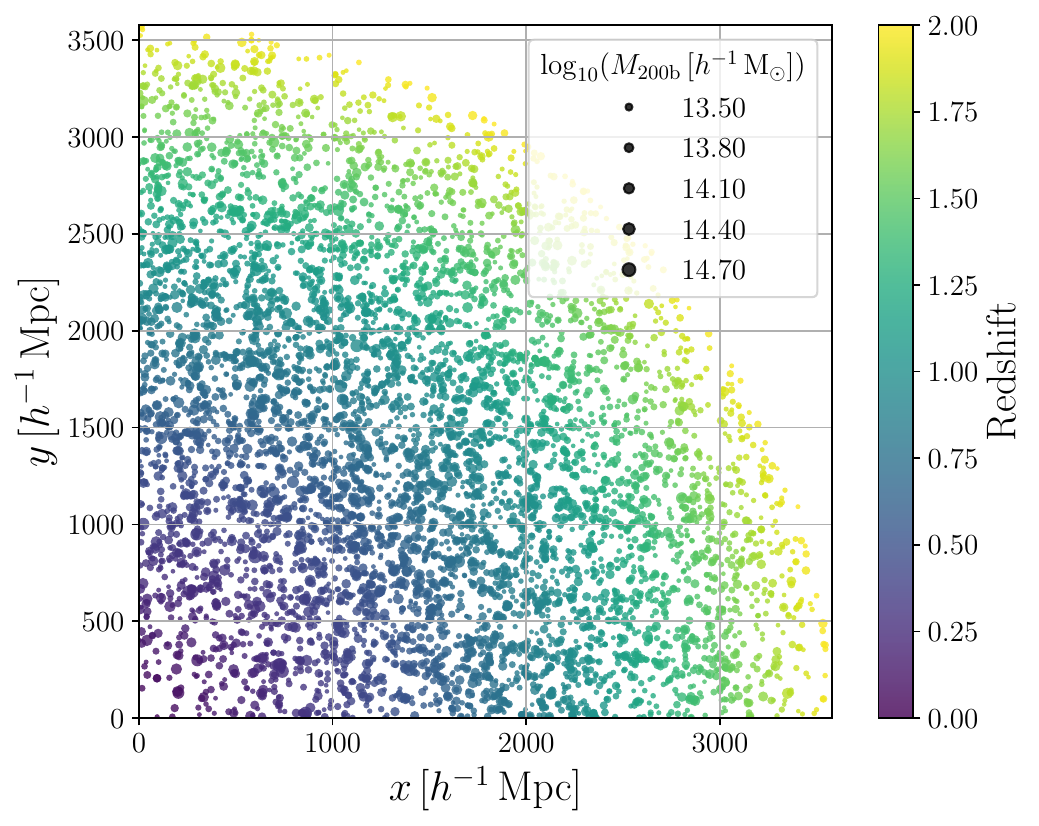}
\caption{The projected 2D lightcone halo distribution in the \GION simulation suite. The halos are projected along the $z$-axis in the range $89^\circ < \theta < 91^\circ$, $0^\circ < \phi < 90^\circ$, and $0 < z < 2$.}
\label{fig:lc_halo_2d}
\end{figure}

The lightcone particle shells are then passed to the halo finding code \texttt{MPI-Rockstar} \cite{Tokuue2026} \footnote{\url{https://github.com/Tomoaki-Ishiyama/mpi-rockstar}} and the mass and momentum shell generation code \texttt{assign\_healpix}.
\texttt{MPI-Rockstar} is the massively parallelized version of \texttt{Rockstar} \cite{Behroozi2013} and we modify the code so that it can accept the lightcone particle shell.
The resultant lightcone halo catalog contains the basic information such as halo mass and angular momentum, which are computed by default in \texttt{MPI-Rockstar}.
In addition, all six components of the inertia tensor computed based on the iterative algorithm and the parent-child relation based on inclusion relation within the virial radius \footnote{This approach is originally implemented as \texttt{find\_parents} program bundled in \texttt{Rockstar}.} are stored for each halo.
Hereafter, we define ``host halos'' as distinct halos that are not subhalos of any other halo according to the parent-child relation, and ``subhalos'' as halos that are subhalos of any other halo.
The lightcone halo catalog is saved in HDF5 format.
The minimum halo mass is set as $50 \times m_\mathrm{part} = 6.38 \times 10^{10} \, \hMsun$.
Next, the lightcone particle shells are used to generate density shells and momentum shells by \texttt{assign\_healpix} code.
The density and momentum shells are required to simulate weak lensing and the kSZ effect, respectively.
The \texttt{assign\_healpix} code projects particle mass and line-of-sight velocity in the \texttt{HEALPix} format \cite{Gorski2005} for every $25 \, \hMpc$ shell.
The resolution parameter of the \texttt{HEALPix} maps is $N_\mathrm{side} = 8192$ and the approximate pixel size, which is the square root of the area of a pixel, is $0.429 \, \mathrm{arcmin}$.
Once these Level 1 DPs are produced, the lightcone particle shell at the step before the current step is deleted (the shell at the previous step is used by \texttt{combine\_lc\_output}) and \texttt{GINKAKU} restarts for the next step.
These lightcone modules start from $z = 5.15$ and are iterated until $z = 0$.
There is a caveat about the maximum redshift of the lightcone outputs.
The halo finder runs only from $z = 5.0$ to $z = 0$, and thus the lightcone halo catalogs are available only in this redshift range while the density and momentum shells are available from $z = 5.14$ to $z = 0$.
To illustrate the lightcone halo distributions, Figure~\ref{fig:lc_halo_3d} shows the 3D distributions of halos with $M_\mathrm{200b} > 10^{13} \, \hMsun$ in a part of the sky ($89^\circ < \theta < 91^\circ$, $0^\circ < \phi < 90^\circ$, and $0 < z < 2$), and Figure~\ref{fig:lc_halo_2d} displays the projected distribution with the same halo sample.

All the calculations, $N$-body simulations and lightcone modules, are conducted on the supercomputer Fugaku at the RIKEN Center for Computational Science in Japan.
The calculations are run with 8,192 A64FX computing nodes, i.e., 393,216 cores.
All the codes are implemented in MPI and OpenMP hybrid parallelization.
For every run, 8,192 MPI processes (1 process per node) and 48 OpenMP threads are used.
Table~\ref{tab:walltime} shows the wall-clock time of each task.
The $N$-body simulations by \texttt{GINKAKU} account for the dominant part, but a comparable wall-clock time is required for halo finding by \texttt{MPI-Rockstar} due to the large volume of the lightcone outputs.

\begin{table}[htbp]
    \caption{Wall-clock time for each program in the lightcone simulation.
    The time is measured for one realization of the \GION simulation suite.}
    \label{tab:walltime}
    \begin{ruledtabular}
        \begin{tabular}{ccc}
            Program & Wall-clock time [$\mathrm{s}$] & Fraction \\
            \hline
            \texttt{GINKAKU} & 29093 & 37\% \\
            \texttt{combine\_lc\_output} & 5024 & 6\% \\
            \texttt{assign\_healpix} & 15703 & 20\% \\
            \texttt{MPI-Rockstar} & 28447 & 36\% \\
            \hline
            Total & 78267 & 100\% \\
        \end{tabular}
    \end{ruledtabular}
\end{table}

\section{Validation at fixed redshifts}
\label{sec:validation}
In this section, we validate the fundamental statistics of LSS from the \GION simulation suite against theoretical predictions at fixed redshifts using the particle snapshots and halo catalogs described in Section~\ref{sec:numerical_methods}.
We focus on three quantities: the matter power spectrum, the halo mass function, and the halo power spectra, i.e., the halo--halo auto spectrum and the halo--matter cross spectrum.
The statistics in this section are measured from the single realization of the \GION simulation suite as the particle snapshots and halo catalogs for the wide range of redshifts $z = 0 \text{--} 5$ are stored only for one realization due to the large volume of data.
The Level 1 and Level 2 DPs constructed from the lightcone outputs are validated separately in Section~\ref{sec:lightcone_outputs}.

\subsection{Matter power spectrum}
We first validate the clustering of the underlying dark matter distribution.
The matter density contrast $\delta(\bm{x})$ is defined as $\delta(\bm{x}) = \rho(\bm{x}) / \bar\rho_\mathrm{m} - 1$, where $\bm{x}$ is the comoving position, $\rho (\bm{x})$ is the density field, $\bar\rho_\mathrm{m}$ is the mean comoving matter density, and the matter power spectrum $P_\mathrm{mm}(k)$ is defined through
\begin{equation}
  \label{eq:pk_def}
  \langle \tilde\delta(\bm{k}) \tilde\delta(\bm{k}') \rangle = (2\pi)^3 \delta_\mathrm{D}(\bm{k} + \bm{k}') \, P_\mathrm{mm}(k),
\end{equation}
where $\tilde\delta(\bm{k})$ is the Fourier transform of $\delta(\bm{x})$, $\bm{k}$ is the wave-vector, and $\delta_\mathrm{D}$ is the Dirac delta function, and $\langle \cdot \rangle$ denotes ensemble average.
First, we construct the density grid with the number of grid cells per dimension $N_\mathrm{grid} = 3072$ from the particle snapshots with the triangular shaped cloud assignment scheme \cite{Hockney1981}.
Then, we apply a fast Fourier transform to the density grid and estimate the power spectrum with aliasing correction \cite{Jing2005}.
In this configuration, the fundamental mode is $k_\mathrm{F} = 2 \pi /L = 0.0042 \, \hMpcinv$ and the Nyquist mode is $k_\mathrm{Ny} = N_\mathrm{grid} \pi / L = 6.43 \, \hMpcinv$.
The power spectrum is binned with $200$ logarithmically spaced bins between $k_\mathrm{F}$ and $k_\mathrm{Ny}$ and measured from the particle snapshots at $z = 4, 3, 2, 1, 0.5$, and $0$.
Since there is a single snapshot for each redshift, it is not possible to measure the covariance of the power spectrum from a single realization.
Instead, we estimate the statistical uncertainty of the power spectrum measurement with Gaussian covariance:
\begin{equation}
  \mathrm{Cov} [P_\mathrm{mm} (k_i), P_\mathrm{mm} (k_j)]
  = \delta_{ij} \frac{2}{N_\mathrm{mode}(k_i)} P^2_\mathrm{mm}(k_i) ,
\end{equation}
where $N_\mathrm{mode}(k_i)$ is the number of independent Fourier modes in the $i$-th bin and $\delta_{ij}$ is the Kronecker delta. Note that the Gaussian covariance is a good approximation on large scales, but it underestimates the true covariance on small scales due to the non-Gaussian contribution \cite{Scoccimarro1999, Meiksin1999, Takahashi2011}.

\begin{figure}
\includegraphics[width=\columnwidth]{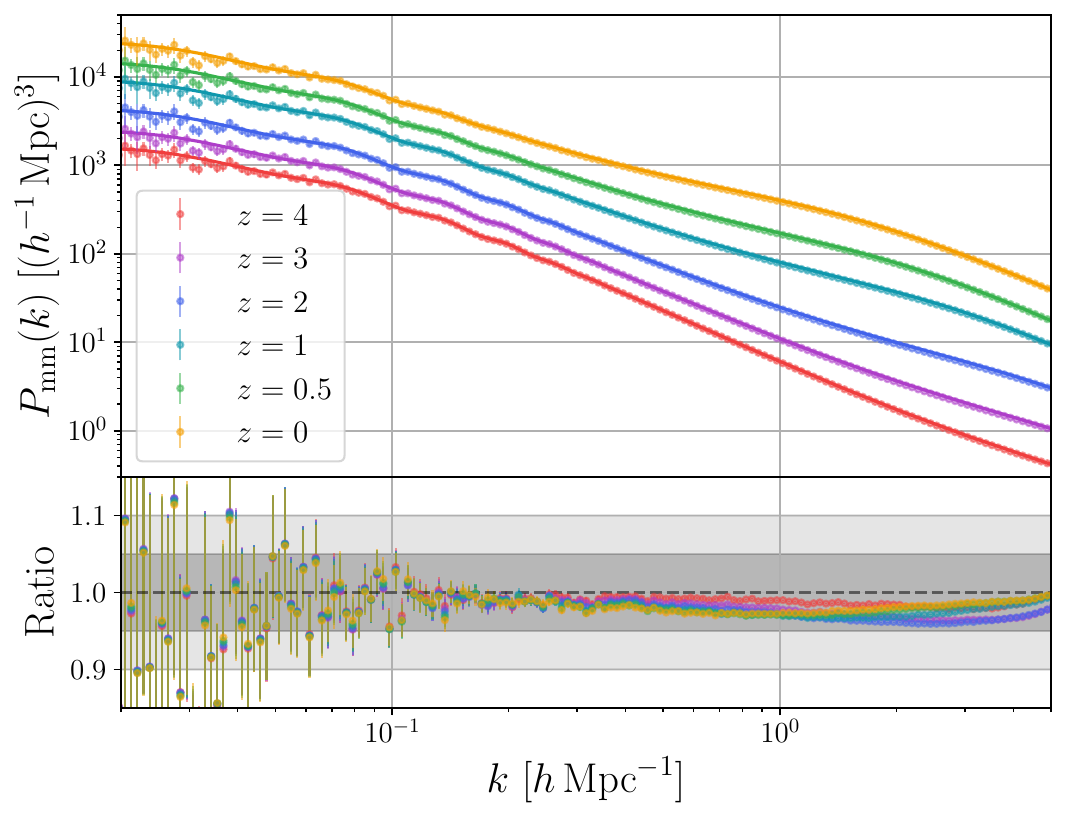}
\caption{The matter power spectrum $P_\mathrm{mm}(k)$ measured from the particle snapshots of the \GION simulation suite (points) at $z = 4, 3, 2, 1, 0.5$, and $0$, compared with the non-linear matter power spectrum computed with \texttt{DarkEmulator2} (solid lines).
The error bars show the statistical uncertainty estimated with Gaussian covariance.
The bottom panel shows the ratio of the measured power spectrum to the theoretical prediction.
The light (dark) shaded region corresponds to 10\% (5\%) deviation from the theoretical prediction.}
\label{fig:pkmm}
\end{figure}

The result is compared with the non-linear matter power spectrum computed with \texttt{DarkEmulator2} \cite{Tanaka2026}, which is a neural network emulator trained on a massive suite of $N$-body simulations, shown by the solid lines in Figure~\ref{fig:pkmm}.
There is one caveat in the comparison: \texttt{DarkEmulator2} is trained on the simulations where the linear neutrino perturbations are incorporated, while the \GION simulation suite treats the massive neutrino species as CDM.
To take into account the effect at the linear level, we correct the matter power spectrum from \texttt{DarkEmulator2} by multiplying the ratio of the linear matter power spectrum with and without massive neutrinos, i.e.,
\begin{multline}
  \label{eq:pkmm_correction}
  P_\mathrm{mm} (k, z) \to P_\mathrm{mm} (k, z; m_\nu = 0 \, \mathrm{eV}) \\
  \times \frac{P_\mathrm{lin}(k, z=0; m_\nu = 0.06 \, \mathrm{eV})}{P_\mathrm{lin} (k, z=0; m_\nu = 0 \, \mathrm{eV})} ,
\end{multline}
where $P_\mathrm{lin}(k, z)$ is the linear matter power spectrum computed with the linear power spectrum emulator in \texttt{DarkEmulator2}.
In these power spectra calculations, the total matter density is fixed to $\Omega_\mathrm{m} = 0.3158$ by adjusting the CDM density to absorb the massive neutrino density.
As shown in the bottom panel of Figure~\ref{fig:pkmm}, the measured power spectrum agrees with the theoretical prediction to within a few percent over the range $k \gtrsim 0.1 \, \hMpcinv$ at all redshifts.
The residual can be attributed to the limited accuracy of the neutrino correction.
On large scales, $k \lesssim 0.1 \, \hMpcinv$, the ratio shows larger scatter due to the limited number of independent Fourier modes available within the finite simulation volume $(1.5 \, \hGpc)^3$, i.e., the sample variance.

\subsection{Halo mass function}
Next, we validate the abundance of dark matter halos identified by \texttt{MPI-Rockstar}.
Similarly to the matter power spectrum, only one of the realizations is used for validation of the halo mass function.
The halos are counted in $80$ logarithmically spaced mass bins in the range of $[10^{10}, 10^{16}] \, \hMsun$.
We also compute the statistical uncertainty of the halo mass function by assuming the halo counts in each mass bin follow the Poisson distribution.
When computing mass functions, only host halos are considered.

\begin{figure}
\includegraphics[width=\columnwidth]{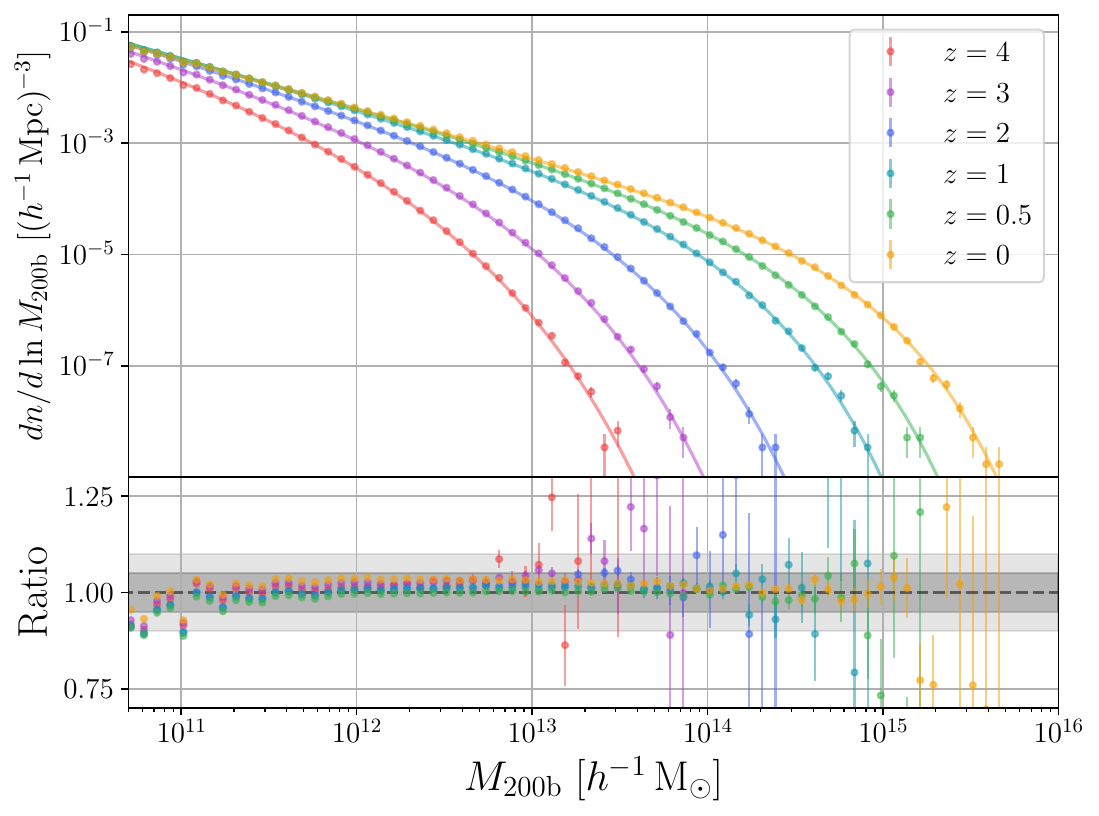}
\caption{The halo mass function $dn/d\ln M_\mathrm{200b}$ measured from the halo catalogs of the \GION simulation suite (points) at $z = 4, 3, 2, 1, 0.5$, and $0$, compared with the fitting function \cite{Fiorilli2026} (solid lines).
The error bars show the statistical uncertainty estimated with Poisson noise.
The bottom panel shows the ratio of the measured halo mass function to the fitting function.
The light (dark) shaded region corresponds to 10\% (5\%) deviation from the theoretical prediction.}
\label{fig:mfunc}
\end{figure}

Motivated by the pioneering work of Ref.~\cite{Press1974}, the halo mass function is parameterized in terms of the peak height $\nu \equiv \delta_\mathrm{cr}/\sigma (M, z)$:
\begin{equation}
  \label{eq:mfunc}
  \frac{dn}{d\ln M} = f(\nu) \, \frac{\bar\rho_\mathrm{m}}{M} \, \frac{d\ln\nu}{d\ln M} ,
\end{equation}
where $\delta_\mathrm{cr}=3/5 (3\pi /2)^{2/3}\simeq 1.686$ is the overdensity threshold in spherical collapse in the Einstein--de Sitter (EdS) Universe \footnote{In the $\Lambda$CDM Universe, the difference of the threshold from the EdS one is at the sub-percent level \cite{Kitayama1996}.
We adopt the EdS value throughout this paper.}, $f(\nu)$ is the halo multiplicity function and $\sigma(M, z)$ is the root-mean-square fluctuation of the linear matter density field smoothed with a real-space top-hat filter $W$ enclosing the mass $M$ at Lagrangian radius $R = (3M / 4\pi\bar\rho_\mathrm{m})^{1/3}$,
\begin{equation}
  \label{eq:sigmaM}
  \sigma^2(M, z) = \int_0^\infty d \ln k \, \frac{k^3 P_\mathrm{lin}(k, z)}{2\pi^2} \, |W(kR)|^2 ,
\end{equation}
with $P_\mathrm{lin}(k, z)$ the linear matter power spectrum computed with \texttt{CLASS} and $W(x) = 3(\sin x - x\cos x)/x^3$ the Fourier transform of the top-hat filter.
Here, we adopt the fitting formula calibrated with $N$-body simulations presented in Ref.~\cite{Fiorilli2026}.
Figure~\ref{fig:mfunc} shows the halo mass function $dn/d\ln M_\mathrm{200b}$ as a function of $M_\mathrm{200b}$ at $z = 4, 3, 2, 1, 0.5$, and $0$, together with the prediction of the fitting formula.
The measured halo mass function agrees with the fitting formula to within $5\%$ for $2 \times 10^{11} \, \hMsun \lesssim M_\mathrm{200b} \lesssim 10^{13} \, \hMsun$ at all redshifts.
In the lowest mass bin, $M_\mathrm{200b} \simeq 10^{11} \, \hMsun$, the measurement falls $\sim 10\%$ below the prediction, which we attribute to the finite mass resolution.
At the high-mass end, the ratio for the high-redshift samples ($z = 3, 4$) systematically exceeds unity by $10$--$25\%$, while for the lower-redshift samples the deviations scatter around unity.
We attribute the former to the limited accuracy of the fitting function outside the mass and redshift range used for its calibration, and the latter to the Poisson noise associated with the rarity of massive halos.
The mass resolution of the \GION simulation suite, $m_\mathrm{part} = 1.28 \times 10^9 \, \hMsun$, allows us to resolve halos down to $M_\mathrm{200b} \sim 10^{11} \, \hMsun$ with $\sim 80$ particles, which is sufficient to capture the relevant halo population for the Stage-IV cosmological surveys.

\subsection{Halo power spectrum}
Finally, we validate the clustering of halos as a function of halo mass by measuring the halo--matter cross power spectrum $P_\mathrm{hm}(k)$ and the halo--halo auto power spectrum $P_\mathrm{hh}(k)$ in eight mass bins spanning $11.0 \le \log_{10}(M_\mathrm{200b}/\hMsun) < 15.0$ in logarithmically equal steps of $0.5 \, \mathrm{dex}$, at $z = 4, 3, 2, 1, 0.5$, and $0$.
The halo number density field is constructed with the same grid size and assignment scheme as the matter density field, and the power spectra are measured with the same procedure as for the matter power spectrum.
The shot noise contribution $P_\mathrm{shot} = 1/\bar{n}_\mathrm{h}$ is subtracted when measuring the halo--halo auto power spectrum, where $\bar{n}_\mathrm{h}$ is the number density of halos in the mass bin.
As an uncertainty estimate, we adopt the Gaussian covariance for the halo--matter cross power spectrum and the halo--halo auto power spectrum:
\begin{multline}
  \mathrm{Cov} [P_\mathrm{hm} (k_i), P_\mathrm{hm} (k_j)]
  = \delta_{ij} \frac{1}{N_\mathrm{mode}(k_i)} \\
  \times \left[ P_\mathrm{hm}^2(k_i) + P_\mathrm{mm}(k_i)
  (P_\mathrm{hh}(k_i)+ P_\mathrm{shot}) \right] ,
\end{multline}
\begin{multline}
  \mathrm{Cov} [P_\mathrm{hh} (k_i), P_\mathrm{hh} (k_j)]
  = \delta_{ij} \frac{2}{N_\mathrm{mode}(k_i)} \\
  \times \left[ P_\mathrm{hh}(k_i) + P_\mathrm{shot} \right]^2 .
\end{multline}
The most massive bins are progressively displayed only at lower redshifts \footnote{The halo power spectra are not measured for the samples with less than $100$ halos because the scarcity leads to a noisy measurement.}, reflecting the hierarchical growth of structure and the limited volume available for the rarest halos at high redshift.

The measured power spectra are compared with the linear halo bias model,
\begin{equation}
  \label{eq:linear_bias_model}
  P_\mathrm{hm}(k, z) = \bar b_\mathrm{h}(z) \, P_\mathrm{lin}(k, z) , \,
  P_\mathrm{hh}(k, z) = \bar b_\mathrm{h}^2(z) \, P_\mathrm{lin}(k, z) ,
\end{equation}
where $\bar b_\mathrm{h}(z)$ is the effective linear halo bias of the sample.
For a mass bin $[M_1, M_2]$, $\bar b_\mathrm{h}(z)$ is the number-weighted average of the halo bias $b_\mathrm{h}(M, z)$,
\begin{equation}
  \label{eq:bias_avg}
  \bar b_\mathrm{h}(z) = \frac{\displaystyle\int_{M_1}^{M_2} dM \, \frac{dn}{dM} \, b_\mathrm{h}(M, z)}{\displaystyle\int_{M_1}^{M_2} dM \, \frac{dn}{dM}} .
\end{equation}
We adopt the fitting function of the halo bias of Ref.~\cite{Comparat2017}, which is a function of the peak height calibrated with $N$-body simulations.

\begin{figure*}
\includegraphics[width=\textwidth]{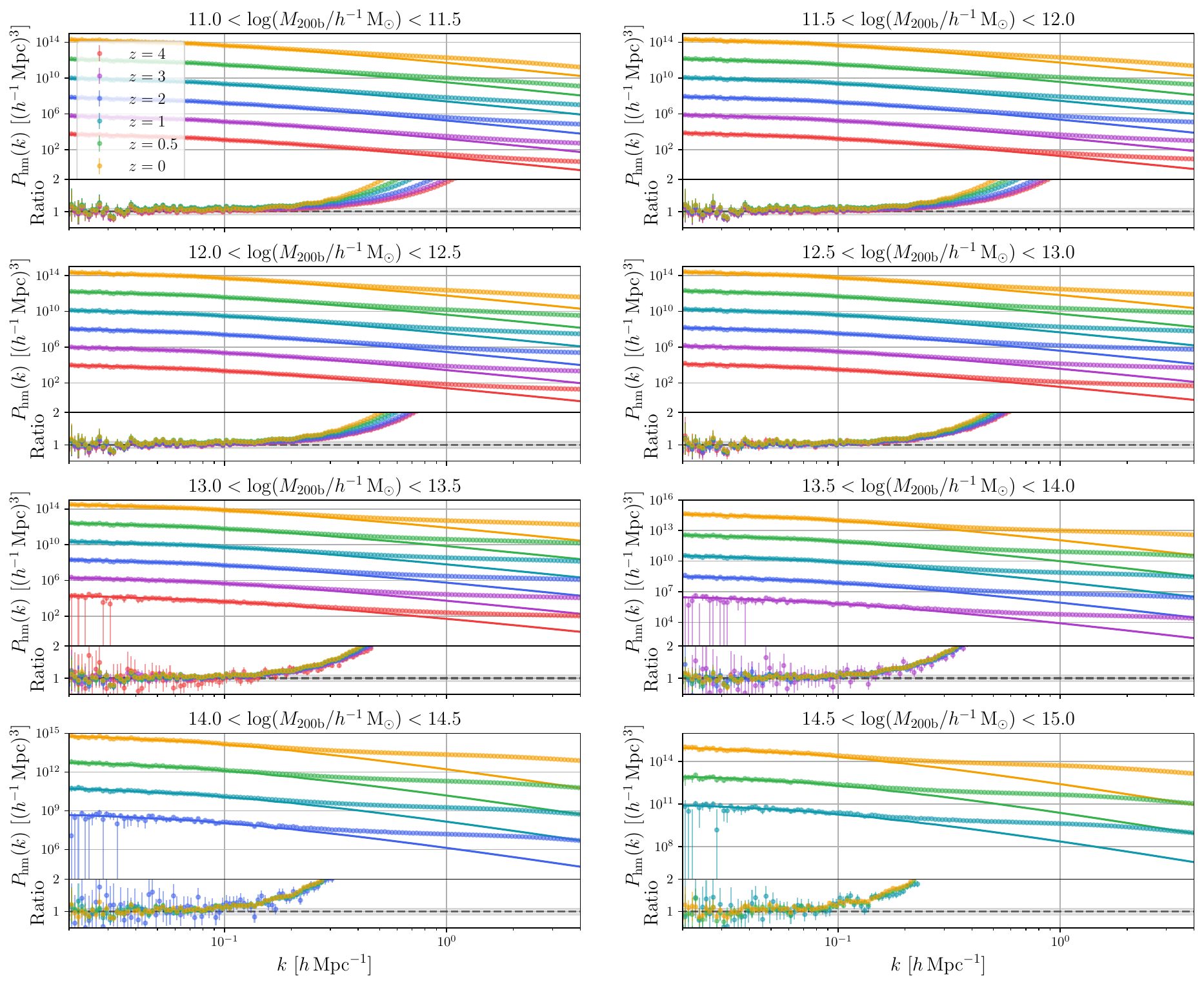}
\caption{The halo--matter cross power spectrum $P_\mathrm{hm}(k, z)$ measured from the \GION simulation suite (points) in eight mass bins of $M_\mathrm{200b}$ at $z = 4, 3, 2, 1, 0.5$, and $0$, compared with the linear halo bias model $P_\mathrm{hm}(k, z) = \bar{b}_\mathrm{h}(z) \, P_\mathrm{lin}(k, z)$ (solid lines).
The error bars show the statistical uncertainty estimated with Gaussian covariance.
The bottom panel of each mass bin shows the ratio of the measured power spectrum to the model.
The shaded region corresponds to 10\% deviation from the theoretical prediction.
For visual clarity, the amplitude is shifted by a factor of $100^n$ for the $n$-th redshift, with $n = 0, 1, \dots, 5$ from the highest ($z=4$) to the lowest ($z=0$) redshift.}
\label{fig:pkhm}
\end{figure*}

\begin{figure*}
\includegraphics[width=\textwidth]{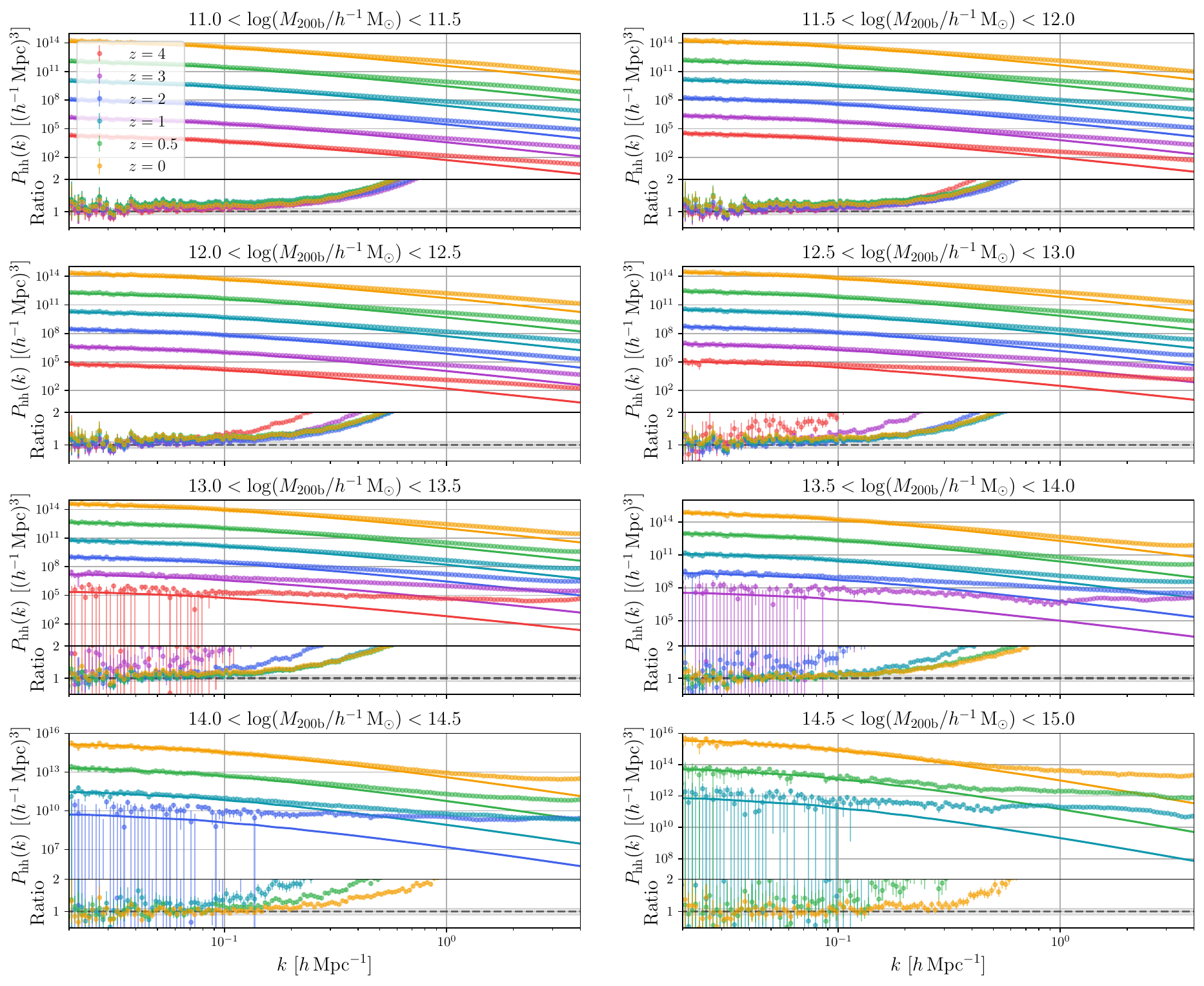}
\caption{Same as Figure~\ref{fig:pkhm}, but for the halo--halo auto power spectrum $P_\mathrm{hh}(k, z)$, compared with the linear halo bias model $P_\mathrm{hh}(k, z) = \bar{b}_\mathrm{h}(z)^2 \, P_\mathrm{lin}(k, z)$ (solid lines).
Similarly to the halo--matter cross-spectrum, the amplitude is shifted with respect to the redshift.
The shot noise is subtracted from the measured power spectrum.}
\label{fig:pkhh}
\end{figure*}

Figures~\ref{fig:pkhm} and \ref{fig:pkhh} show the halo--matter cross-spectrum and halo--halo auto-spectrum, respectively.
The linear bias model reproduces the measured power spectra to within $\sim 10\%$ on large scales, $k \lesssim 0.1 \, h \, \mathrm{Mpc}^{-1}$, for all mass bins and redshifts, except for the rarest samples at high redshift (e.g., $13.5 < \log_{10}(M_\mathrm{200b}/\hMsun) < 14.0$ at $z = 3$), for which the measurement is noisy and the ratio already exceeds unity by a few tens of percent at $k \sim 0.1 \, \hMpcinv$.
On smaller scales, the ratio between the measured and predicted power spectra increases monotonically with $k$, doubling at $k \sim 0.2 \text{--} 1 \, h \, \mathrm{Mpc}^{-1}$, reflecting the well-known scale dependence of the halo bias and the breakdown of the linear bias approximation in the non-linear regime \cite{Mead2021}.
This scale dependence sets in at progressively larger scales (i.e., smaller $k$) for more massive halos, consistent with the larger Lagrangian radius of massive halos.
For the most massive bins, where the number of halos within the simulation volume is small, the measured power spectra show larger scatter, particularly for the auto-power spectrum, which is additionally affected by the shot noise associated with the discreteness of the halo distribution.

\section{Lightcone outputs}
\label{sec:lightcone_outputs}
In this section, we present the map making algorithms and validation of the lightcone Level 1 and Level 2 DPs introduced in Section~\ref{sec:numerical_methods}, which extend over the full sky from $z = 0$ to $z = 5$.
We first validate the lightcone halo catalog itself by comparing the redshift evolution of the halo number counts with theoretical predictions.
We then demonstrate the construction of four representative LSS observables from the Level 1 DPs: weak lensing convergence maps, a mock galaxy catalog, and tSZ and kSZ maps.
For each observable, we compute the angular power spectrum and compare it with the corresponding theoretical prediction, averaged over the $10$ realizations of the \GION suite.
We also consider the cross-power spectrum between the weak lensing and tSZ maps as a representative example of cross-correlation between different observables.

\subsection{Lightcone halo number counts}
\begin{figure}
\includegraphics[width=\columnwidth]{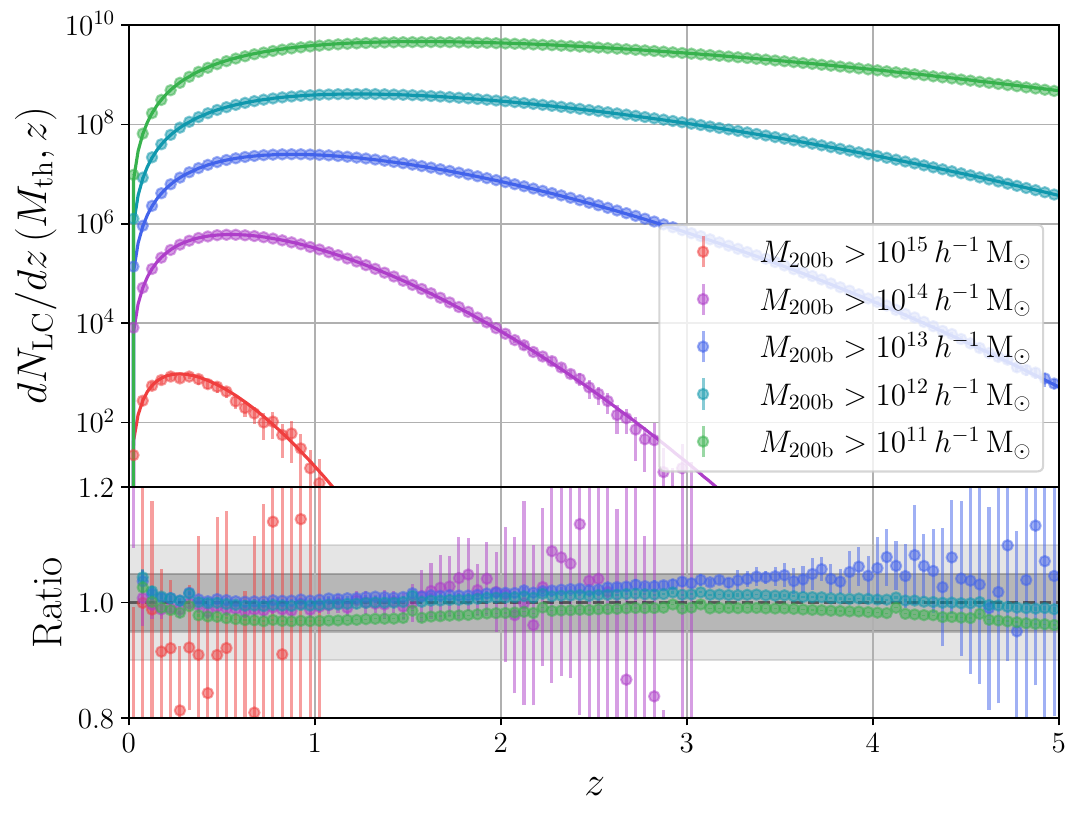}
\caption{The differential number counts $dN_\mathrm{LC}/dz$ of halos in the lightcone of the \GION simulation suite (points) above five mass thresholds with the error bars estimated as the standard deviation of the $10$ realizations, compared with the prediction obtained by integrating the halo mass function over the full-sky comoving volume element (solid lines).
The bottom panel shows the ratio of the measured to the predicted number counts.
The light (dark) shaded region corresponds to 10\% (5\%) deviation from the theoretical prediction.}
\label{fig:lc_number}
\end{figure}

\begin{figure}
\includegraphics[width=\columnwidth]{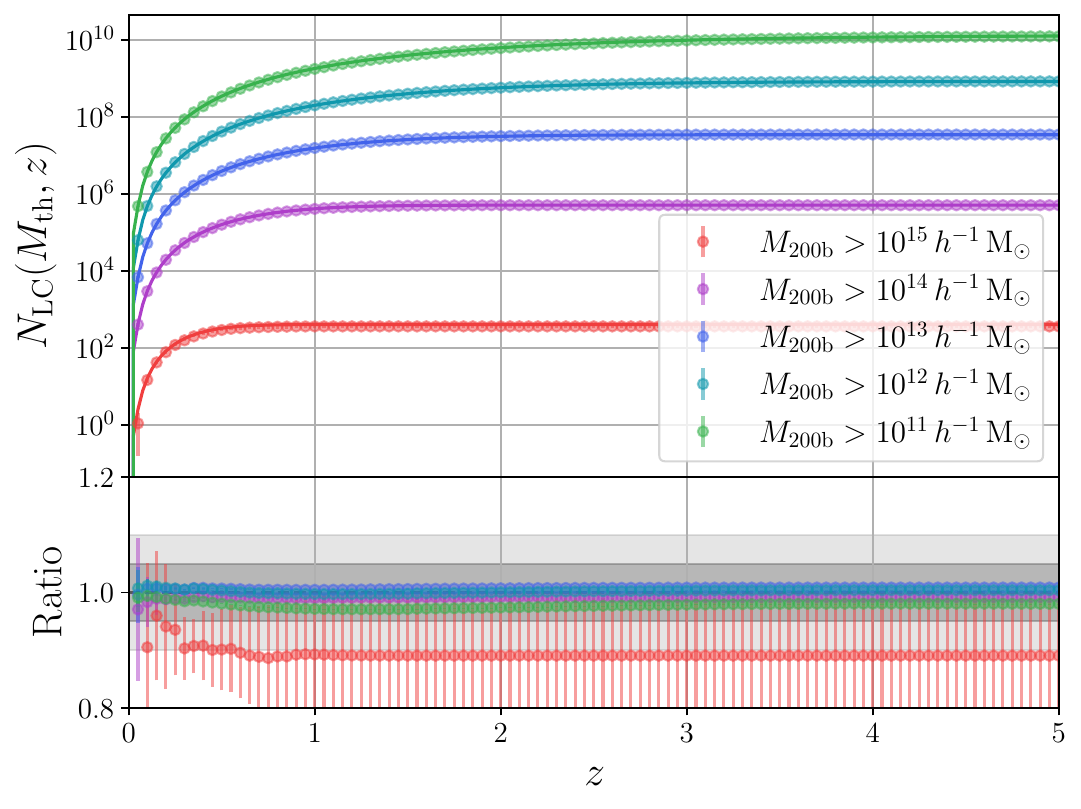}
\caption{Same as Figure~\ref{fig:lc_number}, but for the cumulative number counts $N_\mathrm{LC}$.}
\label{fig:lc_number_cumul}
\end{figure}

We first validate the lightcone halo catalog by counting the number of halos above five mass thresholds, $M_\mathrm{200b} > 10^{11}, 10^{12}, 10^{13}, 10^{14}$, and $10^{15} \, \hMsun$, as a function of redshift.
Here, only host halos are selected for consistency with the halo mass function.
Figure~\ref{fig:lc_number} shows the differential number counts $dN_\mathrm{LC}/dz$, and Figure~\ref{fig:lc_number_cumul} shows the cumulative number counts $N_\mathrm{LC}$, both measured from the lightcone halo catalog and compared with the fitting formula obtained by integrating the halo mass function \cite{Fiorilli2026}, shown in Figure~\ref{fig:mfunc}, over the full-sky comoving volume.
For a mass threshold $M_\mathrm{th}$, the differential number count is
\begin{equation}
  \label{eq:dNdz}
  \frac{dN_\mathrm{LC}}{dz} (M_\mathrm{th}, z) = \frac{dV}{dz} \int_{M_\mathrm{th}}^{\infty} dM \, \frac{dn}{dM}(M, z) ,
\end{equation}
and the cumulative count is given as
\begin{equation}
  N_\mathrm{LC}(M_\mathrm{th}, z) = \int_0^z dz' \, \frac{dN_\mathrm{LC}}{dz} (M_\mathrm{th}, z') .
\end{equation}
The full-sky comoving volume element per unit redshift is
\begin{equation}
  \label{eq:dVdz}
  \frac{dV}{dz} = 4\pi \, \chi^2(z) \, \frac{c}{H(z)} ,
\end{equation}
with $\chi(z)$ the comoving distance, $H(z)$ the Hubble parameter, and $c$ the speed of light.

For the three lowest mass thresholds, $M_\mathrm{200b} > 10^{11}, 10^{12}$, and $10^{13} \, \hMsun$, the measured $dN_\mathrm{LC}/dz$ agrees with the theoretical prediction to within $5\%$ over the entire redshift range, apart from the noisy measurements at $z \gtrsim 3.5$ for the $10^{13} \, \hMsun$ threshold, which scatter around the prediction within the statistical uncertainty.
The counts above $10^{11} \, \hMsun$ are $2$--$4\%$ lower than the prediction, reflecting the deficit in the lowest mass bins of the halo mass function, while those above $10^{13} \, \hMsun$ show a few-percent excess at $z \gtrsim 3$, consistent with the excess at the high-mass end of the halo mass function at $z = 3, 4$ (Figure~\ref{fig:mfunc}).
For $M_\mathrm{200b} > 10^{14} \, \hMsun$, the agreement is at the percent level for $z \lesssim 1.5$, beyond which the counts become noisy due to the rarity of these massive halos.
For $M_\mathrm{200b} > 10^{15} \, \hMsun$, the number counts are dominated by Poisson noise, and the measured counts are consistent with the theoretical prediction within the statistical uncertainty over the full redshift range.
The cumulative number counts in Figure~\ref{fig:lc_number_cumul} agree with the prediction to within $\sim 2\%$ over the entire redshift range for the thresholds $10^{12}$--$10^{14} \, \hMsun$, while those above $10^{11} \, \hMsun$ are $2$--$3\%$ lower than the prediction, consistent with the deficit in the differential counts.
For $M_\mathrm{200b} > 10^{15} \, \hMsun$, the cumulative counts converge to $\sim 10\%$ below the prediction at $z \gtrsim 0.7$, which is within the statistical uncertainty.
These results demonstrate that the on-the-fly lightcone construction and the lightcone halo finding with \texttt{MPI-Rockstar} reproduce the expected halo abundance and its evolution with redshift.

\subsection{Weak lensing}
As a first example, we generate weak lensing convergence maps using the lightcone density shells.
For a source at redshift $z_\mathrm{s}$, the convergence field $\kappa (\bm{\theta}; z_\mathrm{s})$ is computed in the Born approximation by integrating the density field weighted by the lensing kernel:
\begin{equation}
  \label{eq:kappa_born}
  \kappa (\bm{\theta}; z_\mathrm{s}) = \int_0^{\chi_\mathrm{s}} d\chi \, W_\kappa(\chi, \chi_\mathrm{s}) \, \delta(\chi, \bm{\theta}, z(\chi)) ,
\end{equation}
where $\bm{\theta}$ is the angular position on the sky, $\chi_\mathrm{s} = \chi(z_\mathrm{s})$ is the comoving distance to the source and the lensing kernel is given as
\begin{equation}
  \label{eq:lensing_kernel}
  W_\kappa(\chi, \chi_\mathrm{s}) = \frac{3 \Omega_\mathrm{m} H_0^2}{2c^2} \frac{\chi (\chi_\mathrm{s} - \chi)}{a(\chi) \chi_\mathrm{s}} ,
\end{equation}
with $a = 1/(1 + z)$ the scale factor and $H_0$ the present-day Hubble parameter.

\subsubsection{Map making algorithm}
The lensing convergence field is computed from the lightcone density shells by discretizing the integral in Eq.~\eqref{eq:kappa_born} into a sum over the shells.
The direct output from \texttt{assign\_healpix} is the projected mass of the $i$-th shell of thickness $\Delta\chi = 25 \, \hMpc$,
\begin{equation}
  M^{(i)} (\bm{\theta}) = \sum_{p \in \mathcal{P}^{(i)} (\bm{\theta})} m_\mathrm{part} ,
  \label{eq:mass_shell}
\end{equation}
where $\mathcal{P}^{(i)} (\bm{\theta})$ denotes the set of particles in the $i$-th shell whose angular positions fall within the \texttt{HEALPix} pixel with the angular position $\bm{\theta}$, i.e., $M^{(i)} (\bm{\theta})$ is the particle count in the pixel multiplied by the particle mass $m_\mathrm{part}$.
The maps are constructed in \texttt{HEALPix} format with the resolution parameter $N_\mathrm{side} = 8192$.
Then, the angular surface mass density field $\Sigma^{(i)} (\bm{\theta})$ is given as $\Sigma^{(i)} (\bm{\theta}) = M^{(i)} (\bm{\theta}) / A_\mathrm{pix}$, where $A_\mathrm{pix} = 4\pi / N_\mathrm{pix}$ is the area of a pixel and $N_\mathrm{pix} = 12 N_\mathrm{side}^2$ is the total number of pixels.
Finally, the convergence field is expressed as \cite{Ferlito2023}
\begin{align}
  \kappa (\bm{\theta}; z_\mathrm{s}) = \frac{4 \pi G}{c^2} \frac{1}{\chi_\mathrm{s}}
  \sum_i & \frac{(\chi_\mathrm{s} - \chi^{(i)})}{\chi^{(i)} a^{(i)}}
  \nonumber \\
  & \times \left[ \Sigma^{(i)} (\bm{\theta}) - \bar{\Sigma}^{(i)} \right],
\end{align}
where $G$ is the gravitational constant and $\bar{\Sigma}^{(i)}$ is the mean surface mass density of the $i$-th shell.
The summation runs over the shells before the source redshift.
The comoving distance of the $i$-th shell $\chi^{(i)}$ is the cone-volume-weighted distance of the $i$-th shell \cite{Shirasaki2015},
\begin{equation}
  \label{eq:chi_shell}
  \chi^{(i)} = \frac{3}{4} \frac{(\chi^{(i)}_\mathrm{max})^4 - (\chi^{(i)}_\mathrm{min})^4}{(\chi^{(i)}_\mathrm{max})^3 - (\chi^{(i)}_\mathrm{min})^3},
\end{equation}
where $\chi^{(i)}_\mathrm{min}$ and $\chi^{(i)}_\mathrm{max}$ are the comoving distances to the near and far edges of the shell, respectively, and the scale factor $a^{(i)} = a(z(\chi^{(i)}))$.
In practical implementation, there are $216$ shells extending up to the comoving distance of $5400 \, \hMpc$.
Then, we generate $216$ convergence maps, for each of which the density field is integrated up to the edge of the shells.
For validation, we use convergence maps with five source redshifts, $z_\mathrm{s} = 0.5, 1, 2, 3$, and $4$, which are constructed by summing over the shells up to the 53rd, 92nd, 144th, 176th, and 198th shells, respectively.

For validation of generated convergence maps, we compute the angular auto-power spectrum of the convergence field $C_{\kappa \kappa} (\ell)$ and compare it with the theoretical prediction.
The angular power spectrum is defined as
\begin{equation}
  \label{eq:cl_def}
  \langle \tilde\kappa(\bm{\ell}) \tilde\kappa(\bm{\ell}') \rangle = (2\pi)^2 \delta_\mathrm{D}(\bm{\ell} + \bm{\ell}') \, C_{\kappa \kappa}(\ell),
\end{equation}
where $\tilde\kappa(\bm{\ell})$ is the Fourier transform of $\kappa(\bm{\theta})$ and $\bm{\ell}$ is the angular wave-vector.
We compute the angular power spectrum from the convergence maps using the \texttt{anafast} routine in \texttt{HEALPix} and deconvolve the pixel window function.
The measured angular power spectrum is binned into $100$ multipole bins, whose edges $\ell_i$ ($i = 0, 1, \dots, 100$) are the $101$ logarithmically equally spaced values between $\ell_\mathrm{min} = 1$ and $\ell_\mathrm{max} = 8192$ truncated to integers, and the $i$-th bin covers the half-open interval $\ell_i \leq \ell < \ell_{i+1}$.
Within each bin, the representative multipole and the binned power spectrum are computed as the mode-number weighted averages:
\begin{align}
  \label{eq:cl_binning}
  \bar{\ell}_i &= \frac{\sum_{\ell \in b_i} (2\ell + 1) \, \ell}{\sum_{\ell \in b_i} (2\ell + 1)}, \\
  \bar{C}_{\kappa \kappa} (\bar{\ell}_i) &= \frac{\sum_{\ell \in b_i} (2\ell + 1) \, C_{\kappa \kappa} (\ell)}{\sum_{\ell \in b_i} (2\ell + 1)},
\end{align}
where the sums run over the integer multipoles belonging to the $i$-th bin, $b_i = \{ \ell \, | \, \ell_i \leq \ell < \ell_{i+1} \}$, and the weight $2\ell + 1$ is the number of modes at a given multipole $\ell$.
The binned spectra are averaged over the $10$ realizations of the \GION simulation suite.
The error bars are estimated as the standard deviation of the $10$ realizations.

\subsubsection{Validation of convergence angular power spectrum}
The convergence field is a weighted projection of the matter density field along the line-of-sight, and under the Limber approximation \cite{Limber1953, LoVerde2008}, its angular power spectrum is given as
\begin{equation}
  \label{eq:cl_kappa_limber}
  C_{\kappa \kappa}(\ell) = \int_0^{\chi_\mathrm{s}} \frac{d\chi}{\chi^2} \, W_\kappa^2(\chi, \chi_\mathrm{s}) \, P_\mathrm{mm} \left( k = \frac{\ell + 1/2}{\chi}, z(\chi) \right) ,
\end{equation}
where $P_\mathrm{mm} (k, z)$ is the non-linear matter power spectrum.
The Limber approximation is accurate at small scales (i.e., large $\ell$) but breaks down at the largest angular scales.
Thus, we compute the theoretical prediction of the convergence angular power spectrum using the non-Limber expression (Eq.~\ref{eq:cl_kappa_nonlimber}) for $\ell < 100$ and Eq.~\eqref{eq:cl_kappa_limber} for $\ell \geq 100$ with the non-linear matter power spectrum obtained with \texttt{DarkEmulator2}.
The non-Limber expression and the accuracy of the Limber approximation are described in detail in Appendix~\ref{sec:exact_model}.
Figure~\ref{fig:cl_kappa} shows the resulting convergence angular power spectrum for five source redshifts, $z_\mathrm{s} = 0.5, 1, 2, 3$, and $4$, together with the theoretical prediction.
The measured and predicted power spectra agree to within a few percent over the range $30 \lesssim \ell \lesssim 5000$ for all source redshifts, and at $\ell \lesssim 30$, the measurements are consistent with the prediction within the statistical uncertainty, demonstrating that the density shells produced by \texttt{assign\_healpix} provide an accurate representation of the projected matter distribution for weak lensing applications.

\begin{figure*}
\includegraphics[width=0.9\textwidth]{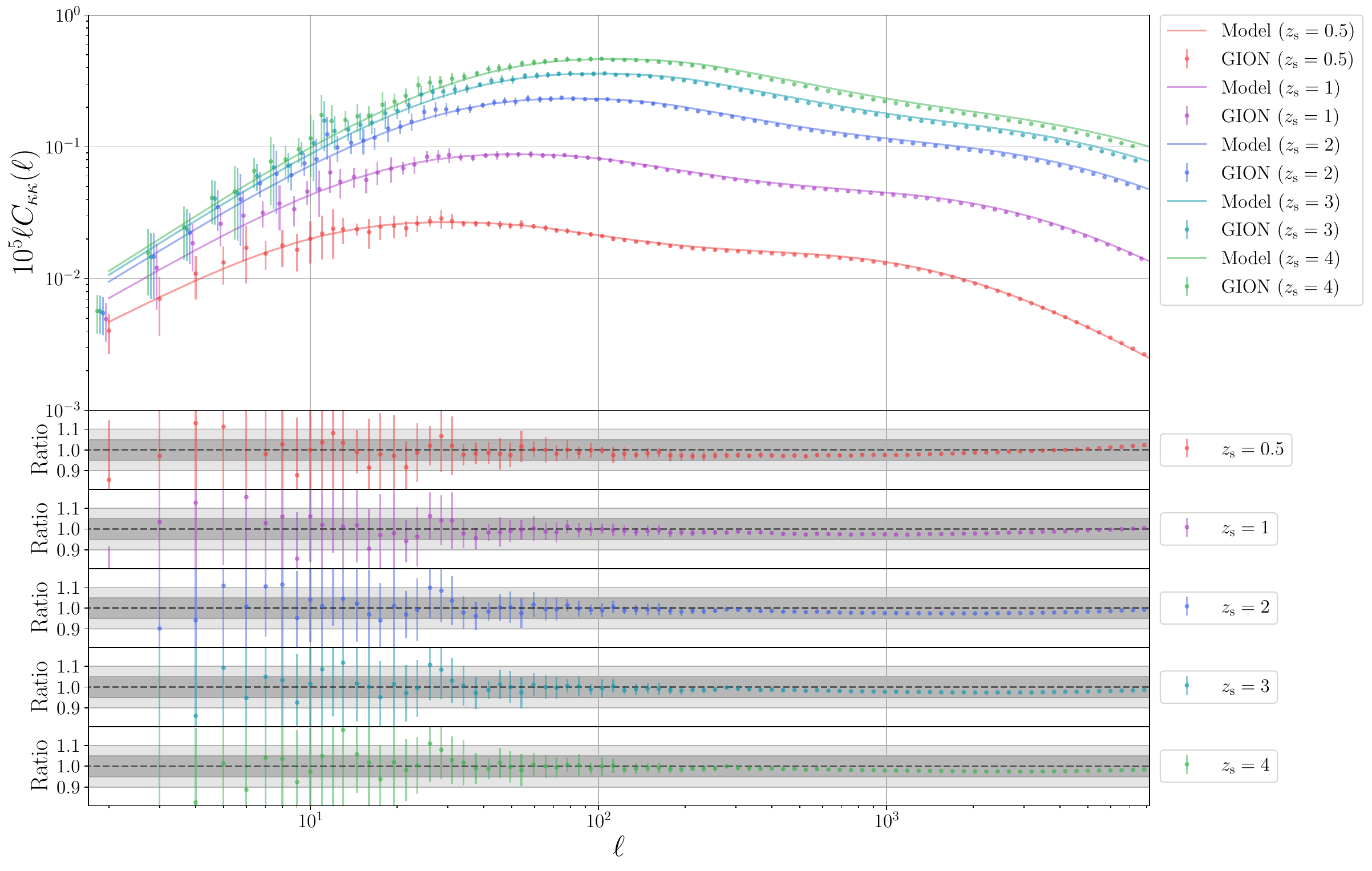}
\caption{The weak lensing convergence angular power spectrum $C_{\kappa \kappa}(\ell)$ measured from the lightcone density shells of the \GION simulation suite (points) for source redshifts $z_\mathrm{s} = 0.5, 1, 2, 3$, and $4$, compared with the theoretical prediction (solid lines).
The points are slightly shifted along the horizontal axis for visual clarity.
The bottom panels show the ratio of the measured to the predicted angular power spectra for each source redshift.
The light (dark) shaded region corresponds to 10\% (5\%) deviation from the theoretical prediction.}
\label{fig:cl_kappa}
\end{figure*}

\subsection{Galaxy clustering}
As a second example of Level 2 DPs, we construct a mock galaxy catalog from the lightcone halo catalog.
Populating halos with galaxies requires galaxy--halo connection modeling \cite{Wechsler2018} such as halo occupation distribution \cite{Seljak2000} or subhalo abundance matching \cite{Vale2004}.
Here, the objective is to validate the lightcone halo catalog mimicking the observed galaxy distributions in terms of their redshift and halo mass distributions.
Thus, we employ the simplest approach: one-to-one mapping between halos and galaxies.
We consider galaxy samples in the DESI survey: luminous red galaxies (LRGs) and emission line galaxies (ELGs).
The former is hosted by relatively older and massive halos, and in contrast, the latter is hosted by younger and less massive halos.
Following DESI samples \cite{DESI-DR1-2026}, we define three LRG samples, LRG1 ($0.4 < z < 0.6$), LRG2 ($0.6 < z < 0.8$), and LRG3 ($0.8 < z < 1.1$), and two ELG samples, ELG1 ($0.8 < z < 1.1$) and ELG2 ($1.1 < z < 1.6$), and after selecting halos in the redshift range, we further select halos in the halo mass range, $M_\mathrm{vir} \in [10^{12.5}, 10^{13.5}] \, \hMsun$ for LRGs and $M_\mathrm{vir} \in [10^{11.5}, 10^{12.5}] \, \hMsun$ for ELGs \cite{Favole2026}.
Note that galaxies are assigned only to host halos, and subhalos are excluded, and the redshift space distortion effect is not included.
Table~\ref{tab:galaxy_samples} summarizes the properties of the resulting galaxy samples.
The resulting redshift distributions are shown in Figure~\ref{fig:dNdz_halo}.

\begin{table*}[htbp]
    \caption{Properties of the galaxy samples constructed from the lightcone halo catalog.
    The redshift and halo mass ranges are chosen to mimic the observed DESI galaxy samples.}
    \label{tab:galaxy_samples}
    \begin{ruledtabular}
        \begin{tabular}{ccccc}
            Sample & Redshift & Halo mass & Number \footnotemark[1] & Comoving density $[(\hMpc)^{-3}]$ \footnotemark[1] \\
            \hline
            LRG1 & $0.4 < z < 0.6$ & $10^{12.5} < M_\mathrm{vir}/(\hMsun) < 10^{13.5}$ & $1.198 \times 10^6$ & $1.216 \times 10^{-4}$ \\
            LRG2 & $0.6 < z < 0.8$ & $10^{12.5} < M_\mathrm{vir}/(\hMsun) < 10^{13.5}$ & $1.714 \times 10^6$ & $1.123 \times 10^{-4}$ \\
            LRG3 & $0.8 < z < 1.1$ & $10^{12.5} < M_\mathrm{vir}/(\hMsun) < 10^{13.5}$ & $3.088 \times 10^6$ & $9.804 \times 10^{-5}$ \\
            ELG1 & $0.8 < z < 1.1$ & $10^{11.5} < M_\mathrm{vir}/(\hMsun) < 10^{12.5}$ & $3.358 \times 10^7$ & $1.066 \times 10^{-3}$ \\
            ELG2 & $1.1 < z < 1.6$ & $10^{11.5} < M_\mathrm{vir}/(\hMsun) < 10^{12.5}$ & $6.650 \times 10^7$ & $9.767 \times 10^{-4}$ \\
        \end{tabular}
    \end{ruledtabular}
    \footnotetext[1]{The values are the mean of $10$ realizations of the \GION simulation suite.}
\end{table*}

\subsubsection{Map making algorithm}
For each galaxy sample, we first compute the angular auto-power spectrum of the halo number density field $C_{\mathrm{hh}}(\ell)$.
To construct the halo number density field, we pixelate the halo number counts in \texttt{HEALPix} format with $N_\mathrm{side} = 8192$ and compute the number density contrast $\delta^\mathrm{2D}_\mathrm{h} (\bm{\theta}) = n^{\mathrm{2D}}_\mathrm{h} / \bar{n}^{\mathrm{2D}}_\mathrm{h} - 1$, where $n^{\mathrm{2D}}_\mathrm{h}$ is the halo number density in each pixel and $\bar{n}^{\mathrm{2D}}_\mathrm{h}$ is the mean number density of the sample.
We then compute the power spectrum using the \texttt{anafast} algorithm with the pixel window function deconvolved, and bin the power spectrum with the same configuration as the convergence power spectrum.
Next, we also compute the cross-power spectrum between the halo number density field and the lensing convergence field $C_{\mathrm{h}\kappa}(\ell)$.
Since the halo number density field is correlated with the underlying matter distribution, the non-zero cross-power spectrum with the convergence field is expected.
In observations, the cross-power spectrum corresponds to galaxy--galaxy lensing, which is a powerful probe of the galaxy--halo connection and the underlying matter distribution.
The cross-power spectrum is computed in a similar manner to the auto-power spectrum, using the same \texttt{HEALPix} maps for the halo number density field and the convergence field.

\subsubsection{Theoretical prediction}
Here, we describe the theoretical prediction of the angular power spectra based on halo model formalism \cite{Cooray2002, Asgari2023}, where all matter is associated with halos and various statistics are computed from halo clustering.
We begin with the three-dimensional halo--halo auto-spectra and halo--matter cross-spectra, $P_\mathrm{hh}(k, z)$ and $P_\mathrm{hm}(k, z)$, which are then projected along the line-of-sight to obtain the angular power spectra.
The three-dimensional power spectra are expressed as the sum of clustering within halos (1-halo term) and clustering between different halos (2-halo term).
For halo--halo auto-spectra, the 1-halo term is the shot noise associated with the discreteness of the halo distribution, and the 2-halo term is the linear power spectrum multiplied by the square of the effective halo bias since on large scales, the halos are linearly biased tracers of the underlying matter distribution.
As a result, the halo--halo auto-spectrum is expressed as
\begin{align}
  \label{eq:pkhh_halomodel}
  P_\mathrm{hh} (k, z) &= P_\mathrm{hh}^\mathrm{1h} (k, z) + P_\mathrm{hh}^\mathrm{2h} (k, z), \\
  P_\mathrm{hh}^\mathrm{1h} (k, z) &= \frac{1}{\bar n_\mathrm{h}}, \\
  P_\mathrm{hh}^\mathrm{2h} (k, z) &= \bar b_\mathrm{h}^2 \, P_\mathrm{lin}(k, z) ,
\end{align}
where $\bar n_\mathrm{h}(z) = \int_{M_1}^{M_2} dM \, (dn/dM)$ is the sample number density and $\bar b_\mathrm{h}(z)$ is the effective bias from Eq.~\eqref{eq:bias_avg}.
For the halo--matter cross-spectrum, the 1-halo term is the average matter density profile of halos in the sample, and the 2-halo term is the large-scale cross-spectrum between halos and matter, which is also described by the linear bias prescription.
Similarly, the halo--matter cross-spectrum is expressed as
\begin{align}
  \label{eq:pkhm_halomodel}
  P_\mathrm{hm}(k, z) &= P_\mathrm{hm}^\mathrm{1h} (k, z) + P_\mathrm{hm}^\mathrm{2h} (k, z) , \\
  P_\mathrm{hm}^\mathrm{1h} (k, z) &= \frac{1}{\bar n_\mathrm{h}} \int_{M_1}^{M_2} dM \, \frac{dn}{dM} \frac{M}{\bar\rho_\mathrm{m}} \, u(k ; M, z) , \\
  P_\mathrm{hm}^\mathrm{2h} (k, z) &= \bar b_\mathrm{h} \, P_\mathrm{lin}(k, z)
  \int_{0}^{\infty} dM \, \frac{dn}{dM}
  \nonumber \\
  & \quad \times b_\mathrm{h}(M, z) \frac{M}{\bar\rho_\mathrm{m}} \, u (k ; M, z) ,
\end{align}
where $u(k ; M, z)$ is the Fourier transform of the normalized matter density profile of halos $\rho(r)/M$ with mass $M$ at redshift $z$.
We assume the halo follows the Navarro--Frenk--White (NFW) density profile \cite{Navarro1997} and adopt the fitting formula of the concentration--mass relation in Ref.~\cite{Ishiyama2021b}.
We assume the NFW profile truncated at the virial radius, and the explicit expression for the Fourier transform of the normalized NFW density profile is given in Refs.~\cite{Scoccimarro2001, Oguri2011}.
The mass integration range needs special care.
In the 2-halo term, the mass integration takes into account the entire mass range.
However, in practical implementations, the mass range is truncated at some minimum and maximum masses $[M_\mathrm{min}, M_\mathrm{max}]$.
In addition, in the large scale limit, the following condition must be satisfied to ensure the large-scale matter normalization:
\begin{equation}
  \int_0^\infty dM \, M b_\mathrm{h} (M, z) \frac{dn}{dM} = \bar\rho_\mathrm{m} .
\end{equation}
The fitting functions of the halo mass function and halo bias are not generally guaranteed to satisfy this condition.
To enforce this condition with the finite mass integration range $[M_\mathrm{min}, M_\mathrm{max}]$, following the prescription of Refs.~\cite{Schmidt2016,Mead2020,Mead2021}, we modify the halo mass function in the 2-halo term as
\begin{equation}
  n(M) \to n(M) \Theta(M-M_\mathrm{min}) +
  \frac{A (M_\mathrm{min}) \delta_\mathrm{D} (M-M_\mathrm{min})}{b_\mathrm{h}(M_\mathrm{min})M_\mathrm{min}/\bar{\rho}_\mathrm{m}},
\end{equation}
where $\Theta$ is the Heaviside step function and
\begin{equation}
  A(M_\mathrm{min}) = 1 - \frac{1}{\bar{\rho}_\mathrm{m}} \int_{M_\mathrm{min}}^{M_\mathrm{max}} M b_\mathrm{h}(M, z) n(M) \, dM .
\end{equation}
This modification is applied to the 2-halo term and we adopt the mass integration range $[M_\mathrm{min}, M_\mathrm{max}] = [10^{11}, 10^{16}] \, \hMsun$.

\begin{figure}
\includegraphics[width=\columnwidth]{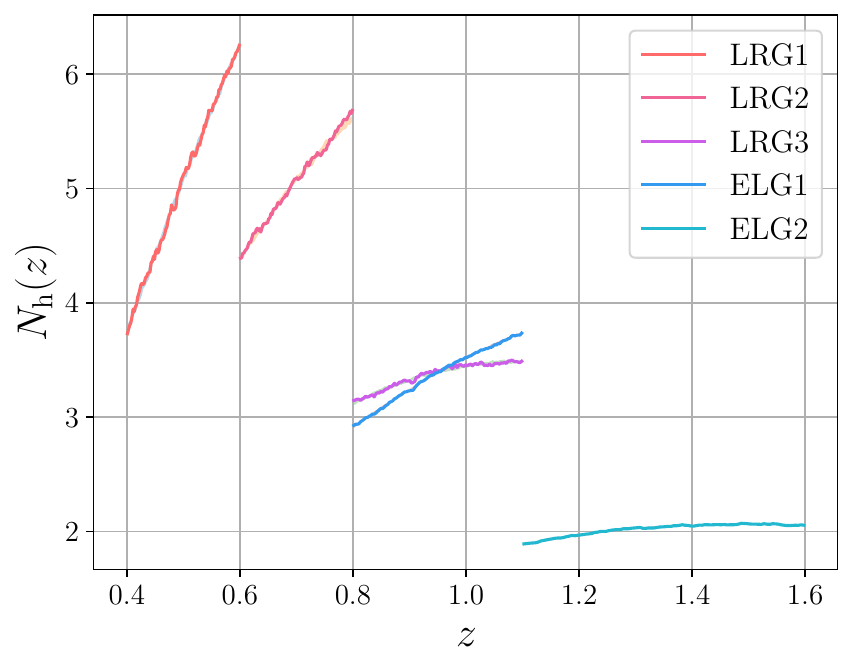}
\caption{The redshift distributions $N_\mathrm{h} (z)$ of the mock LRG (LRG1, LRG2, and LRG3) and ELG (ELG1 and ELG2) samples constructed from the lightcone halo catalog of the \GION simulation suite.
Note the redshift distributions are normalized to unity.}
\label{fig:dNdz_halo}
\end{figure}

Under the Limber approximation, the angular power spectra are given as
\begin{align}
  C_\mathrm{hh} (\ell) &= \int_0^{\chi_H} \frac{d\chi}{\chi^2} \, W_\mathrm{h}^2 (\chi)
  \nonumber \\
  & \quad \times P_{\mathrm{hh}} \left( k = \frac{\ell + 1/2}{\chi}, z(\chi) \right) ,
  \label{eq:halo_model_hh} \\
  C_{\mathrm{h}\kappa} (\ell) &= \int_0^{\chi_H} \frac{d\chi}{\chi^2} \, W_\mathrm{h} (\chi) \, W_\kappa (\chi, \chi_\mathrm{s})
  \nonumber \\
  & \quad \times P_{\mathrm{hm}} \left( k = \frac{\ell + 1/2}{\chi}, z(\chi) \right) ,
  \label{eq:halo_model_hk}
\end{align}
where $\chi_H$ is the comoving distance to the horizon, and $W_\mathrm{h} (\chi)$ is the radial kernel of the halo number density field.
The kernel is $W_\mathrm{h}(\chi) = [H(z)/c] \, N_\mathrm{h}(z)$, where $N_\mathrm{h}(z)$ is the normalized redshift distribution of the sample ($\int dz \, N_\mathrm{h}(z) = 1$) shown in Figure~\ref{fig:dNdz_halo}, and the lensing kernel $W_\kappa$ is given in Eq.~\eqref{eq:lensing_kernel}.
The Limber approximation and the flat-sky projection of individual halos implicitly assumed in the 1-halo term become inaccurate at the largest angular scales.
Thus, we compute the theoretical predictions using the exact expressions presented in Appendix~\ref{sec:exact_model} for $\ell < 100$ and Eqs.~\eqref{eq:halo_model_hh} and \eqref{eq:halo_model_hk} for $\ell \geq 100$.

\begin{figure*}
\includegraphics[width=\textwidth]{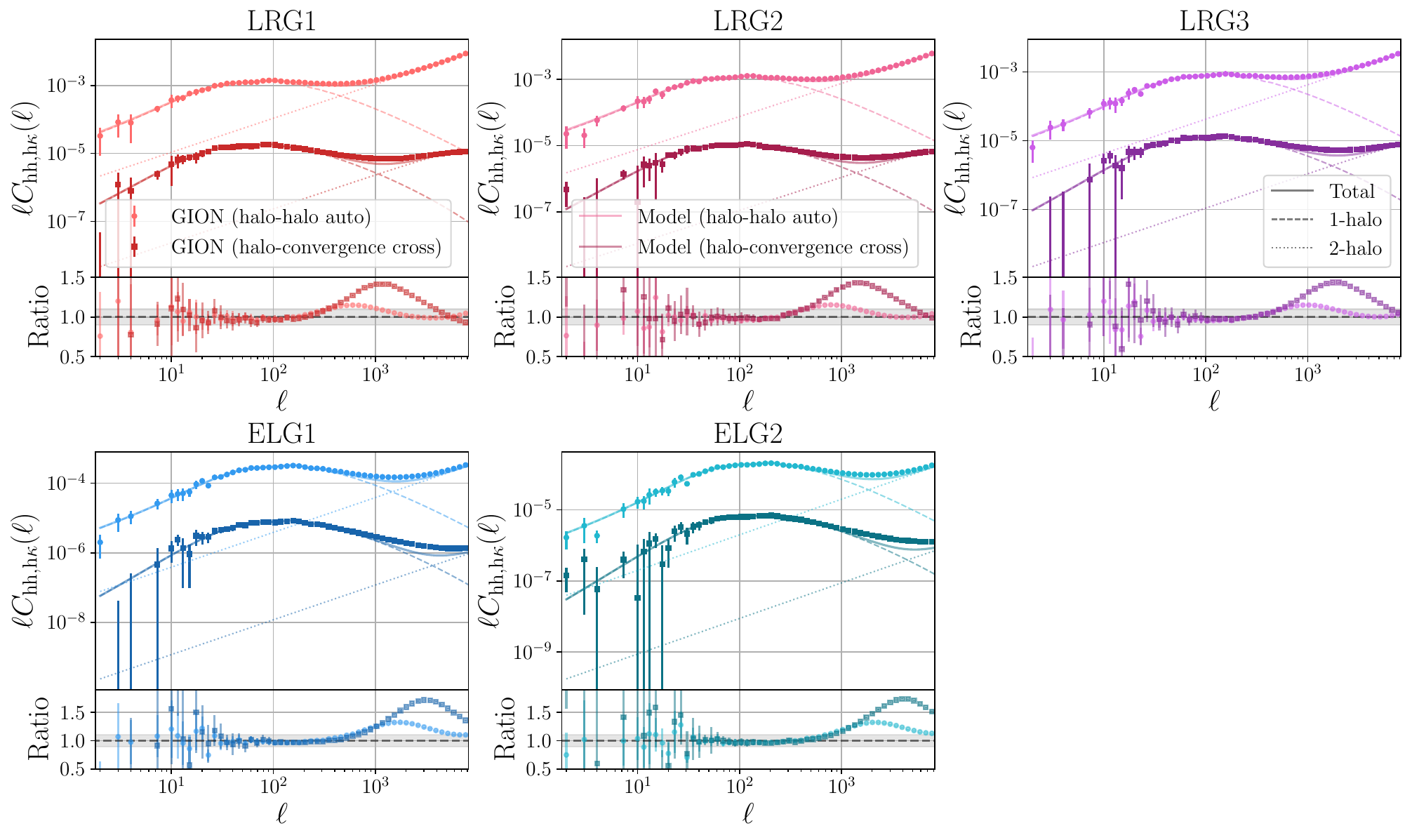}
\caption{The halo--halo auto ($C_{\mathrm{hh}}$) and halo--convergence cross ($C_{\mathrm{h}\kappa}$) angular power spectra for the five mock galaxy samples shown in Figure~\ref{fig:dNdz_halo} (points), compared with the halo-model prediction (solid lines).
The 1-halo and 2-halo contributions to the halo-model prediction are shown as dashed and dotted lines, respectively.
The bottom panel of each sample shows the ratio of the measured to the predicted angular power spectra.
The shaded region corresponds to 10\% deviation from the theoretical prediction.}
\label{fig:cl_halo}
\end{figure*}

Figure~\ref{fig:cl_halo} shows the auto- and cross-power spectra for the five samples, together with the corresponding theoretical predictions.
The measured and predicted angular power spectra agree to within $\sim 10\%$ over the range $10^1 \lesssim \ell \lesssim 10^2$ for all five samples.
At $\ell \lesssim 10$, the agreement is limited by the sample variance associated with the small number of available modes, while at $\ell \gtrsim 10^2$, the measured power spectra exceed the model prediction, reflecting the same non-linear bias effect identified in Figures~\ref{fig:pkhm} and \ref{fig:pkhh}.
The excess peaks at $\ell \sim (1\text{--}2) \times 10^3$, corresponding to the transition regime between 1-halo and 2-halo contributions, where it reaches $\sim 40\text{--}80\%$ for the halo--convergence cross-power spectra and $\sim 10\text{--}30\%$ for the halo auto-power spectra, and the ratio decreases again toward smaller scales, where the 1-halo term dominates.

\subsection{Thermal Sunyaev--Zel'dovich effect}
As a third example, we consider the tSZ effect, which is one of the secondary anisotropies of CMB temperature due to the inverse Compton scattering of CMB photons by hot electrons in the intracluster medium (ICM) of galaxy clusters \cite{Carlstrom2002, Kitayama2014}.
The temperature variation due to the tSZ effect is given as
\begin{equation}
  \frac{\Delta T_\mathrm{tSZ}}{T_\mathrm{CMB}} = g_\nu (x) y(\bm{\theta}) ,
\end{equation}
where the frequency-dependent part is
\begin{equation}
  g_\nu (x) = x \coth(x/2) - 4, \quad x = \frac{h\nu}{k_\mathrm{B} T_\mathrm{CMB}} ,
\end{equation}
$\nu$ is the frequency of CMB photons, $h$ is the Planck constant, $k_\mathrm{B}$ is the Boltzmann constant, and $T_\mathrm{CMB} = 2.725 \, \mathrm{K}$ is the CMB temperature.
The tSZ effect has a unique frequency dependence, which allows it to be separated from other CMB anisotropies.
The amplitude of the temperature variation is proportional to the Compton-$y$ parameter $y(\bm{\theta})$, which is the line-of-sight projection of the electron pressure $P_\mathrm{e}$:
\begin{equation}
  \label{eq:compton_y}
  y(\bm{\theta}) = \frac{\sigma_\mathrm{T}}{m_\mathrm{e} c^2} \int_0^{l_\mathrm{rec}} dl \, P_\mathrm{e}(l, \bm{\theta}) ,
\end{equation}
where $\sigma_\mathrm{T}$ is the Thomson cross section, $m_\mathrm{e}$ the electron mass, and $l$ the proper distance along the line-of-sight direction, and $l_\mathrm{rec}$ is the proper distance to the last scattering surface.
Assuming that gas is fully ionized, the electron pressure is related to the thermal gas pressure $P_\mathrm{th}$:
\begin{equation}
  P_\mathrm{e} = \frac{\mu}{\mu_\mathrm{e}} P_\mathrm{th}
  = \frac{2X+2}{5X+3} P_\mathrm{th} ,
\end{equation}
where $X = 0.76$ is the primordial hydrogen mass fraction, $\mu = 4/(5X+3)$ is the mean molecular weight, and $\mu_\mathrm{e} = 2/(X+1)$ is the mean molecular weight per electron.
The tSZ effect is a powerful probe of the thermal energy content of the Universe and can be used to study the properties of galaxy clusters and the LSS of the Universe.

\subsubsection{Map making algorithm}
The Compton-$y$ parameter is sourced from the gas distribution, which is not directly modeled in the \GION simulation suite.
Instead, we generate full-sky Compton-$y$ maps by painting pressure profiles onto the halos in the lightcone halo catalog using the halo-based pasting approach of the \texttt{Baryon Pasting} \cite{Osato2023} method.
Since the main contributor of the tSZ signal is the hot gas within galaxy clusters, it suffices to take into account the gas within halos and the contributions from diffuse gas are subdominant.
The pressure profile of the ICM is modeled as a function of halo mass and redshift, and we consider three different models: the semi-analytic model of Ref.~\cite{Shaw2010} and generalized NFW (GNFW) profiles calibrated with tSZ observations \cite{Planck2013V} and hydrodynamical simulations \cite{Battaglia2012}.
The semi-analytic model solves the spherical ICM profiles of the gas density and pressure based on the fluid equations and considers the energy injection due to dynamical friction with dark matter and supernovae and active galactic nuclei (AGN).
In addition, the non-thermal pressure components originating mainly from turbulent motion in galaxy clusters are also taken into account.
This model introduces free parameters to control the efficiency of the energy transfer and non-thermal pressure contribution and these parameters are calibrated with hydrodynamical simulations and X-ray observations.
We refer to this semi-analytic model profile as the ``Shaw profile''.
We also employ the fitting functions of the pressure profile based on the tSZ observations and hydrodynamical simulations.
In contrast to the semi-analytic model, the GNFW profile is a phenomenological model but it has been shown to provide a good fit to the observed pressure profiles of galaxy clusters.
We use the two GNFW pressure profiles, one calibrated with the tSZ observations of \textit{Planck} \cite{Planck2013V} and the other with the hydrodynamical simulations of Ref.~\cite{Battaglia2012}.
We refer to these two GNFW profiles as ``Planck profile'' and ``Battaglia profile'', respectively.

In practical implementation, we first compute the pressure profile of each halo in the lightcone halo catalog using the three pressure profile models.
The Compton-$y$ profile is then projected onto a \texttt{HEALPix} map with $N_\mathrm{side} = 8192$.
The technical details are found in Section~3.3 of Ref.~\cite{Osato2023}.
We process halos with $M_\mathrm{200b} > 5 \times 10^{12} \, \hMsun$ and $z > 0.005$: the former because the contribution from less massive halos is negligible and the pressure profile of such less massive halos is not well calibrated against either observations or hydrodynamical simulations, and the latter to mimic the masking of nearby clusters in observations.
The resulting Compton-$y$ maps are then used to compute the tSZ angular power spectrum, which is compared with the halo-model prediction.

\subsubsection{Halo model prediction}
Next, we describe the theoretical prediction of the tSZ angular power spectrum based on the halo model formalism \cite{Komatsu1999}.
We begin with the contribution from a single halo, which is expressed in Fourier space:
\begin{multline}
  \tilde y_\ell(M, z) = \frac{4\pi r_s}{\ell_s^2} \frac{\sigma_\mathrm{T}}{m_\mathrm{e} c^2}
  \int_0^{x_f} dx \, x^2 P_\mathrm{e}(x; M, z) \\
  \times \frac{\sin [(\ell + 1/2) x / \ell_s]}{(\ell + 1/2) x / \ell_s} ,
  \label{eq:y_ell}
\end{multline}
where $r_s$ is the arbitrary scale radius, $x = r / r_s$, $P_\mathrm{e} (x; M, z)$ is the electron pressure profile, $\ell_s = d_A(z) / r_s$ is the characteristic multipole, and $d_A (z)$ is the angular diameter distance.
We consider three pressure profile models described above: Shaw, Battaglia, and Planck profiles.
The integration is performed up to $x_f = r_f / r_s$, where $r_f$ is the boundary of the gas distribution, and for the Shaw profile, the boundary is analytically computed, and for the Battaglia and Planck profiles, we adopt $r_f = 5 \, R_\mathrm{500c}$.
The tSZ angular power spectrum is the sum of the 1-halo and 2-halo terms:
\begin{align}
  C_{yy} (\ell) &= C_{yy}^{\mathrm{1h}} (\ell) + C_{yy}^{\mathrm{2h}} (\ell) , \\
  \label{eq:cl_yy_1h}
  C_{yy}^{\mathrm{1h}} (\ell) &= \int_{z_\mathrm{min}}^{z_\mathrm{max}} dz \, \frac{d^2V}{dzd\Omega} \int dM \, \frac{dn}{dM} \, \tilde y_\ell^2(M, z) , \\
  C_{yy}^{\mathrm{2h}} (\ell) &= \int_{z_\mathrm{min}}^{z_\mathrm{max}} dz \, \frac{d^2V}{dzd\Omega} \, P_\mathrm{lin}
  \left(k = \frac{\ell+1/2}{\chi}, z \right)
  \nonumber \\
  & \times \left[ \int dM \, \frac{dn}{dM} \, b_\mathrm{h}(M, z) \, \tilde y_\ell(M, z) \right]^2 ,
  \label{eq:cl_yy_2h}
\end{align}
where $z_\mathrm{min} = 0.005$ and $z_\mathrm{max} = 5$ are the minimum and maximum redshift, respectively, and $d^2V/dzd\Omega = \chi^2 (z) \, c / H(z)$ is the comoving volume element per unit redshift and per steradian.
The mass integration range is taken to be $[M_\mathrm{min}, M_\mathrm{max}] = [5 \times 10^{12}, 10^{16}] \, \hMsun$ to match the mass range of halos used to construct the Compton-$y$ maps.
Note that the power spectrum at large scales strongly depends on the minimum redshift $z_\mathrm{min}$ because the nearest massive halos contribute significantly to the tSZ signal at large scales.
In the real observations, the tSZ signal from nearby massive halos is often masked to reduce the sample variance and the contamination from foregrounds.
As described above, the nearby halos in the lightcone halo catalog with $z < z_\mathrm{min}$ are not used.
These expressions rely on the Limber approximation for the 2-halo term and the flat-sky projection for the profile of individual halos (Eq.~\ref{eq:y_ell}), both of which become inaccurate at the largest angular scales.
Thus, we adopt the exact expressions presented in Appendix~\ref{sec:exact_model} for $\ell < 100$ and the expressions above for $\ell \geq 100$.

\begin{figure}
\includegraphics[width=\columnwidth]{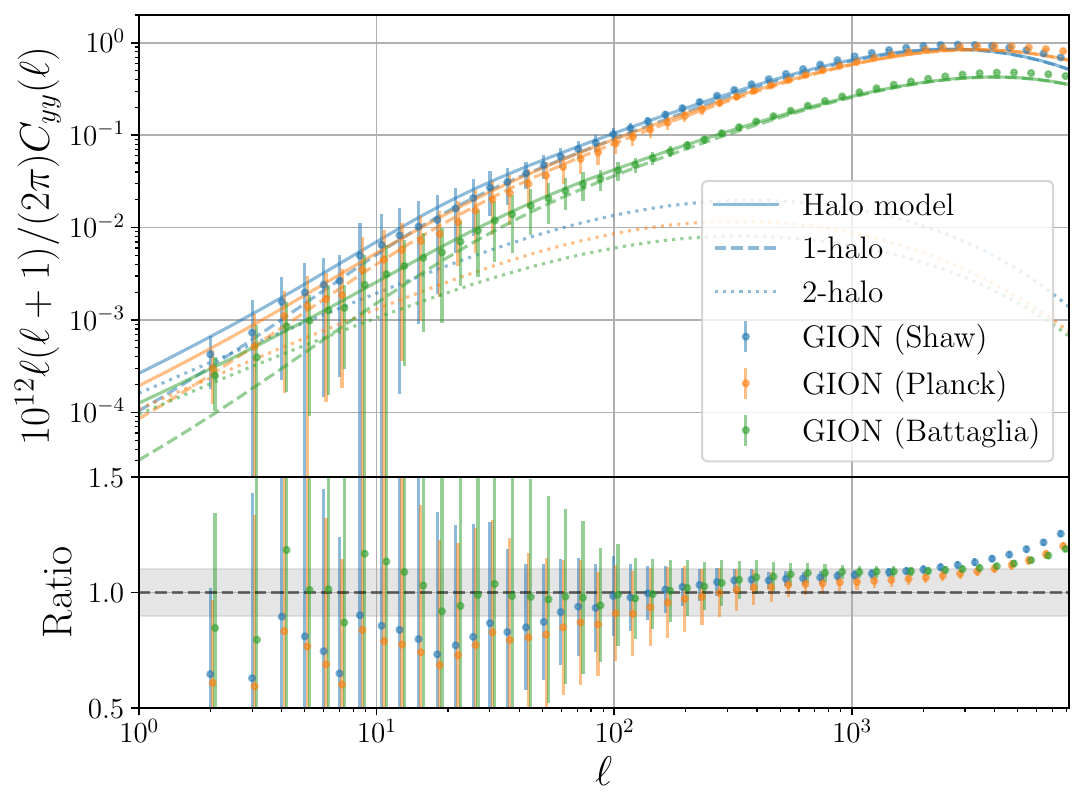}
\caption{The thermal SZ angular power spectrum measured from the \GION simulation suite (points) using Shaw, Planck, and Battaglia pressure profiles, compared with the corresponding halo-model predictions (solid lines).
The 1-halo and 2-halo contributions to the halo-model prediction are shown as dashed and dotted lines, respectively.
The bottom panel shows the ratio of the measured to the predicted angular power spectra.
The shaded region corresponds to 10\% deviation from the theoretical prediction.}
\label{fig:cl_tSZ}
\end{figure}

Figure~\ref{fig:cl_tSZ} shows the resulting tSZ angular power spectrum for the three pressure profile models, together with the corresponding halo-model predictions.
At $100 \lesssim \ell \lesssim 2000$, all models agree with their respective predictions to within $\sim 10\%$, while at $\ell \gtrsim 3000$, the measurements exceed the predictions, by up to $\sim 25\%$ at $\ell \simeq 8000$.
Although the amplitude for the Battaglia profile is about half of those for the Shaw and Planck profiles, the ratios of the measured to the predicted power spectra are nearly identical for the three pressure profile models, suggesting that the residual originates from the halo model ingredients common to the three models, e.g., the halo mass function and the halo bias, rather than from the pressure profiles.
At $\ell \lesssim 100$, the measured power spectra are still consistent with the predictions within the $1\text{-}\sigma$ level, but large statistical fluctuations are observed.
At these large scales, the covariance matrix is dominated by the connected non-Gaussian term from the tSZ trispectrum \cite{Osato2021} and the signal is affected by the sample variance associated with the small number of massive nearby halos.

\subsection{Kinematic Sunyaev--Zel'dovich effect}
In addition to the tSZ effect, the kSZ effect is another secondary anisotropy of CMB temperature due to the bulk motion of free electrons.
The kSZ temperature fluctuation is sourced by the line-of-sight component of the ionized-gas momentum:
\begin{align}
  \label{eq:ksz_b}
  \frac{\Delta T_\mathrm{kSZ}}{T_\mathrm{CMB}} (\bm{\theta}) &= -b(\bm{\theta}), \\
  b(\bm{\theta}) &= \frac{\sigma_\mathrm{T}}{c} \int_0^{l_\mathrm{rec}} dl \,
  e^{-\tau (l)} n_\mathrm{e}(l, \bm{\theta}) v_r(l, \bm{\theta}) ,
\end{align}
where $b(\bm{\theta})$ is the Doppler-$b$ parameter, $n_\mathrm{e}$ is the electron number density, $v_r$ is the line-of-sight peculiar velocity, which is positive if gas is receding from the observer, $l$ is the proper distance along the line-of-sight direction, and $\tau$ is the mean optical depth of Thomson scattering up to the redshift $z$,
\begin{equation}
  \label{eq:tau}
  \tau(z) = \sigma_\mathrm{T} \int_0^z \frac{c \, dz'}{H(z')} \, (1 + z')^2 \, \bar{n}_\mathrm{e,0} ,
\end{equation}
where $\bar{n}_\mathrm{e,0} = \Omega_\mathrm{b}\rho_\mathrm{cr,0} / (\mu_\mathrm{e} m_\mathrm{p})$ is the mean electron number density in the present Universe and $m_\mathrm{p}$ is the proton mass.
The kSZ effect is a powerful probe of the large-scale velocity field and the distribution of ionized gas in the Universe.
In contrast to the tSZ effect, the kSZ effect is independent of the electron temperature and thus can probe the diffuse gas in the outskirts of galaxy clusters and in the intergalactic medium.
The kSZ effect is also sensitive to the reionization history of the Universe, as it is sourced by free electrons along the line-of-sight.
The kSZ effect is challenging to measure because of its low amplitude and its blackbody spectrum, which matches that of the primary CMB in thermodynamic temperature units, but recent observations have detected the kSZ signal from galaxy clusters and LSS \cite{Hand2012, Soergel2016, Sugiyama2018}.

\subsubsection{Map making algorithm}
Similarly to the tSZ effect, the kSZ effect is sourced by the gas distribution, which is not directly modeled in the \GION simulation suite.
Here, we rely on the simplest approach: we assume that the ionized gas traces the underlying matter distribution, and thus the electron number density is proportional to the matter density.
The \texttt{assign\_healpix} code outputs the projected momentum field of the $i$-th shell, $q^{(i)}(\bm{\theta}) = \sum_{p \in \mathcal{P}^{(i)} (\bm{\theta})} m_\mathrm{part} v_{r,p}$, where $v_{r,p}$ is the line-of-sight velocity of the $p$-th particle and the set of particles $\mathcal{P}^{(i)} (\bm{\theta})$ is defined in the same manner as in Eq.~\eqref{eq:mass_shell}.
The Doppler-$b$ parameter map is then constructed by summing the contributions from all shells along the line-of-sight direction \cite{Schaye2023}:
\begin{equation}
  b(\bm{\theta}) = \sum_i e^{-\tau(z^{(i)})}
  \frac{f_\mathrm{b} \sigma_\mathrm{T}}{c \mu_\mathrm{e} m_\mathrm{p}} \frac{q^{(i)}(\bm{\theta})}{S_\mathrm{pix}},
\end{equation}
where $S_\mathrm{pix}$ is the proper area of a pixel at the redshift of the shell, and $f_\mathrm{b} = \Omega_\mathrm{b}/\Omega_\mathrm{m}$ is the universal cosmic baryon fraction.
The redshift of the shell $z^{(i)}$ is determined from the cone-volume-weighted comoving distance defined in Eq.~\eqref{eq:chi_shell}.
In contrast to the tSZ effect, the kSZ effect is constructed from the momentum shells.
Note that the momentum shells extend up to $z_\mathrm{max} = 5.14$, which is higher than the maximum redshift of the lightcone halo catalog ($z_\mathrm{max} = 5$) used to construct the Compton-$y$ maps.

\subsubsection{Theoretical prediction}
Unlike the tSZ effect, which is sourced by the hot gas confined in virialized halos and is therefore well described by the halo model, the kSZ effect receives a substantial contribution from the diffuse gas residing outside halos, and hence a perturbative description of the ionized-gas momentum field $\bm{q} = (1 + \delta_\mathrm{e}) \bm{v}$, where $\delta_\mathrm{e}$ is the free electron density contrast and $\bm{v}$ is the velocity field, is more appropriate than the halo model prescription.
Since $b(\bm{\theta})$ is linear in $\bm{q}$, the signal vanishes at the leading order in the density perturbation: the velocity field is curl-free in linear theory, and the contributions from the coherent flows toward and away from the observer cancel out along the line-of-sight direction.
A non-vanishing power spectrum therefore arises only at the second order, through the modulation of the velocity field by the density fluctuations, which was first pointed out by Refs.~\cite{Ostriker1986, Vishniac1987} and is now referred to as the Ostriker--Vishniac (OV) power spectrum.
Decomposing the momentum into the longitudinal and transverse components, $\bm{q} = \bm{q}_\parallel + \bm{q}_\perp$, the cancellation applies to the longitudinal mode, and thus the small-scale signal is dominated by the transverse mode.
Subsequent works \cite{Ma2002, Shaw2012, Park2016} extended the OV power spectrum by incorporating the non-linear evolution of the density and velocity fields, which is essential for quantitative predictions at arcminute scales.

Under the Limber approximation, the angular power spectrum sourced by the transverse momentum is given as \cite{Ma2002, Park2016}
\begin{align}
  C_{bb}^\perp (\ell) = \frac{1}{2} \left(\frac{\sigma_\mathrm{T}
  \bar n_\mathrm{e,0}}{c}\right)^2 & \int_0^{\chi_\mathrm{max}}
  d\chi \, \frac{(1 + z)^4}{\chi^2} e^{-2\tau(z)}
  \nonumber \\
  & \times P_{q_\perp} \left( k = \frac{\ell+1/2}{\chi}, z \right) ,
  \label{eq:cl_bb}
\end{align}
where $\chi_\mathrm{max} = \chi(z = z_\mathrm{max})$ is the comoving distance to the highest redshift of the lightcone, and the transverse momentum power spectrum $P_{q_\perp}(k, z)$ is given as
\begin{align}
  P_{q_\perp}(k, z) &= \left[ a H(z) f(z) \right]^2 \int \frac{d^3 k'}{(2\pi)^3}
  \nonumber \\
  & \times \Biggl[
  \frac{1 - \mu^2}{k'^2} \, P_{\delta_\mathrm{e} \delta_\mathrm{e}} (|\bm{k} - \bm{k}'|, z) \, P_{\theta\theta}(k', z)
  \nonumber \\
  & - \frac{1 - \mu^2}{|\bm{k} - \bm{k}'|^2} \, P_{\delta_\mathrm{e} \theta}(|\bm{k} - \bm{k}'|, z) \, P_{\delta_\mathrm{e} \theta}(k', z)
  \Biggr] ,
  \label{eq:pqperp}
\end{align}
where $f (z) = d\ln D_+ / d\ln a$ is the linear growth rate with $D_+ (z)$ the linear growth factor, $\mu = (\bm{k} \cdot \bm{k}') / (k k')$ is the directional cosine, $\theta \equiv -\nabla \cdot \bm{v} / (a H f)$ is the normalized velocity divergence, $P_{\delta_\mathrm{e} \delta_\mathrm{e}}$ and $P_{\theta\theta}$ are the auto power spectra of the electron density contrast and the normalized velocity divergence, respectively, and $P_{\delta_\mathrm{e} \theta}$ is their cross power spectrum.
Here, we assume no velocity bias between electrons and total matter.
The right-hand side does not depend on the direction of $\bm{k}$, and we set $\bm{k} = (0, 0, k)$ when evaluating the integral.

If the electrons trace the total matter distribution, $\delta_\mathrm{e} = \delta$, and both the density and the velocity fields obey linear theory, $\delta = \theta$ and $P_{\delta\delta} = P_{\delta\theta} = P_{\theta\theta} = P_\mathrm{lin}$, Eq.~\eqref{eq:pqperp} reduces to the OV power spectrum,
\begin{equation}
  \label{eq:pqperp_ov}
  P_{q_\perp}^\mathrm{OV}(k, z) = \left[ a H f \right]^2 \!\! \int \! \frac{d^3 k'}{(2\pi)^3} P_\mathrm{lin}(|\bm{k} - \bm{k}'|) P_\mathrm{lin}(k') I(k, k') ,
\end{equation}
where the mode-coupling kernel $I(k, k')$ is given as
\begin{equation}
  \label{eq:ov_kernel}
  I(k, k') = \frac{k (k - 2 k' \mu) (1 - \mu^2)}{k'^2 \left( k^2 + k'^2 - 2 k k' \mu \right)} .
\end{equation}

Beyond the linear order, the density and velocity divergence fields differ from the linear field due to the non-linear evolution.
We therefore evaluate the non-linear (NL) transverse momentum power spectrum by substituting the non-linear spectra into Eq.~\eqref{eq:pqperp},
\begin{align}
  \label{eq:pk_nl_set}
  P_{\delta_\mathrm{e} \delta_\mathrm{e}}(k, z) &= P_\mathrm{mm}(k, z) , \\
  P_{\delta_\mathrm{e} \theta}(k, z) &= b_{\delta\theta}(k, z) \, P_\mathrm{mm}(k, z) , \\
  P_{\theta\theta}(k, z) &= b_{\theta\theta}(k, z) \, P_\mathrm{mm}(k, z) ,
\end{align}
where $P_\mathrm{mm}(k, z)$ is the non-linear matter power spectrum computed with \texttt{DarkEmulator2}.
For density-velocity cross-spectrum and velocity auto-spectrum correction terms, $b_{\delta\theta}$ and $b_{\theta\theta}$, we adopt the fitting formulas of Ref.~\cite{Bel2019}, which are calibrated with $N$-body simulations.
In addition, here we have set $\delta_\mathrm{e} = \delta$, which is the assumption adopted in our map making as well, since the momentum shells are constructed from the dark matter particles and the gas is assumed to trace the total matter distribution.

Finally, we consider the longitudinal mode.
Its contribution is strongly suppressed by the line-of-sight cancellation and is negligible at small scales, but it becomes dominant over the transverse mode at large scales, $\ell \lesssim 100$ \cite{Park2016, Alvarez2016}.
Since our lightcone is truncated at the finite redshift $z_\mathrm{max}$, the cancellation is incomplete at the boundary, and the residual surface term contributes to the power spectrum at large scales.
The contribution from the longitudinal mode is given as \cite{Alvarez2016, Websky}:
\begin{equation}
  \label{eq:cl_bb_para}
  C_{bb}^\parallel (\ell) = \frac{2}{\pi} \bar u_\mathrm{max}^2 \int_0^\infty \frac{dk}{k^2} j_\ell^2(k \chi_\mathrm{max}) \, P_\mathrm{lin}(k, z_\mathrm{max}) ,
\end{equation}
where $j_\ell$ is the spherical Bessel function and $\bar u_\mathrm{max} = (\sigma_\mathrm{T} \bar n_\mathrm{e,0} / c) (1 + z_\mathrm{max})\, H(z_\mathrm{max}) \, f(z_\mathrm{max}) \, e^{-\tau(z_\mathrm{max})}$.
The total prediction is the sum of the transverse and longitudinal contributions:
\begin{equation}
  C_{bb} (\ell) = C_{bb}^\perp (\ell) + C_{bb}^\parallel (\ell) .
\end{equation}

\begin{figure}
\includegraphics[width=\columnwidth]{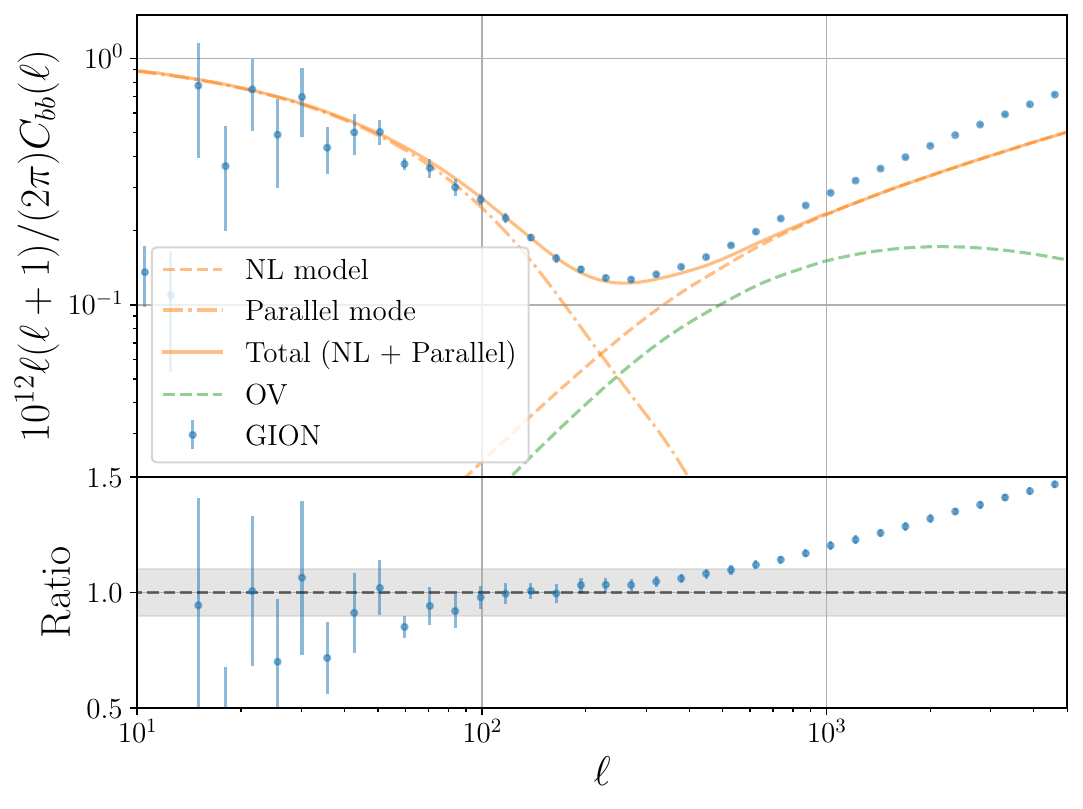}
\caption{The kinematic SZ angular power spectrum measured from the lightcone momentum shells of the \GION simulation suite (points), together with the non-linear (NL) transverse and parallel mode contributions and their sum (total prediction), and the linear OV contribution for comparison.
The bottom panel shows the ratio of the measured to the total predicted angular power spectra. 
The shaded region corresponds to 10\% deviation from the theoretical prediction.}
\label{fig:cl_kSZ}
\end{figure}

Figure~\ref{fig:cl_kSZ} shows the resulting kSZ angular power spectrum together with an analytic decomposition into the NL transverse contribution computed from the non-linear density and velocity divergence power spectra and the contribution from the line-of-sight parallel mode velocity field, whose sum gives the total prediction.
For comparison, the linear OV contribution is also shown.
Note that the kSZ effect at large scales is sourced by the large-scale mode of the density and velocity fields, and thus, the power spectrum is more sensitive to the replication effect of the simulation box than other statistics explored so far.
The power spectrum at the largest angular scales is significantly suppressed by the replication effect: the lowest multipole bin shown in Figure~\ref{fig:cl_kSZ}, $\ell \simeq 10$, already falls far below the prediction, and the multipoles $\ell < 10$, which are not shown, are suppressed even more strongly.
The OV contribution is considerably smaller than the NL one, which demonstrates that the non-linear growth at small scales propagates to the kSZ power spectrum at larger angular scales through the mode coupling in the momentum power spectrum (Eq.~\ref{eq:pqperp}).
The measured power spectrum agrees with the total prediction within $\sim 10\%$ at $100 \lesssim \ell \lesssim 300$, and at $\ell \lesssim 100$, the measurement is consistent with the prediction within the large statistical uncertainty.
On the other hand, at small scales ($\ell \gtrsim 300$), the measured power spectrum exceeds the total prediction, by $\sim 20\%$ at $\ell \sim 10^3$ and up to $\sim 45\%$ at $\ell \simeq 5000$.
This excess on small scales is commonly attributed to the non-linear evolution of the velocity field and to the contribution from the internal, virial motions of halos, which are not captured by the perturbative terms included in the analytic decomposition.

\subsection{Cross-correlation of weak lensing and the tSZ effect}
\label{sec:cross_correlation}
As a final demonstration, we measure the cross-correlation between the weak lensing convergence and the tSZ effect.
The cross-correlation is a powerful tool to study the baryonic effects on the matter distribution and to constrain the gas physics in galaxy clusters and has the potential to break the parameter degeneracy when combined with the auto-correlations.
The cross-correlation has been measured in various surveys \cite{Hojjati2017, Osato2020, Gatti2022, Pandey2022}, and it has been shown that the cross-correlation signal is sensitive to the pressure profile of the ICM and the feedback processes that affect the gas distribution.
As covered in previous sections, the \GION simulation suite is an ideal dataset that allows us to measure the cross-correlation between multiple LSS observables.

Within the same halo model formalism used for the tSZ auto spectrum, the cross power spectrum is the sum of the 1-halo and 2-halo terms:
\begin{align}
  C_{y\kappa} (\ell) &= C_{y\kappa}^{\mathrm{1h}} (\ell) + C_{y\kappa}^{\mathrm{2h}} (\ell) , \\
  \label{eq:cl_yk_1h}
  C_{y\kappa}^{\mathrm{1h}} (\ell) &= \int_{z_\mathrm{min}}^{z_\mathrm{max}} dz \, \frac{d^2 V}{dzd\Omega} \int dM \, \frac{dn}{dM} \, \tilde y_\ell(M, z) \, \tilde\kappa_\ell(M, z) , \\
  C_{y\kappa}^{\mathrm{2h}} (\ell) &= \int_{z_\mathrm{min}}^{z_\mathrm{max}} dz \, \frac{d^2 V}{dzd\Omega} \,
  P_\mathrm{lin}\left(k = \frac{\ell+1/2}{\chi}, z \right)
  \nonumber \\
  & \quad \times \left[ \int dM \, \frac{dn}{dM} \, b_\mathrm{h}(M, z) \, \tilde y_\ell(M, z) \right]
  \nonumber \\
  & \quad \times \left[ \int dM \, \frac{dn}{dM} \, b_\mathrm{h}(M, z) \, \tilde\kappa_\ell(M, z) \right] ,
  \label{eq:cl_yk_2h}
\end{align}
where $\tilde y_\ell$ is the Compton-$y$ profile of Eq.~\eqref{eq:y_ell} and $\tilde\kappa_\ell$ is the Fourier transform of the convergence profile of a halo \cite{Oguri2011}:
\begin{equation}
  \tilde\kappa_\ell(M, z) = \frac{4\pi G}{c^2} \frac{M}{a}
  \frac{\chi_\mathrm{s} - \chi}{\chi \chi_\mathrm{s}}
  u \left( \frac{\ell+1/2}{\chi} ; M, z \right) .
  \label{eq:kappa_ell}
\end{equation}
Here, we adopt the Planck profile for the pressure profile.
The gas distribution of galaxy clusters extends beyond the virial radius and the gas in the outskirts has non-negligible contributions to the cross-correlation signal.
Thus, to incorporate the matter distribution at cluster outskirts, we adopt the smoothly truncated NFW profile of Ref.~\cite{Baltz2009} instead of the standard NFW profile.
The Fourier transform of the density profile $u(k; M, z)$ can analytically be computed and the expressions are found in Ref.~\cite{Baltz2009}.
The mass and redshift integration ranges are the same as those of the tSZ auto spectrum, $[M_\mathrm{min}, M_\mathrm{max}] = [5 \times 10^{12}, 10^{16}] \, \hMsun$ and $[z_\mathrm{min}, z_\mathrm{max}] = [0.005, 5]$.
As in the tSZ auto spectrum, we adopt the exact expressions presented in Appendix~\ref{sec:exact_model} for $\ell < 100$ and the Limber and flat-sky expressions above for $\ell \geq 100$.

\begin{figure}
\includegraphics[width=\columnwidth]{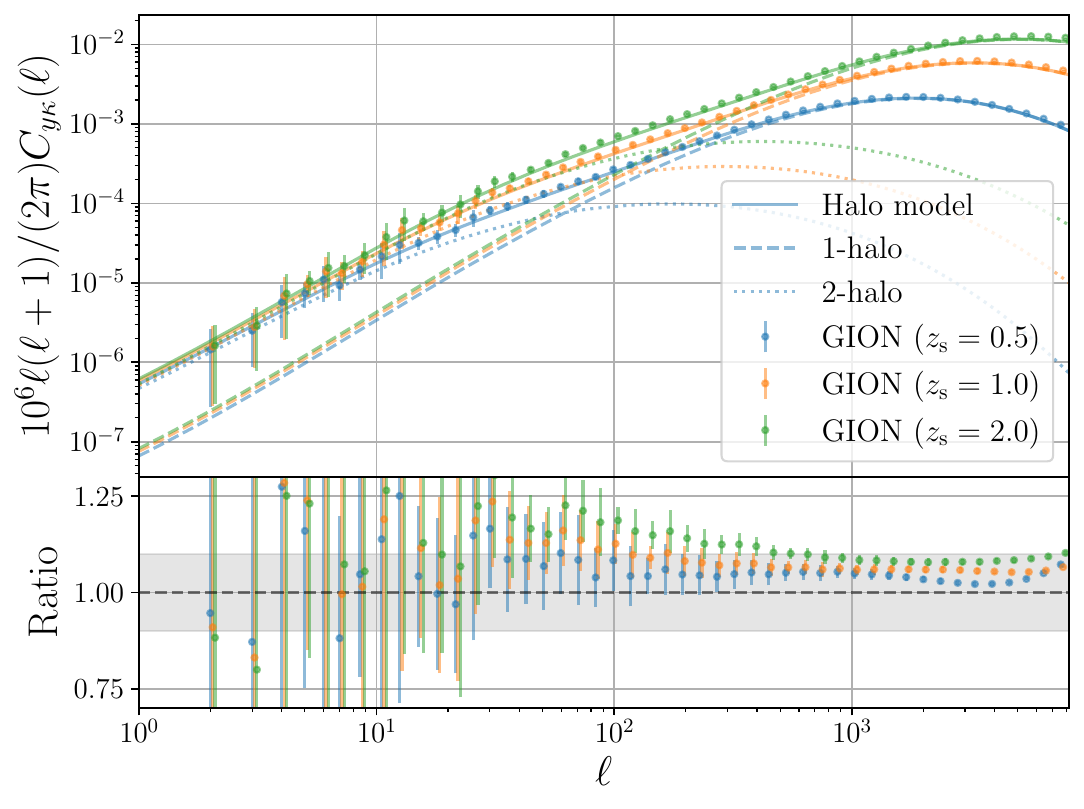}
\caption{The cross power spectrum between the weak lensing convergence and the Compton-$y$ parameter $C_{y\kappa}(\ell)$ measured from the \GION simulation suite (points) for source redshifts $z_\mathrm{s} = 0.5, 1$, and $2$, together with the halo model prediction (solid lines) and its 1-halo (dashed lines) and 2-halo (dotted lines) terms.
The Planck pressure profile is adopted for the Compton-$y$ maps and the prediction.
The error bars correspond to the standard deviation among the $10$ realizations.
The bottom panel shows the ratio of the measured to the predicted angular power spectra.
The shaded region corresponds to 10\% deviation from the theoretical prediction.}
\label{fig:cl_WL_x_tSZ}
\end{figure}

Figure~\ref{fig:cl_WL_x_tSZ} shows the resulting cross power spectrum measured for three source redshifts, $z_\mathrm{s} = 0.5, 1$, and $2$, together with the halo model prediction and its decomposition into the 1-halo and 2-halo terms.
The error bars correspond to the standard deviation among the $10$ realizations.
The 2-halo term dominates at $\ell \lesssim 100$, where the signal arises from the correlation between the pressure of one halo and the matter distribution of the surrounding halos, while the 1-halo term takes over at smaller scales and sets the broad peak of the spectrum at $\ell \sim$ a few $\times 10^3$.
The transition scale is nearly independent of the source redshift, whereas the overall amplitude increases with $z_\mathrm{s}$.
The measured cross power spectra agree with the halo model prediction to within $\sim 10 \%$ over the range $30 \lesssim \ell \lesssim 5000$ for $z_\mathrm{s} = 0.5$ and $1$, with the measurements systematically exceeding the prediction at $\ell \gtrsim 100$; the offset grows with the source redshift, from a few percent for $z_\mathrm{s} = 0.5$ to $\sim 15\%$ for $z_\mathrm{s} = 2$ at $100 \lesssim \ell \lesssim 1000$.
Since the convergence power spectrum measured from the same density shells is accurate at the percent level (Figure~\ref{fig:cl_kappa}), this residual is attributed to the halo model ingredients on the tSZ side, i.e., the fitting formulae of the halo mass function, the halo bias, and the concentration--mass relation, which are adopted in the prediction while the maps are constructed directly from the simulated halos.
At $\ell \lesssim 30$, the ratio scatters substantially and the error bars become comparable to the signal itself, reflecting the limited number of independent modes available on the full sky, i.e., the sample variance.
These results demonstrate that the Level 2 DPs of the \GION simulation suite provide mutually consistent realizations of multiple observables on the same lightcone, and are therefore directly applicable to the modeling of cross-correlation analyses between CMB secondary anisotropies and galaxy imaging surveys.

\section{Future plans}
\label{sec:future_plans}
In this first paper of the \GION simulation suite, we have covered primary LSS observables, including the halo catalogs, the weak lensing convergence, and the tSZ and kSZ effects.
In future work, we will extend the suite to include additional observables that are relevant for multi-wavelength cosmology.
The current plan for future extensions is summarized as follows.
\begin{description}
  \item[Mock galaxy catalogs]
    We will generate mock galaxy catalogs by populating the lightcone halos with galaxies primarily based on their masses and other properties.
    We will use halo occupation distribution or subhalo abundance matching models.
    The high mass resolution of \GION can resolve halos with mass down to $10^{11} \, \hMsun$, which is suitable to populate ELGs.
    We will generate mock catalogs designed for the upcoming Stage-IV spectroscopic surveys, such as DESI, PFS, \textit{Euclid}, and \textit{Roman}.

  \item[CMB lensing]
    We have demonstrated that the \GION simulation suite can be used to generate weak lensing observables for imaging surveys.
    In addition, it is possible to extend the suite to generate CMB lensing as well.
    The \GION simulation suite has already produced the mass shells up to $z = 5.14$ and the mass shells beyond this boundary to the last scattering surface can be well approximated by the linear Gaussian fields.
    Thus, we can also simulate the CMB lensing convergence and its cross-correlation with other LSS observables, which can be used to study the growth of structure and the expansion history of the Universe.

  \item[CMB foregrounds]
    We will extend the suite to generate CMB foregrounds, such as the cosmic infrared background (CIB) and the radio point sources.
    The CIB is sourced by the thermal emission from dust in star-forming galaxies, and it traces the LSS.
    The radio point sources are mainly blazars and dominant sources at lower frequencies.
    Both sources can be modeled by populating the lightcone halos with galaxies using the halo model \cite{Maniyar2021, Agora}.
    The resulting CIB and radio point source maps can be used to study the contamination of the CMB and the cross-correlation with other LSS observables.

  \item[Weak lensing systematics]
    In simulating weak lensing, we have used the Born approximation and ignored the lens-lens coupling.
    In future work, we will extend the suite to include these effects by performing ray-tracing simulations \cite{Hilbert2009, Ferlito2023}.
    Furthermore, we will also include the intrinsic alignment (IA) of galaxies \cite{Hirata2004, Troxel2015} and the baryonic effects \cite{Osato2015, Semboloni2011, Ferlito2023}, which are important systematics in weak lensing analyses.
    For IA modeling in simulations, the effects can be incorporated based on the non-linear alignment model \cite{Bridle2007}.
    The baryonic effects can also be included at the mass shell level, e.g., through the shell baryonification approach \cite{Fluri2019, Anbajagane2024}.
    The resulting weak lensing maps can be used to study the impact of these systematics on the cosmological parameter estimation and to develop mitigation strategies.

  \item[X-ray emission]
    The X-ray emission from galaxy clusters is another important observable for multi-wavelength cosmology.
    The X-ray emission is mainly sourced by the hot gas in the ICM. The \texttt{Baryon Pasting} approach can also provide the gas density and temperature profiles of the ICM, which can be used to generate mock X-ray maps by combining plasma emission models, e.g., \cite{Smith2001}.
    The X-ray map making pipeline with \texttt{Baryon Pasting} has already been developed in Ref.~\cite{Lau2025}.
    The resulting X-ray maps can be used to study the auto-correlation and cross-correlation with other LSS observables to constrain the gas physics in galaxy clusters.
\end{description}

\section{Conclusions}
\label{sec:conclusions}
We have presented the \GION simulation suite, a set of $10$ independent full-sky cosmological lightcone simulations designed for multi-wavelength large-scale structure cosmology in the Stage-IV survey era.
Each realization evolves $6144^3$ particles in a periodic box of $(1.5 \, \hGpc)^3$ with the \textit{Planck} 2018 cosmology, corresponding to a particle mass of $m_\mathrm{part} = 1.28 \times 10^9 \, \hMsun$, and the lightcone covers the full sky out to $z = 5$.
This combination of volume, mass resolution, and depth resolves the dark matter halos down to $M_\mathrm{200b} \sim 10^{11} \, \hMsun$ even at high redshift, i.e., the halo population that hosts the faint emission-line galaxies targeted by Stage-IV spectroscopic surveys, while the $10$ realizations provide a handle on the sample variance of full-sky observables.

The key element of the suite is the on-the-fly lightcone construction implemented in the TreePM code \texttt{GINKAKU}, which outputs particles as they cross the backward lightcone and thereby captures the continuous evolution of structures along the line-of-sight direction, in contrast to the conventional approach of stacking snapshots at discrete redshifts.
The particle outputs are immediately post-processed into the halo catalogs with \texttt{MPI-Rockstar} and into the projected mass and momentum shells with \texttt{assign\_healpix}, and are then discarded, which keeps the storage requirement manageable despite the large number of particles.
We have organized the resulting data into three levels: the temporary raw outputs (Level 0), the lightcone halo catalogs and mass and momentum shells (Level 1), and the LSS observables constructed from them (Level 2).
The entire workflow, from the $N$-body evolution to the generation of Level 1 DPs, takes about $7.8 \times 10^4 \, \mathrm{s}$ of wall-clock time per realization on $8{,}192$ nodes of the supercomputer Fugaku.

We first validated the fundamental statistics of the simulations at fixed redshifts against theoretical predictions.
The matter power spectrum agrees with the prediction of \texttt{DarkEmulator2} to within a few percent for $k \gtrsim 0.1 \, \hMpcinv$ at $0 \le z \le 4$, with the residual scatter on larger scales set by the finite box volume.
The halo mass function is reproduced to within $5 \%$ over $10^{11} \lesssim M_\mathrm{200b}/\hMsun \lesssim 10^{13}$ at $0 \le z \le 4$, with larger deviations at the high-mass end that are attributable to the limited calibration range of the fitting function and to the Poisson noise.
The halo--matter cross and halo--halo auto power spectra follow the linear bias model to within $\sim 10 \%$ on large scales, $k \lesssim 0.1 \, \hMpcinv$, for all mass bins and redshifts except for the rarest samples at high redshift, and depart from it on smaller scales as expected from the scale dependence of the halo bias.
In Appendix~\ref{sec:convergence_tests}, we further confirmed that these statistics are converged with respect to the particle-mesh parameters, the softening length, and the starting redshift used for the initial conditions.

We then presented the map-making algorithms for the lightcone data products and validated them through their angular power spectra, averaged over the $10$ realizations.
The lightcone halo number counts reproduce the abundance obtained by integrating the halo mass function over the full-sky comoving volume to within $5 \%$ for $M_\mathrm{200b} > 10^{11}$--$10^{13} \, \hMsun$ over $0 < z < 5$, and the residual deviations trace those of the halo mass function itself.
For the mock DESI-like LRG and ELG samples constructed from the lightcone halo catalog by a one-to-one mapping between halos and galaxies, the angular auto power spectra and the cross power spectra with the convergence agree with the halo model prediction to within $\sim 10 \%$ over $10 \lesssim \ell \lesssim 10^2$, while on smaller scales, the measurements exceed the prediction by up to $\sim 50\%$ at $\ell \sim 10^3$ due to the non-linear halo bias, which is not captured by the linear bias prescription.
The weak lensing convergence power spectra measured from the density shells match the theoretical prediction to within a few percent over $30 \lesssim \ell \lesssim 5000$ for all the source redshifts, and are consistent with it within the statistical uncertainty on larger scales, which demonstrates the accuracy of the density shells as a representation of the projected matter distribution.
For the tSZ effect, the Compton-$y$ power spectra obtained by pasting the three pressure profile models onto the lightcone halo catalog agree with the corresponding halo model predictions to within $\sim 10 \%$ over $100 \lesssim \ell \lesssim 2000$, and the measurements exceed the predictions by up to $\sim 25\%$ at $\ell \simeq 8000$, though there is a large statistical uncertainty at large scales because the power spectrum is dominated by the rare nearby massive halos.
For the kSZ effect, the power spectra measured from the momentum shells agree with the perturbative prediction to within $\sim 10 \%$ at $100 \lesssim \ell \lesssim 300$, around the minimum of the spectrum, and exceed it by $\sim 20\%$ at $\ell \sim 10^3$ and up to $\sim 45 \%$ at $\ell \simeq 5000$, where the non-linear evolution of the velocity field and the internal motions of halos are not captured by the perturbative terms.
Finally, as a demonstration of the multi-wavelength capability of the suite, we measured the cross-correlation between the weak lensing convergence and the Compton-$y$ parameter, which agrees with the halo model prediction to within $\sim 10 \%$ over $30 \lesssim \ell \lesssim 5000$ for $z_\mathrm{s} = 0.5$ and $1$, while the excess reaches $\sim 15 \%$ at $100 \lesssim \ell \lesssim 1000$ for $z_\mathrm{s} = 2$.

A few limitations should be kept in mind when using the suite.
The simulations follow only dark matter, and the gas distribution is not evolved self-consistently but is pasted onto the halos or the particles in post-processing.
The SZ observables therefore inherit the assumptions of the adopted ICM models, and the back-reaction of baryons on the total matter distribution is not included.
The tSZ maps constructed with halo-based pasting miss the diffuse gas component outside halos and the contribution from halos below the mass threshold of $5 \times 10^{12} \, \hMsun$, and the kSZ maps assume that the electrons trace the total matter distribution.
On the weak lensing side, the convergence maps are computed in the Born approximation, so that the lens--lens coupling is neglected, and neither the intrinsic alignment of galaxies nor the baryonic effects on the matter distribution are taken into account.
The mock galaxy samples are likewise constructed with the simplest one-to-one mapping between halos and galaxies, which is sufficient for validating the lightcone halo catalog but does not reproduce the detailed galaxy--halo connection.
These restrictions are not intrinsic to the suite, and we plan to relax them in the extensions described in Section~\ref{sec:future_plans}.
In addition, the simulation box is replicated to fill the lightcone volume, so that the modes larger than the box size are absent and the same structures reappear along the line-of-sight direction.
Finally, the suite is run with a single set of cosmological parameters, and the $10$ realizations are sufficient to characterize the mean signal and to give a first estimate of its variance, but not to construct a converged covariance matrix for high-dimensional data vectors.

All the Level 1 and Level 2 DPs of the \GION simulation suite will be made publicly available, and the details of the data release are summarized in Appendix~\ref{sec:data_release}.
As outlined in Section~\ref{sec:future_plans}, the suite will be extended along five directions: mock galaxy catalogs based on halo occupation distribution or subhalo abundance matching models tailored to the Stage-IV spectroscopic surveys, which is enabled by the mass resolution of \GION down to the halos hosting ELGs; CMB lensing maps, obtained by supplementing the mass shells at $z < 5.14$ with Gaussian realizations of the linear density field up to the last scattering surface; CMB foregrounds, i.e., the CIB and the radio point sources, modeled by populating the lightcone halos within the halo model framework; a more complete treatment of weak lensing systematics through ray-tracing beyond the Born approximation, the intrinsic alignment of galaxies, and the baryonic effects incorporated at the mass shell level; and X-ray emission maps of galaxy clusters based on the gas density and temperature profiles provided by \texttt{Baryon Pasting}.
Together with the data products presented in this paper, these extensions will make the \GION simulation suite a testbed for the analysis pipelines of Stage-IV surveys and the associated CMB experiments, in particular for the joint analyses of multiple observables that share the same underlying matter distribution.

\appendix

\section{Data release}
\label{sec:data_release}
In this appendix, we summarize the data products of the \GION simulation suite that will be made publicly available.
The release comprises all the Level 1 and Level 2 DPs defined in Section~\ref{sec:numerical_methods} for each of the $10$ realizations, which are listed in Table~\ref{tab:data_products}.
The Level 0 DPs, i.e., the lightcone particle shells and the particle snapshots, are temporary data and are not part of the release.
The data are hosted at \url{https://github.com/GION-simulations/GION_Data_Release} and distributed under CC BY 4.0.
The total data volume is 10.1 TB per realization.

\begin{table*}[htbp]
    \caption{Data products of the \GION simulation suite included in the public release.
    All the products are provided for each of the $10$ realizations, and all the maps are in \texttt{HEALPix} format with $N_\mathrm{side} = 8192$.
    The density and momentum shells have the thickness of $25 \, \hMpc$.}
    \label{tab:data_products}
    \begin{ruledtabular}
        \begin{tabular}{llll}
            Data product & Format & Number \footnotemark[1] & Content \\
            \hline
            \multicolumn{4}{c}{Level 1 DP} \\
            \hline
            Halo catalog & HDF5 & 103 & Halo catalogs split in the redshift range $0 < z < 5$ \\
            Density shells & HDF5 & 216 & Projected mass $M^{(i)} (\bm{\theta})$ \\
            Momentum shells & HDF5 & 216 & Projected line-of-sight momentum $q^{(i)} (\bm{\theta})$ \\
            \hline
            \multicolumn{4}{c}{Level 2 DP} \\
            \hline
            Projected number density maps & FITS (\texttt{HEALPix}) & 5 & Mock LRG1, LRG2, LRG3, ELG1, and ELG2 samples \\
            Weak lensing maps & FITS (\texttt{HEALPix}) & 216 & Convergence $\kappa (\bm{\theta})$ with the source plane at each shell edge \\
            tSZ maps & FITS (\texttt{HEALPix}) & 3 & Compton-$y$ for Shaw, Planck, and Battaglia profiles \\
            kSZ maps & FITS (\texttt{HEALPix}) & 1 & Doppler-$b$ \\
        \end{tabular}
    \end{ruledtabular}
    \footnotetext[1]{Number of files (catalogs) or maps per realization.}
\end{table*}

\subsection{Level 1 data products}
\paragraph*{Halo catalogs}
The lightcone halo catalogs are generated with \texttt{MPI-Rockstar} and stored in HDF5 format.
The catalogs cover the full sky in the redshift range $0 < z < 5$ and contain all halos and subhalos with at least $50$ particles, i.e., more massive than $6.38 \times 10^{10} \, \hMsun$.
For each halo, the catalog stores the default outputs of \texttt{MPI-Rockstar}, which include the position, the velocity, the halo masses and radii with different overdensity definitions, and the angular momentum, as well as the six components of the inertia tensor and the ID of the parent halo, which distinguishes host halos from subhalos.
The positions are given in comoving coordinates with the observer at the origin, and thus the comoving distance, the angular position, and the redshift of each halo are readily derived from the stored position.
The full list of the stored quantities and their units is given in the documentation accompanying the release.

\paragraph*{Density and momentum shells}
The density and momentum shells are generated with \texttt{assign\_healpix} and are full-sky \texttt{HEALPix} maps with $N_\mathrm{side} = 8192$, i.e., $N_\mathrm{pix} = 12 N_\mathrm{side}^2$ pixels with the pixel size of $0.429 \, \mathrm{arcmin}$.
There are $216$ shells with the thickness of $\Delta \chi = 25 \, \hMpc$ in each realization, and the $i$-th shell ($i = 1, \dots, 216$) covers the comoving distance range $(i - 1) \Delta \chi \leq \chi < i \Delta \chi$, so that the shells extend up to $\chi = 5400 \, \hMpc$ corresponding to $z = 5.14$.
The density shells store the projected mass $M^{(i)} (\bm{\theta}) = \sum_{p \in \mathcal{P}^{(i)} (\bm{\theta})} m_\mathrm{part}$ in each pixel, and the momentum shells store the projected line-of-sight momentum $q^{(i)} (\bm{\theta}) = \sum_{p \in \mathcal{P}^{(i)} (\bm{\theta})} v_{r,p} \, m_\mathrm{part}$, where the sums run over the set $\mathcal{P}^{(i)} (\bm{\theta})$ of particles in the $i$-th shell falling into the pixel labeled by the angular position $\bm{\theta}$, and the line-of-sight velocity $v_{r,p}$ of the $p$-th particle is positive for particles receding from the observer.
The maps are stored in HDF5 format.
Since the shells are stored individually, users can construct projected fields with arbitrary radial kernels, e.g., the convergence field for a given source redshift distribution, by taking weighted sums of the shells.

\subsection{Level 2 data products}

\paragraph*{Projected number density maps}
The projected number density maps of the mock galaxies consist of the five DESI-like samples listed in Table~\ref{tab:galaxy_samples}, which are selected from the host halos in the lightcone halo catalog by the redshift and the halo mass.
Each map stores the number counts of the selected halos in each pixel.
Since the samples are constructed by the one-to-one mapping between halos and galaxies, the corresponding mock galaxy catalogs are subsets of the halo catalog and can be readily reconstructed from the Level 1 halo catalogs with the selection criteria in Table~\ref{tab:galaxy_samples}.

\paragraph*{Weak lensing maps}
The convergence maps are computed from the density shells in the Born approximation.
For each realization, we provide $216$ maps, and the $i$-th map corresponds to the source plane located at the far edge of the $i$-th shell, $\chi_\mathrm{s} = i \Delta \chi$.
The source redshifts thus range from $z_\mathrm{s} = 0.0083$ to $5.14$.

\paragraph*{tSZ maps}
The Compton-$y$ maps are generated by pasting the pressure profiles onto the halos more massive than $5 \times 10^{12} \, \hMsun$ and up to $z = 5$ in the lightcone halo catalog.
We provide three maps for each realization, which correspond to the three pressure profile models: Shaw, Planck, and Battaglia profiles.

\paragraph*{kSZ maps}
The Doppler-$b$ map is computed from the momentum shells under the assumption that the ionized gas traces the total matter distribution.
We provide one map for each realization, where the contributions from all the shells up to $z = 5.14$ are summed.

\subsection{Usage notes}
All the data products of a given realization are constructed from the same lightcone and share the same observer position and coordinate system.
Therefore, the maps and catalogs are mutually consistent and can be directly cross-correlated.
The $10$ realizations are run with different initial conditions and are statistically independent.
Note that the simulation box with the side length of $1.5 \, \hGpc$ is replicated to fill the lightcone volume, and thus the same structures appear multiple times at different redshifts and angular positions.
The cosmological parameters adopted in the simulations are given in Section~\ref{sec:intro}, and we also provide the linear matter power spectrum and the transfer function used to generate the initial conditions in the release.
Along with the data, we provide example Jupyter notebooks to read the data products and to reproduce the measurements presented in this paper at \url{https://github.com/GION-simulations/GION_Data_Release}.
The data products for the extensions outlined in Section~\ref{sec:future_plans} will be added to the release as they become available.

\section{Parameter convergence tests}
\label{sec:convergence_tests}
Here, we present a series of convergence tests that assess how the numerical parameters of the $N$-body simulation and its initial conditions affect the basic clustering statistics of the \GION simulation suite.
We examine three factors: the particle-mesh (PM) parameters of \texttt{GINKAKU}, the gravitational softening length, and the starting redshift together with the order of the Lagrangian perturbation theory used to generate the initial conditions with \texttt{monofonIC}.
For each test we run a set of smaller simulations that share the same cosmological parameters and the same random phases of the initial conditions, while varying one parameter at a time.
For each configuration we measure the matter power spectrum $P_\mathrm{mm}(k)$, the halo auto power spectrum $P_\mathrm{hh}(k)$ in eight mass bins spanning $11.0 \le \log_{10}(M_\mathrm{200b}/\hMsun) < 15.0$ in logarithmically equal steps of $0.5 \, \mathrm{dex}$, and the halo mass function $dn/d\ln M$ from the particle snapshots at $z = 0$.
To isolate the numerical dependence, in each test we take one configuration as the reference and show the fractional difference of the other runs relative to it, rather than comparing to an external theoretical prediction.
For the PM and starting redshift tests, the reference run corresponds to the settings adopted for the main \GION suite (Table~\ref{tab:simulation_parameters}), whereas for the softening length test, we take the most accurate configuration with a smaller time step and the smallest softening length as the reference (see Section~\ref{sec:app_soft}).
The full set of test runs is summarized in Table~\ref{tab:test_runs}.

\begin{table}[htbp]
    \caption{Summary of the convergence test runs.
    In each test the listed parameters are varied, while all other settings match the main \GION suite (Table~\ref{tab:simulation_parameters}).
    All runs within a test share the same cosmological parameters and the same random phases of the initial conditions.}
    \label{tab:test_runs}
    \begin{ruledtabular}
        \begin{tabular}{ll}
            Run & Configuration \\
            \hline
            \multicolumn{2}{c}{Particle mesh ($L = 0.75 \, \hGpc$, $N = 3072^3$)} \\
            \multicolumn{2}{c}{(\texttt{pm\_factor}, \texttt{cutoff\_radius})} \\
            \hline
            \texttt{pm1\_cut3.0} \footnotemark[1] & $(1, 3.0)$ \\
            \texttt{pm1\_cut4.5} & $(1, 4.5)$ \\
            \texttt{pm2\_cut3.0} & $(2, 3.0)$ \\
            \hline
            \multicolumn{2}{c}{Softening length ($L = 0.5 \, \hGpc$, $N = 2048^3$)} \\
            \multicolumn{2}{c}{(\texttt{global\_dt\_factor}, \texttt{pp\_softening\_ratio})} \\
            \hline
            \texttt{softref} & $(0.1, 0.01)$ \\
            \texttt{soft0.01} & $(1.0, 0.01)$ \\
            \texttt{soft0.02} & $(1.0, 0.02)$ \\
            \texttt{soft0.03} \footnotemark[1] & $(1.0, 0.03)$ \\
            \texttt{soft0.05} & $(1.0, 0.05)$ \\
            \texttt{soft0.10} & $(1.0, 0.10)$ \\
            \hline
            \multicolumn{2}{c}{Starting redshift ($L = 0.5 \, \hGpc$, $N = 2048^3$)} \\
            \hline
            \texttt{3LPT\_zs11.5} & 3LPT, $z_\mathrm{start} = 11.5$ \\
            \texttt{3LPT\_zs24} \footnotemark[1] & 3LPT, $z_\mathrm{start} = 24$ \\
            \texttt{3LPT\_zs49} & 3LPT, $z_\mathrm{start} = 49$ \\
            \texttt{3LPT\_zs99} & 3LPT, $z_\mathrm{start} = 99$ \\
            \texttt{2LPT\_zs11.5} & 2LPT, $z_\mathrm{start} = 11.5$ \\
            \texttt{2LPT\_zs24} & 2LPT, $z_\mathrm{start} = 24$ \\
            \texttt{2LPT\_zs49} & 2LPT, $z_\mathrm{start} = 49$ \\
            \texttt{2LPT\_zs99} & 2LPT, $z_\mathrm{start} = 99$ \\
        \end{tabular}
    \end{ruledtabular}
    \footnotetext[1]{Configuration adopted for the main \GION suite.}
\end{table}

\subsection{Particle mesh}
\label{sec:app_pm}
There are two main parameters in the PM part of the TreePM code \texttt{GINKAKU}: the number of PM grids, controlled by the grid factor \texttt{pm\_factor}, and the cutoff radius \texttt{cutoff\_radius} that sets the scale at which the force computation is split between the long-range PM force and the short-range tree force.
The number of PM grids per dimension is $N_\mathrm{grid} = \texttt{pm\_factor} \times N^{1/3}$, where $N$ is the number of particles in the simulation.
The cutoff radius $r_\mathrm{cut}$ is $\texttt{cutoff\_radius} \times (L/N_\mathrm{grid})$ with S2 cutoff function \cite{Hockney1981} adopted.
We compare three runs: the fiducial configuration with $\texttt{pm\_factor} = 1$ and $\texttt{cutoff\_radius} = 3.0$ (\texttt{pm1\_cut3.0}, taken as the reference), a run with a larger cutoff radius \texttt{cutoff\_radius}${} = 4.5$ (\texttt{pm1\_cut4.5}), and a run with a doubled PM grid resolution \texttt{pm\_factor}${} = 2$ (\texttt{pm2\_cut3.0}).
All the test runs have the same mass resolution as the main suite, i.e., the same mean interparticle separation of $0.24 \, \hMpc$.

\begin{figure}
\includegraphics[width=\columnwidth]{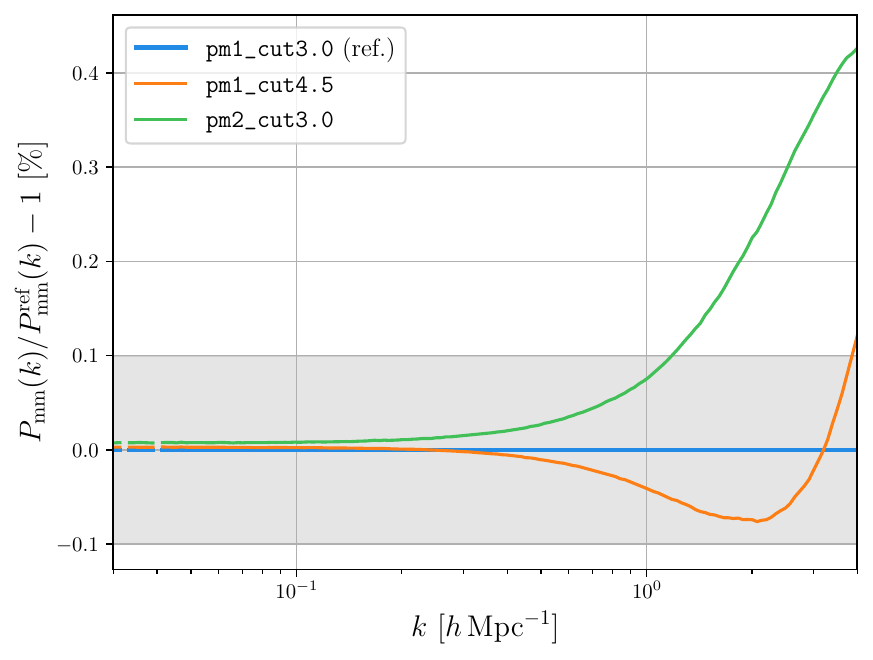}
\caption{Fractional difference of the matter power spectrum $P_\mathrm{mm}(k)$ for the PM parameter tests at $z = 0$, relative to the fiducial run \texttt{pm1\_cut3.0}.
The lines show a run with a larger PM cutoff radius (\texttt{pm1\_cut4.5}) and a run with a doubled PM grid resolution (\texttt{pm2\_cut3.0}).
The shaded region corresponds to $0.1\%$ deviation.}
\label{fig:pkmm_pm_cut}
\end{figure}

\begin{figure*}
\includegraphics[width=0.88\textwidth]{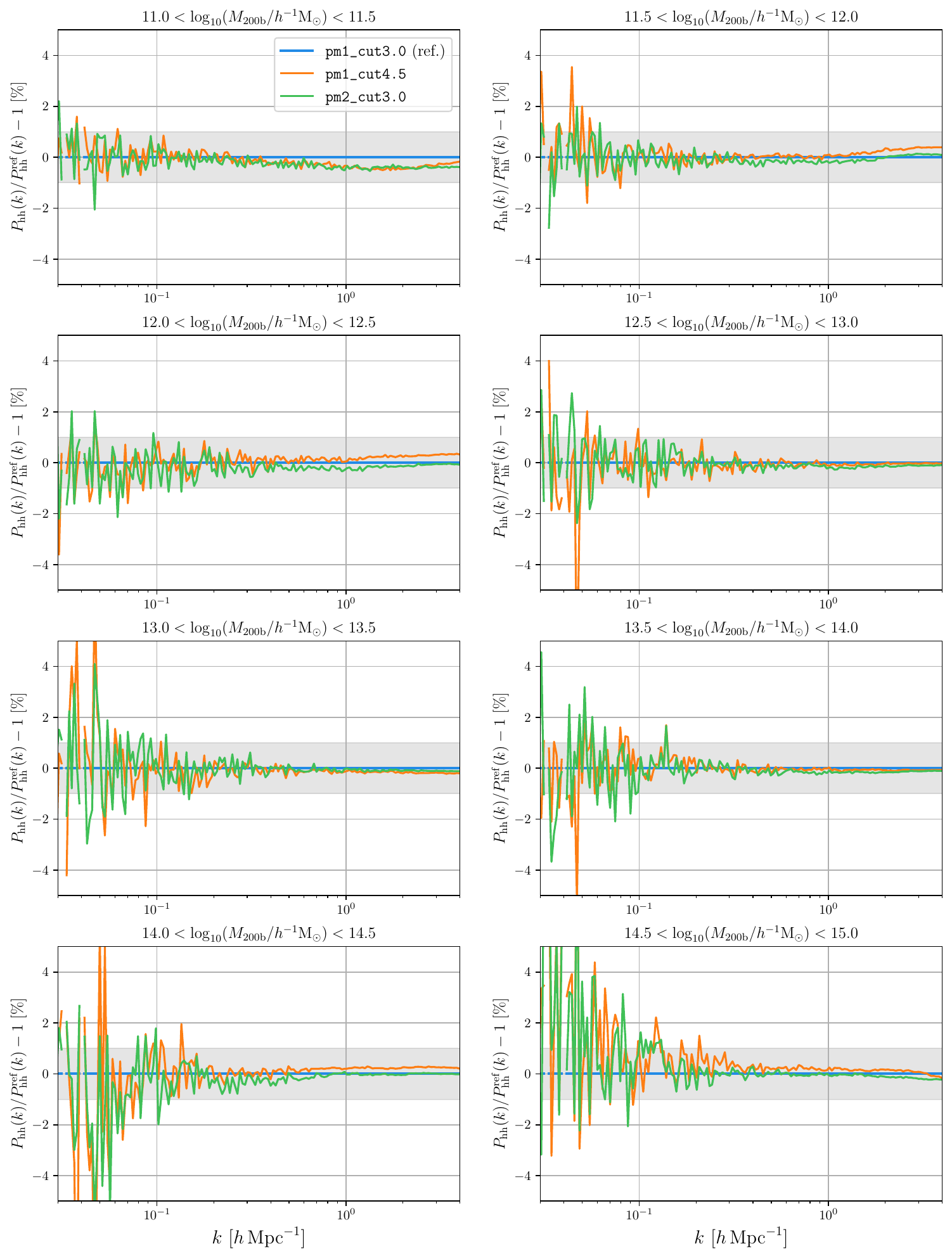}
\caption{Same as Figure~\ref{fig:pkmm_pm_cut}, but for the halo auto power spectrum $P_\mathrm{hh}(k)$ in the eight mass bins $11.0 < \log_{10}(M_\mathrm{200b} / \hMsun) < 15.0$ in logarithmically equal steps of $0.5 \, \mathrm{dex}$.
The shaded region corresponds to $1\%$ deviation.}
\label{fig:pkhh_pm_cut}
\end{figure*}

\begin{figure}
\includegraphics[width=\columnwidth]{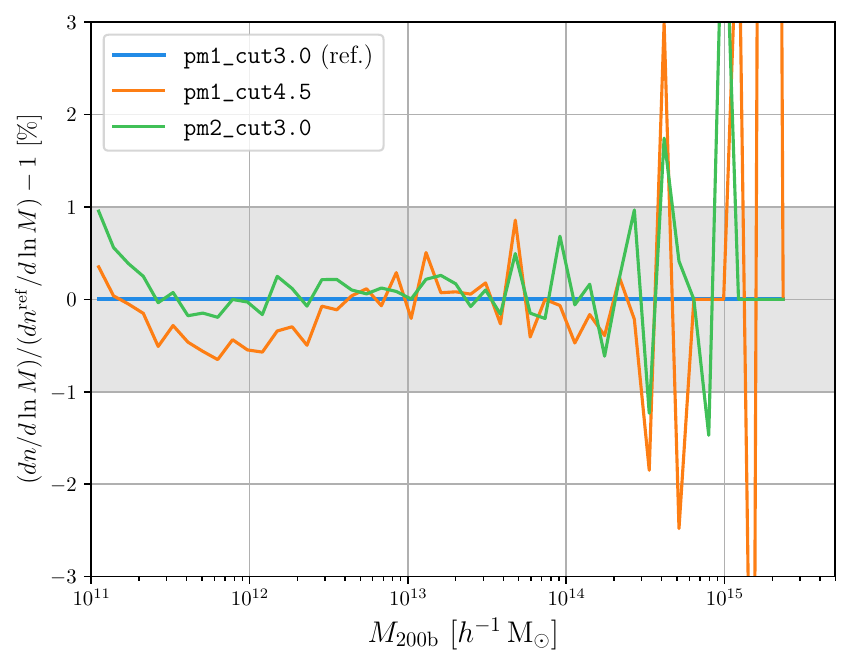}
\caption{Fractional difference of the halo mass function $dn/d\ln M$ for the PM parameter tests at $z = 0$, relative to the fiducial run \texttt{pm1\_cut3.0}.
The shaded region corresponds to $1\%$ deviation.}
\label{fig:mfunc_pm_cut}
\end{figure}

Figure~\ref{fig:pkmm_pm_cut} shows the fractional difference of the matter power spectrum.
At $z = 0$, all the variations agree with the reference to better than $0.5\%$ over the whole range $k \lesssim 4 \, \hMpcinv$.
Enlarging the cutoff radius (\texttt{pm1\_cut4.5}) changes the power spectrum by less than $0.15\%$, while doubling the number of PM grids (\texttt{pm2\_cut3.0}) mildly enhances the power on small scales, by $\sim 0.4\%$ at $k \sim 4 \, \hMpcinv$, owing to the finer PM force resolution.
The halo power spectra, shown in Figure~\ref{fig:pkhh_pm_cut}, agree within about $1\%$ for $k \gtrsim 0.1 \, \hMpcinv$; the larger scatter on large scales is due to the small number of available Fourier modes and the halo shot noise.
The halo mass function, presented in Figure~\ref{fig:mfunc_pm_cut}, is consistent within about $1\%$ up to $M_\mathrm{200b} \sim 10^{13} \, \hMsun$, above which the ratio becomes noisy because of the small number of massive halos in the reduced-volume test boxes.
We therefore conclude that the fiducial PM settings are sufficiently converged.

\subsection{Softening length}
\label{sec:app_soft}
The gravitational softening length regulates the short-range force and thereby the internal structure of halos and the small-scale clustering.
We vary the softening length parameter ($\texttt{pp\_softening\_ratio}$), which corresponds to the Plummer softening length in units of the mean interparticle separation $\epsilon_\mathrm{P} = \texttt{pp\_softening\_ratio}/2.34 = 0.004274, 0.008547, 0.01282, 0.02137$, and $0.04274$ for $\texttt{pp\_softening\_ratio} = 0.01, 0.02, 0.03, 0.05$, and $0.10$, respectively.
In addition, we also run the most accurate configuration with a smaller global time step ($\texttt{global\_dt\_factor} = 0.1$) and the smallest softening length $\texttt{pp\_softening\_ratio} = 0.01$, which is taken as the reference for this test and labeled as \texttt{softref}.
The value adopted for the main suite is $\texttt{pp\_softening\_ratio} = 0.03$.

\begin{figure}
\includegraphics[width=\columnwidth]{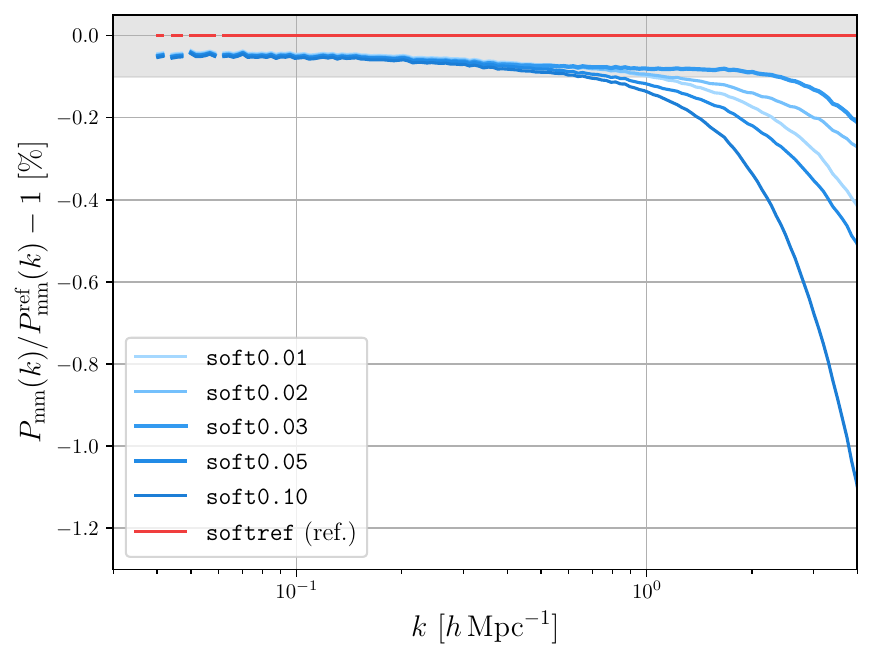}
\caption{Fractional difference of the matter power spectrum $P_\mathrm{mm}(k)$ for the softening length tests at $z = 0$, relative to the reference run \texttt{softref}.
The lines correspond to the softening length in units of the mean interparticle separation, $\texttt{pp\_softening\_ratio} = 0.01, 0.02, 0.03, 0.05$, and $0.10$.
The shaded region corresponds to $0.1\%$ deviation.}
\label{fig:pkmm_soft}
\end{figure}

\begin{figure*}
\includegraphics[width=0.88\textwidth]{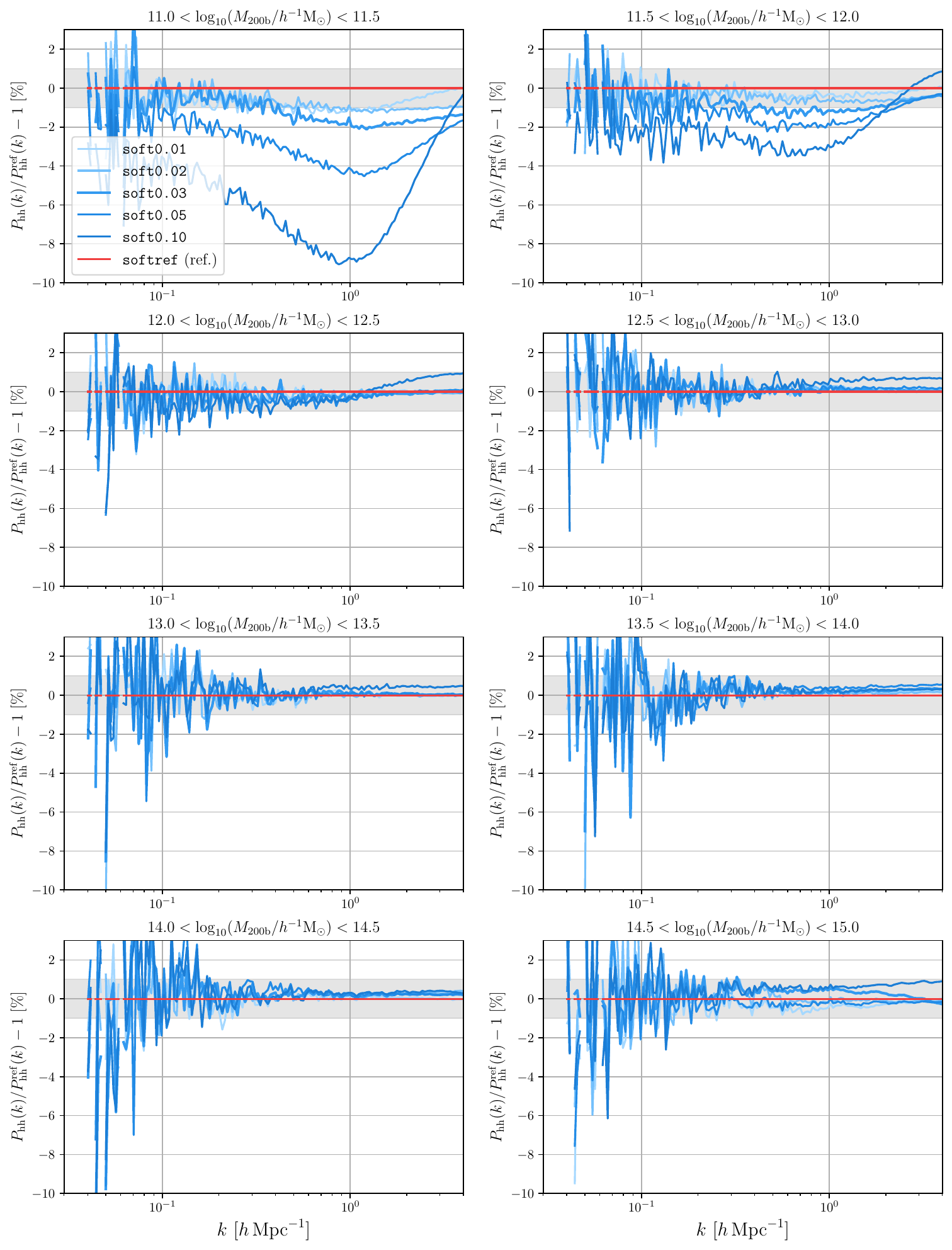}
\caption{Same as Figure~\ref{fig:pkmm_soft}, but for the halo auto power spectrum $P_\mathrm{hh}(k)$ in the eight mass bins $11.0 < \log_{10}(M_\mathrm{200b} / \hMsun) < 15.0$ in logarithmically equal steps of $0.5 \, \mathrm{dex}$.
The shaded region corresponds to $1\%$ deviation.}
\label{fig:pkhh_soft}
\end{figure*}

\begin{figure}
\includegraphics[width=\columnwidth]{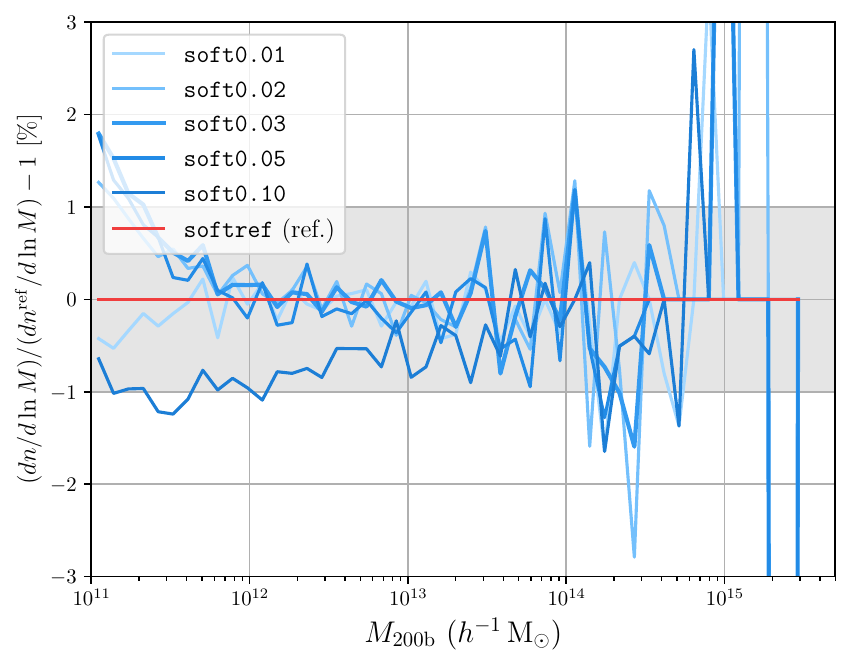}
\caption{Fractional difference of the halo mass function $dn/d\ln M$ for the softening length tests at $z = 0$, relative to the reference run \texttt{softref}.
The shaded region corresponds to $1\%$ deviation.}
\label{fig:mfunc_soft}
\end{figure}

Figure~\ref{fig:pkmm_soft} shows that for $\texttt{pp\_softening\_ratio} \geq 0.03$, increasing the softening length suppresses the small-scale matter power, as expected from the weaker short-range force.
At $z = 0$ the largest softening length suppresses the power by about $1\%$ at $k \sim 4 \, \hMpcinv$, while the fiducial choice $\texttt{pp\_softening\_ratio} = 0.03$ stays within about $0.2\%$.
The smaller softening lengths, $\texttt{pp\_softening\_ratio} = 0.01$ and $0.02$, show slightly larger deviations, $0.3\text{--}0.4\%$ at $k \sim 4 \, \hMpcinv$, which is attributed to the smaller time step adopted only in the reference run.
All choices agree within $0.1\%$ on large scales, $k \lesssim 0.5 \, \hMpcinv$.
The halo power spectra (Figure~\ref{fig:pkhh_soft}) show the same trend: in the lowest mass bin, $11.0 < \log_{10}(M_\mathrm{200b}/\hMsun) < 11.5$, the deviation at $k \sim 1 \, \hMpcinv$ is about $1\%$ for the smaller softening lengths and about $2\%$ for the fiducial value, whereas the larger softening lengths, $\texttt{pp\_softening\_ratio} = 0.05$ and $0.10$, suppress the low-mass halo clustering by $\sim 4\%$ and $\sim 9\%$, respectively, indicating that an excessively large softening impairs the formation of low-mass halos.
In the next mass bin, $11.5 < \log_{10}(M_\mathrm{200b}/\hMsun) < 12.0$, the suppression is milder but still reaches $\sim 2\%$ and $\sim 3.5\%$ for $\texttt{pp\_softening\_ratio} = 0.05$ and $0.10$, respectively, while for halos with $M_\mathrm{200b} \gtrsim 10^{12} \, \hMsun$, all the runs agree within about $1\%$ for $k \gtrsim 0.2 \, \hMpcinv$.
In the halo mass function (Figure~\ref{fig:mfunc_soft}), the largest softening length reduces the abundance of low-mass halos by $\sim 1\%$, while for the other softening lengths, including the fiducial value, the abundance agrees with the reference within $\sim 2\%$, with a slight excess at $M_\mathrm{200b} \lesssim 3 \times 10^{11} \, \hMsun$.
The fiducial softening length $\texttt{pp\_softening\_ratio} = 0.03$ thus provides converged clustering statistics while retaining sufficient force resolution.

\subsection{Starting redshift}
\label{sec:app_zstart}
Finally, we examine the dependence on the starting redshift $z_\mathrm{start}$ and the order of the Lagrangian perturbation theory (LPT) used to set up the initial conditions.
A too low starting redshift introduces transients from the neglected higher-order growing modes, whereas a too high starting redshift can amplify numerical noise from the finite force and mass resolution.
We consider initial conditions generated with second-order LPT (2LPT) and third-order LPT (3LPT), each at $z_\mathrm{start} = 11.5, 24, 49$, and $99$, and take the 3LPT run with $z_\mathrm{start} = 24$ (\texttt{3LPT\_zs24}), corresponding to the main suite, as the reference.

\begin{figure}
\includegraphics[width=\columnwidth]{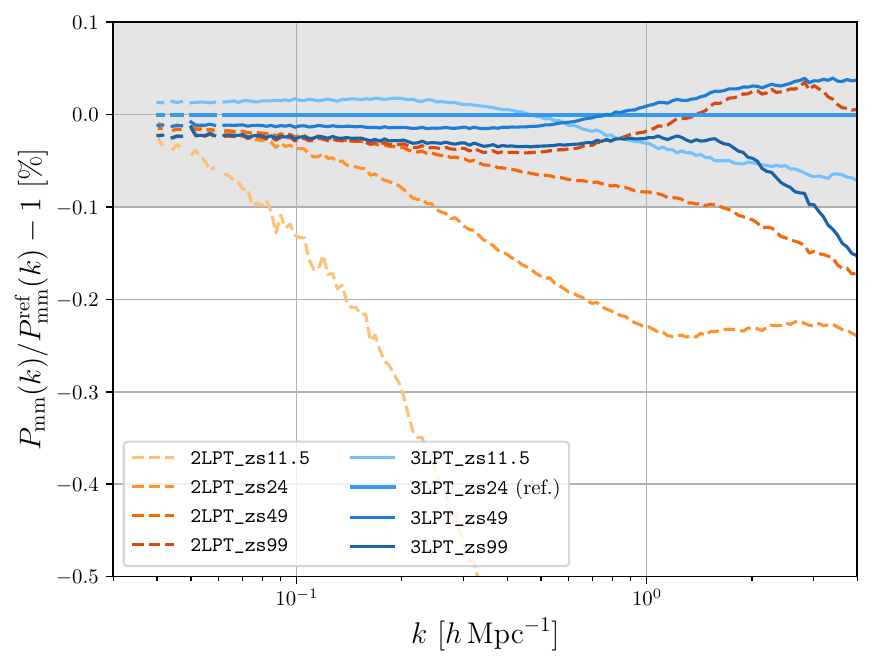}
\caption{Fractional difference of the matter power spectrum $P_\mathrm{mm}(k)$ for the starting redshift tests at $z = 0$, relative to the fiducial run \texttt{3LPT\_zs24}.
The dashed and solid lines correspond to the 2LPT and 3LPT initial conditions, respectively, generated at $z_\mathrm{start} = 11.5, 24, 49$, and $99$.
The shaded region corresponds to $0.1\%$ deviation.}
\label{fig:pkmm_zstart}
\end{figure}

\begin{figure*}
\includegraphics[width=0.88\textwidth]{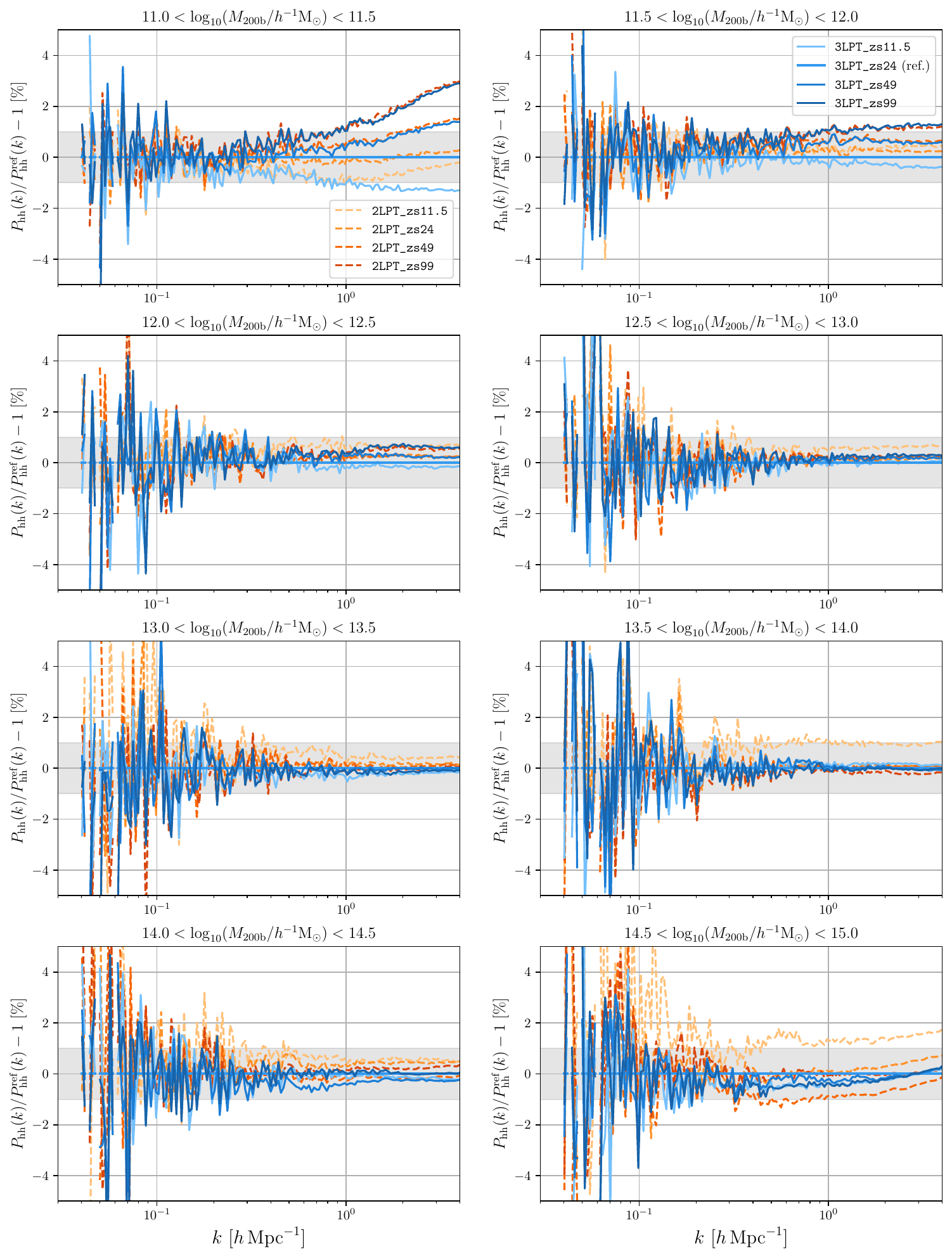}
\caption{Same as Figure~\ref{fig:pkmm_zstart}, but for the halo auto power spectrum $P_\mathrm{hh}(k)$ in the eight mass bins $11.0 < \log_{10}(M_\mathrm{200b} / \hMsun) < 15.0$ in logarithmically equal steps of $0.5 \, \mathrm{dex}$.
The shaded region corresponds to $1\%$ deviation.}
\label{fig:pkhh_zstart}
\end{figure*}

\begin{figure}
\includegraphics[width=\columnwidth]{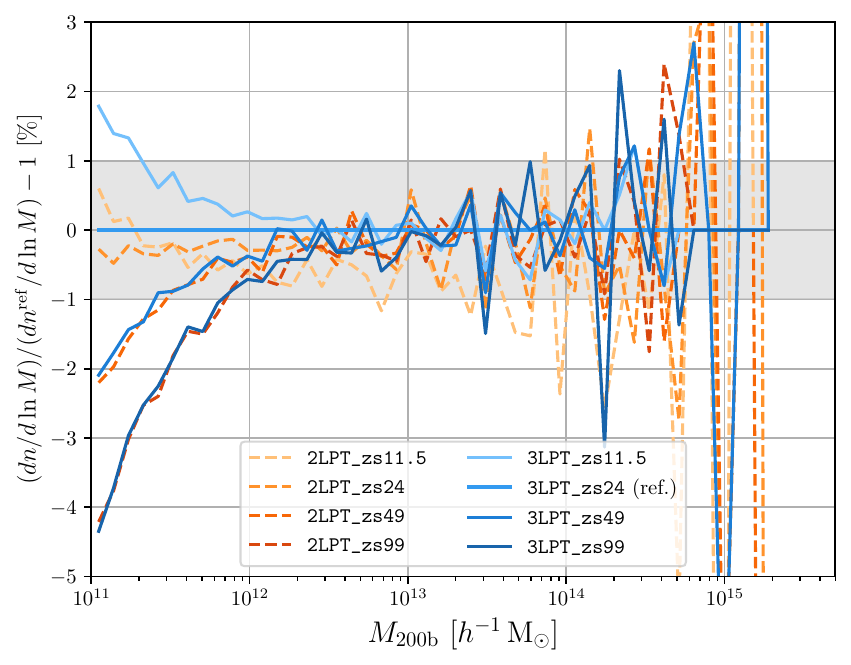}
\caption{Fractional difference of the halo mass function $dn/d\ln M$ for the starting redshift tests at $z = 0$, relative to the fiducial run \texttt{3LPT\_zs24}.
The shaded region corresponds to $1\%$ deviation.}
\label{fig:mfunc_zstart}
\end{figure}

Figure~\ref{fig:pkmm_zstart} shows that the 2LPT run with the lowest starting redshift (\texttt{2LPT\_zs11.5}) exhibits the largest transient errors.
Raising the starting redshift and, in particular, adopting 3LPT substantially reduce these errors: all the 3LPT runs agree with one another to within a few tenths of a percent at $z = 0$.
At $z = 0$, the halo power spectra (Figure~\ref{fig:pkhh_zstart}) of all the runs agree within about $1\%$ for $M_\mathrm{200b} > 10^{11.5} \, \hMsun$, except for \texttt{2LPT\_zs11.5} in the most massive bins ($1\text{--}2\%$), while in the lowest mass bin, the runs with $z_\mathrm{start} = 49$ and $99$ show an excess of up to $\sim 3\%$ at $k \gtrsim 1 \, \hMpcinv$.
The halo mass function at $z = 0$ (Figure~\ref{fig:mfunc_zstart}) agrees within $\sim 1\%$ above $M_\mathrm{200b} \sim 10^{12} \, \hMsun$ for all the runs.
At lower masses, the runs with $z_\mathrm{start} = 49$ and $99$ show a deficit of $2\text{--}4\%$ at $M_\mathrm{200b} \sim 10^{11} \, \hMsun$ irrespective of the LPT order, whereas \texttt{3LPT\_zs11.5} shows an excess of $\sim 2\%$.
These tests demonstrate that 3LPT initial conditions reduce the transient errors in the matter power spectrum compared with 2LPT at the same starting redshift, while the halo statistics disfavor high starting redshifts ($z_\mathrm{start} = 49$ and $99$), for which the clustering and abundance of low-mass halos deviate by a few percent, consistent with the amplified numerical noise expected for a too high starting redshift.
These results justify the fiducial choice of 3LPT initial conditions at $z_\mathrm{start} = 24$ for the \GION simulation suite.

\section{Effect of Limber approximation and flat-sky projection}
\label{sec:exact_model}
In this appendix, we describe the exact expressions of the angular power spectra that are used to quantify the effect of the Limber approximation and the flat-sky projection of individual halos.
In Section~\ref{sec:lightcone_outputs}, the theoretical predictions of the angular power spectra adopt the exact expressions at $\ell < 100$ and the expressions based on the Limber approximation and the flat-sky projection at $\ell \geq 100$.
We present the exact expressions for the convergence auto-power spectrum, the halo auto-power spectrum and halo--convergence cross-power spectrum, and the tSZ auto-power spectrum and tSZ--convergence cross-power spectrum, and discuss the effect of the Limber approximation and the flat-sky projection on these spectra.
Throughout this appendix, $\bm{\theta}$ denotes an angular position on the sky, and $z$ inside the integrands is understood as $z(\chi)$.

\subsection{General expression}
\label{sec:app_exact_general}
We consider two projected fields $X(\bm{\theta}) = \int d\chi \, W_X(\chi) \delta_X(\chi, \bm{\theta}, z(\chi))$ and $Y(\bm{\theta}) = \int d\chi \, W_Y(\chi) \delta_Y(\chi, \bm{\theta}, z(\chi))$, where $W_X$ and $W_Y$ are the radial kernels, and $\delta_X$ and $\delta_Y$ are three-dimensional fields with the unequal-time cross-power spectrum $P_{XY}(k; \chi_1, \chi_2)$.
Expanding the plane waves in spherical harmonics yields the exact full-sky expression of the angular power spectrum:
\begin{align}
  C_{XY}(\ell) = \frac{2}{\pi} & \int_0^\infty dk \, k^2
  \int d\chi_1 \, W_X(\chi_1) \, j_\ell(k\chi_1)
  \nonumber \\
  & \times \int d\chi_2 \, W_Y(\chi_2) \, j_\ell(k\chi_2) \,
  P_{XY}(k; \chi_1, \chi_2) ,
  \label{eq:cl_exact_general}
\end{align}
where $j_\ell$ is the spherical Bessel function of order $\ell$.
All the spectra considered here can be written in the form $P_{XY}(k; \chi_1, \chi_2) = F_X(k, \chi_1)  F_Y(k, \chi_2) P(k; \chi_1, \chi_2)$, where $P$ is either the linear or the non-linear matter power spectrum and $F_X$ and $F_Y$ are the factors given by the halo model ingredients specified below.
The unequal-time power spectrum is approximated by the geometric mean of the power spectra at the two redshifts \cite{Kitching2017a}, $P(k; \chi_1, \chi_2) \approx \sqrt{P(k, z(\chi_1)) \, P(k, z(\chi_2))}$, which is exact for the linear power spectrum with scale-independent growth.
Then, Eq.~\eqref{eq:cl_exact_general} factorizes into
\begin{align}
  C_{XY}(\ell) &= \frac{2}{\pi} \int_0^\infty dk \, k^2 \, I^X_\ell(k) \, I^Y_\ell(k) ,
  \label{eq:cl_exact_factorized} \\
  I_\ell(k) &= \int d\chi \, W(\chi) F(k, z) \sqrt{P(k, z)} j_\ell(k\chi) .
  \label{eq:I_ell}
\end{align}
The Limber approximation follows from the replacement $j_\ell(x) \to \sqrt{\pi/(2\ell+1)} \, \delta_\mathrm{D}(\ell + 1/2 - x)$, which simplifies the radial integrals to
\begin{align}
  C_{XY}(\ell) & \simeq \int d\chi \, \frac{W_X(\chi) \, W_Y(\chi)}{\chi^2}
  \nonumber \\
  & \quad \times F_X (k, z) F_Y (k, z)  P\left( k = \frac{\ell+1/2}{\chi}, z \right) .
  \label{eq:cl_limber_general}
\end{align}
Note that each of the two radial integrals in Eq.~\eqref{eq:cl_exact_factorized} runs over the support of its own kernel.
For instance, for the cross-power spectrum between a narrow redshift bin and the broad lensing kernel, the integral for the lensing field extends from the observer to the source plane, whereas the Limber expression (Eq.~\ref{eq:cl_limber_general}) receives contributions only from the overlap of the two kernels.

\subsection{Weak lensing}
\label{sec:app_exact_kappa}
In the Born approximation, the lensing potential is given as
\begin{equation}
  \psi(\bm{\theta}) = \frac{2}{c^2} \int_0^{\chi_\mathrm{s}} d\chi \,
  \frac{\chi_\mathrm{s} - \chi}{\chi \, \chi_\mathrm{s}} \, \Phi(\chi, \bm{\theta}, z(\chi)) ,
  \label{eq:lensing_potential}
\end{equation}
where $\Phi$ is the gravitational potential, and the convergence is defined as $\kappa = \nabla_{\bm{\theta}}^2 \psi / 2$, i.e., $\kappa_{\ell m} = -\ell(\ell+1) \, \psi_{\ell m} / 2$ in harmonic space.
The gravitational potential is related to the matter density contrast through the Poisson equation, $\nabla^2 \Phi = (3/2) \, \Omega_\mathrm{m} H_0^2 \, \delta / a$, i.e., $\tilde{\Phi} = -(3/2) \, \Omega_\mathrm{m} H_0^2 \tilde{\delta} / (a k^2)$ in Fourier space.
Therefore, the lensing potential is the projected field with $W_\psi F_\psi = -2 W_\kappa(\chi, \chi_\mathrm{s}) / (\chi^2 k^2)$, where $W_\kappa$ is the lensing kernel of Eq.~\eqref{eq:lensing_kernel}, and Eq.~\eqref{eq:cl_exact_factorized} multiplied by $[\ell(\ell+1)/2]^2$ leads to the exact expression of the convergence auto-power spectrum \cite{Kilbinger2017}:
\begin{align}
  C_{\kappa \kappa}(\ell) &= \frac{2}{\pi} \ell^2 (\ell+1)^2
  \int_0^\infty \frac{dk}{k^2}
  \nonumber \\
  & \quad \times \left( \int_0^{\chi_\mathrm{s}} \frac{d\chi}{\chi^2} W_\kappa (\chi, \chi_\mathrm{s})
  \sqrt{P_\mathrm{mm} (k, z)} \, j_\ell (k\chi) \right)^2 ,
  \label{eq:cl_kappa_nonlimber}
\end{align}
where $P_\mathrm{mm}$ is the non-linear matter power spectrum.
Since this expression involves the spherical Bessel functions, which oscillate rapidly, we use the \texttt{pylevin} code \cite{Reischke2025} \footnote{\url{https://github.com/rreischke/levin_bessel}} based on Levin's method to efficiently compute the integral.
Applying the Limber approximation to Eq.~\eqref{eq:cl_kappa_nonlimber} formally yields $[\ell(\ell+1)]^2 / (\ell+1/2)^4$ times the standard Limber expression (Eq.~\ref{eq:cl_kappa_limber}).
This prefactor deviates from unity only at the order of $\ell^{-2}$, which is the same order as the error of the Limber approximation itself \cite{LoVerde2008}, and thus we adopt Eq.~\eqref{eq:cl_kappa_limber} as the Limber expression.

\begin{figure}
\includegraphics[width=\columnwidth]{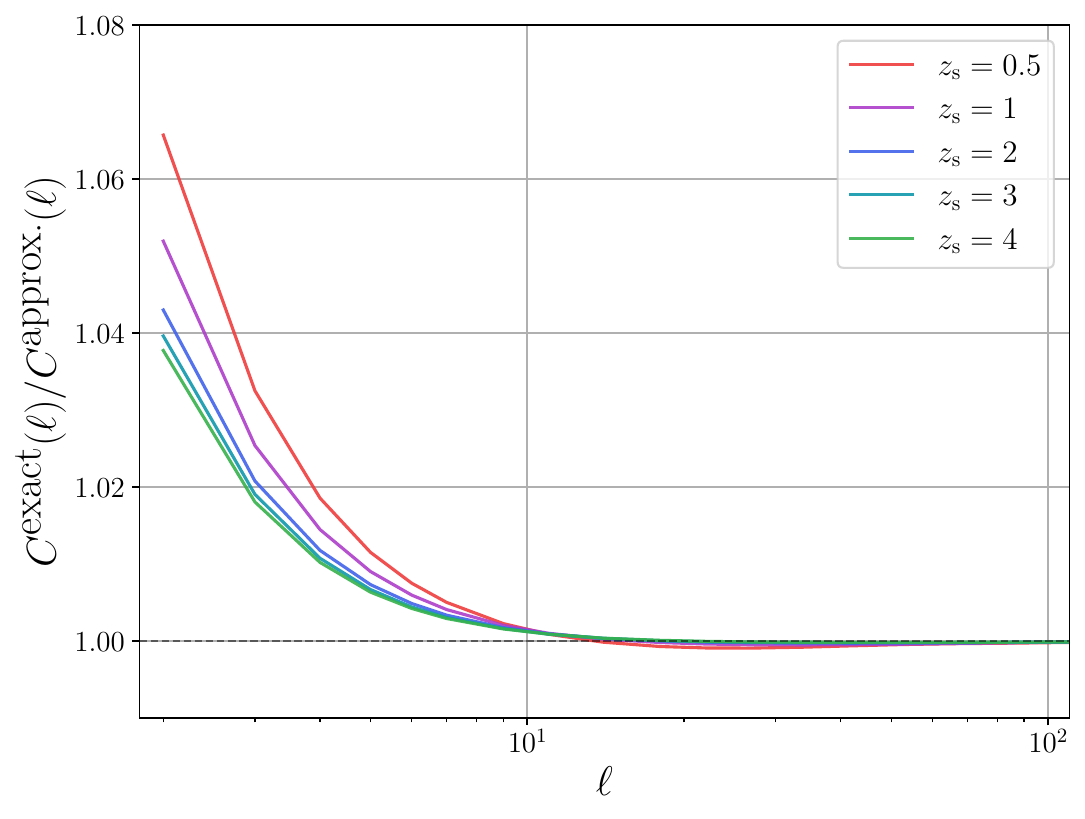}
\caption{Ratio of the exact (non-Limber) convergence angular power spectrum $C_{\kappa\kappa}(\ell)$ (Eq.~\ref{eq:cl_kappa_nonlimber}) to its Limber approximation (Eq.~\ref{eq:cl_kappa_limber}) for source redshifts $z_\mathrm{s} = 0.5, 1, 2, 3$, and $4$.
The horizontal dashed line indicates unity.}
\label{fig:nonlimber_ratio_WL}
\end{figure}

Figure~\ref{fig:nonlimber_ratio_WL} shows the ratio of the exact convergence power spectrum (Eq.~\ref{eq:cl_kappa_nonlimber}) to its Limber approximation (Eq.~\ref{eq:cl_kappa_limber}) for the fiducial model.
The ratio is larger for lower source redshifts, and is within $1\%$ of unity for $\ell \gtrsim 5$.

\subsection{Halo-halo auto- and halo--convergence cross-power spectra}
\label{sec:app_exact_halo}
For the halo samples, the radial kernel is $W_\mathrm{h}(\chi) = [H(z)/c] \, N_\mathrm{h}(z)$, and the three-dimensional power spectra are given in Eqs.~\eqref{eq:pkhh_halomodel} and \eqref{eq:pkhm_halomodel}.
We write the 2-halo term of the halo--matter cross-spectrum as $P_\mathrm{hm}^\mathrm{2h}(k, z) = \bar b_\mathrm{h} \, \mathcal{M}(k, z) \, P_\mathrm{lin}(k, z)$, where
\begin{align}
  \mathcal{M}(k, z) &= \int_{M_\mathrm{min}}^{M_\mathrm{max}} dM \, \frac{dn}{dM} \, b_\mathrm{h}(M, z) \,
  \frac{M}{\bar\rho_\mathrm{m}} \, u(k; M, z)
  \nonumber \\
  & \quad + A(M_\mathrm{min}) \, u(k; M_\mathrm{min}, z)
\end{align}
includes the correction for the finite mass integration range, so that $\mathcal{M} \to 1$ on large scales.
The exact 2-halo terms follow from Eq.~\eqref{eq:cl_exact_factorized} with $F_\mathrm{h} = \bar b_\mathrm{h}$ and $F_\mathrm{m} = \mathcal{M}$:
\begin{align}
  C_\mathrm{hh}^\mathrm{2h}(\ell) &= \frac{2}{\pi} \int_0^\infty dk \, k^2 \left[ I^\mathrm{h}_\ell(k) \right]^2 , \\
  C_{\mathrm{h}\kappa}^\mathrm{2h}(\ell) &= \frac{2}{\pi} \int_0^\infty dk \, k^2 \, I^\mathrm{h}_\ell(k) \, I^\mathrm{m}_\ell(k) ,
\end{align}
with
\begin{align}
  I^\mathrm{h}_\ell(k) &= \int d\chi \, W_\mathrm{h}(\chi) \, \bar b_\mathrm{h}(z) \sqrt{P_\mathrm{lin}(k, z)} \, j_\ell(k\chi) , \\
  I^\mathrm{m}_\ell(k) &= \int_0^{\chi_\mathrm{s}} d\chi \, W_\kappa(\chi, \chi_\mathrm{s}) \, \mathcal{M}(k, z)
  \sqrt{P_\mathrm{lin}(k, z)} \, j_\ell(k\chi) .
\end{align}

The 1-halo terms do not require the non-Limber treatment in the same sense.
The 1-halo term of the halo auto-power spectrum is pure shot noise, i.e., a Dirac delta function in real space, for which the Limber expression $C_\mathrm{hh}^\mathrm{1h} = \int d\chi \, W_\mathrm{h}^2 / (\chi^2 \, \bar n_\mathrm{h})$ is already exact.
The 1-halo term of the halo--convergence cross-power spectrum correlates the position of a halo with its own matter profile:
\begin{equation}
  C_{\mathrm{h}\kappa}^\mathrm{1h}(\ell) = \int d\chi \, \frac{W_\mathrm{h}(\chi)}{\bar n_\mathrm{h}(z)}
  \int_{M_1}^{M_2} dM \, \frac{dn}{dM} \, \tilde\kappa_\ell(M, z) ,
\end{equation}
where $\tilde\kappa_\ell(M, z)$ is the harmonic transform of the convergence profile of a single halo.
Its flat-sky expression is given by Eq.~\eqref{eq:kappa_ell} with the NFW profile truncated at the virial radius, which is equivalently written as $\tilde\kappa_\ell(M, z) = [W_\kappa(\chi, \chi_\mathrm{s}) / \chi^2] \, (M / \bar\rho_\mathrm{m}) \, u((\ell+1/2)/\chi; M, z)$.
There, the approximation lies not in the Limber approximation but in the flat-sky projection of a single halo, which we replace by the exact curved-sky projection.
For a spherically symmetric halo at the comoving distance $\chi$, the projected profile depends only on the angle $\theta$ from the halo center, and its exact harmonic transform is the Legendre transform with respect to $\mu = \cos\theta$:
\begin{align}
  \tilde\kappa_\ell(M, z) &= 2\pi \int_{-1}^{1} d\mu \, P_\ell(\mu) \, \kappa(\theta; M, z) ,
  \label{eq:kappa_ell_legendre} \\
  \kappa(\theta; M, z) &= \int d\chi' \, \frac{W_\kappa(\chi', \chi_\mathrm{s})}{\bar\rho_\mathrm{m}} \,
  a_\mathrm{h}^3 \, \rho(a_\mathrm{h} r; M, z) ,
  \label{eq:kappa_projected}
\end{align}
where $P_\ell$ is the Legendre polynomial, $a_\mathrm{h}$ is the scale factor at the halo redshift, $r = \sqrt{\chi^2 + \chi'^2 - 2\chi\chi'\mu}$ is the comoving distance from the halo center, $\rho$ is the density profile, and the line-of-sight integral runs over the extent of the profile.
In the flat-sky limit $\theta \ll 1$, Eq.~\eqref{eq:kappa_ell_legendre} reduces to Eq.~\eqref{eq:kappa_ell}.
In addition to the curvature of the sky, the exact expression also retains the variation of the lensing kernel $W_\kappa(\chi', \chi_\mathrm{s})$ across the halo.

\begin{figure}
\includegraphics[width=\columnwidth]{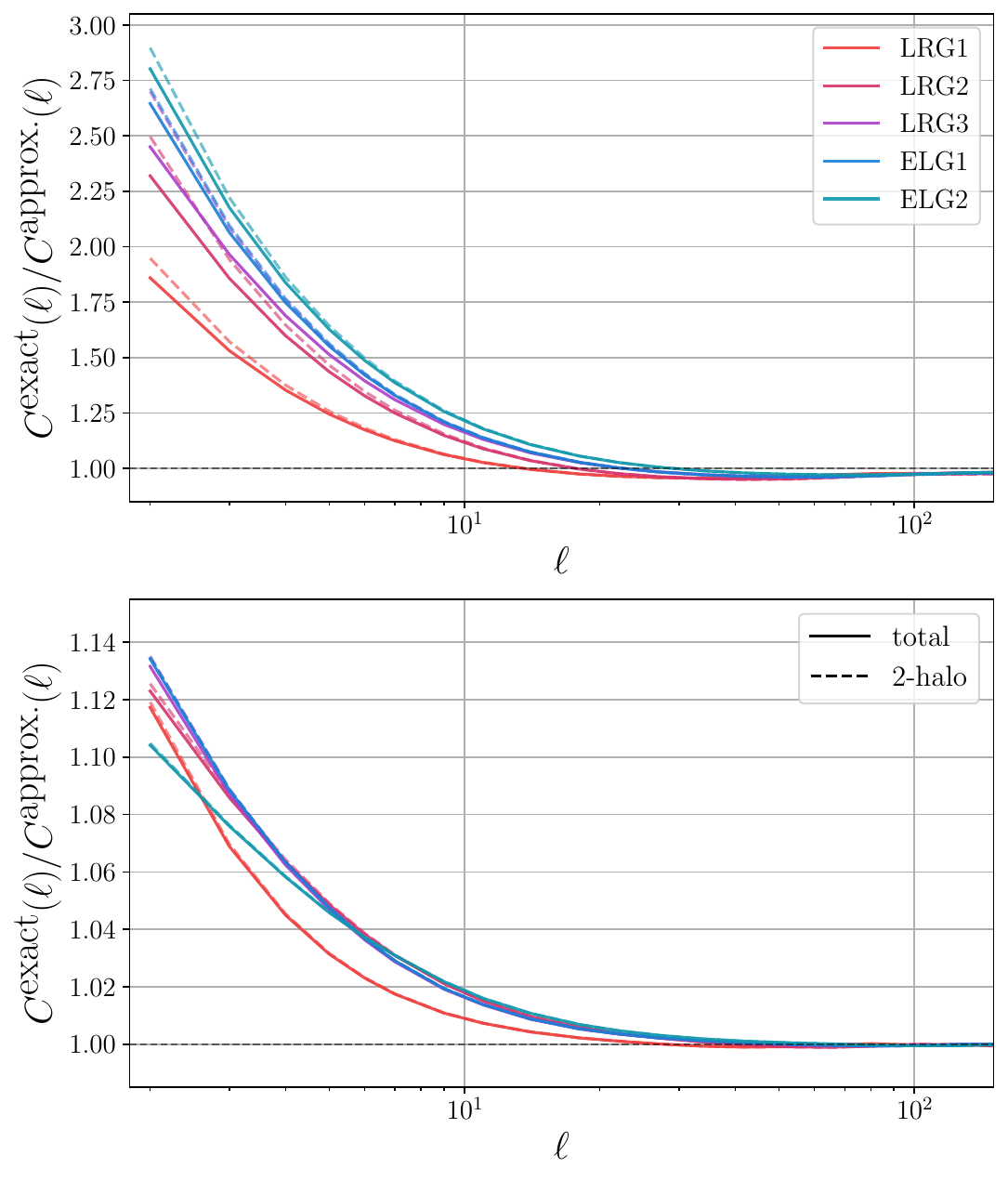}
\caption{Ratio of the exact to the Limber and flat-sky angular power spectra for the five mock galaxy samples: the halo auto-power spectrum $C_\mathrm{hh}(\ell)$ (top) and the halo--convergence cross-power spectrum $C_{\mathrm{h}\kappa}(\ell)$ (bottom).
The solid and dashed lines show the ratios for the total and the 2-halo term, respectively.
The horizontal dashed line indicates unity.}
\label{fig:nonlimber_ratio_halo}
\end{figure}

Figure~\ref{fig:nonlimber_ratio_halo} shows the ratios of the exact to the Limber and flat-sky angular power spectra for the five halo samples.
For the halo auto-power spectrum, the exact 2-halo term exceeds its Limber approximation by a factor of $2 \text{--} 3$ at $\ell = 2$, because the redshift distributions of the samples are narrow top-hat functions.
The ratio crosses unity at $\ell \approx 10\text{--}30$, depending on the sample, and then stays slightly below unity, by $2\text{--}4\%$, up to $\ell \sim 150$.
The 1-halo (shot-noise) term is exact.
For the halo--convergence cross-power spectrum, the ratio is within $1\%$ of unity for $\ell \gtrsim 15$, and the curved-sky correction to the 1-halo term is below $10^{-4}$.

\subsection{Cross-correlation of the thermal SZ effect and weak lensing}
\label{sec:app_exact_tsz}
The Compton-$y$ parameter (Eq.~\ref{eq:compton_y}) is expressed as $y(\bm{\theta}) = \int d\chi \, a(\chi) \, y_\mathrm{3D}(\chi, \bm{\theta}, z(\chi))$ with $y_\mathrm{3D} = (\sigma_\mathrm{T} / m_\mathrm{e} c^2) \, P_\mathrm{e}$.
For a halo with mass $M$ at redshift $z$, we define the Fourier transform of the profile in comoving coordinates,
\begin{equation}
  \tilde y_\mathrm{3D}(k; M, z) = 4\pi \int dr \, r^2 \, \frac{\sin kr}{kr} \,
  y_\mathrm{3D}(a r; M, z) ,
\end{equation}
where $r$ is the comoving radius and $ar$ is the physical radius at which the pressure profile is evaluated.
With this definition, the flat-sky profile of Eq.~\eqref{eq:y_ell} is written as $\tilde y_\ell(M, z) = (a / \chi^2) \, \tilde y_\mathrm{3D}((\ell+1/2)/\chi; M, z)$ \cite{Komatsu2002, Hill2013}, and similarly, Eq.~\eqref{eq:kappa_ell} is written as $\tilde\kappa_\ell(M, z) = [W_\kappa(\chi, \chi_\mathrm{s}) / \chi^2] \, (M / \bar\rho_\mathrm{m}) \, u((\ell+1/2)/\chi; M, z)$ with the truncated NFW profile of Ref.~\cite{Baltz2009}.
The exact 2-halo terms, the first of which corresponds to Eq.~(A23) of Ref.~\cite{Hill2013} in our notation, are
\begin{align}
  C_{yy}^\mathrm{2h}(\ell) &= \frac{2}{\pi} \int_0^\infty dk \, k^2 \left[ I^y_\ell(k) \right]^2 , \\
  C_{y\kappa}^\mathrm{2h}(\ell) &= \frac{2}{\pi} \int_0^\infty dk \, k^2 \, I^y_\ell(k) \, I^\kappa_\ell(k) ,
\end{align}
with
\begin{align}
  I^y_\ell(k) &= \int d\chi \, a(\chi) \, B_y(k, z) \sqrt{P_\mathrm{lin}(k, z)} \, j_\ell(k\chi) , \\
  I^\kappa_\ell(k) &= \int_0^{\chi_\mathrm{s}} d\chi \, W_\kappa(\chi, \chi_\mathrm{s}) \, B_\kappa(k, z)
  \sqrt{P_\mathrm{lin}(k, z)} \, j_\ell(k\chi) ,
\end{align}
and the bias-weighted profiles are defined as
\begin{align}
  B_y(k, z) &= \int dM \, \frac{dn}{dM} \, b_\mathrm{h}(M, z) \, \tilde y_\mathrm{3D}(k; M, z) , \\
  B_\kappa(k, z) &= \int dM \, \frac{dn}{dM} \, b_\mathrm{h}(M, z) \, \frac{M}{\bar\rho_\mathrm{m}} \, u(k; M, z) .
\end{align}
The Limber limits of these expressions are Eqs.~\eqref{eq:cl_yy_2h} and \eqref{eq:cl_yk_2h}.
The 1-halo terms (Eqs.~\ref{eq:cl_yy_1h} and \ref{eq:cl_yk_1h}) again involve no Limber approximation, but the flat-sky profiles $\tilde y_\ell$ and $\tilde\kappa_\ell$ are replaced by their exact curved-sky counterparts.
For the convergence profile, we use Eqs.~\eqref{eq:kappa_ell_legendre} and \eqref{eq:kappa_projected} with the truncated NFW profile of Ref.~\cite{Baltz2009}, and the Compton-$y$ profile is obtained in the same manner by replacing the integrand of the line-of-sight projection:
\begin{align}
  \tilde y_\ell(M, z) &= 2\pi \int_{-1}^{1} d\mu \, P_\ell(\mu) \, y(\theta; M, z) ,
  \label{eq:y_ell_legendre} \\
  y(\theta; M, z) &= \int d\chi' \, a(\chi') \, y_\mathrm{3D}(a_\mathrm{h} r; M, z) .
  \label{eq:y_projected}
\end{align}
Equation~\eqref{eq:y_ell_legendre} is equivalent to Eq.~(A15) of Ref.~\cite{Hill2013}, whose $k$-integral is the completeness relation of the spherical Bessel functions.
Evaluating that double-Bessel expression directly requires the oscillatory integrand to reproduce the Dirac delta function and is numerically unreliable, whereas the Legendre transform involves only the positive and sharply peaked projected profile.
In the flat-sky limit, Eq.~\eqref{eq:y_ell_legendre} reduces to Eq.~\eqref{eq:y_ell}.

\begin{figure}
\includegraphics[width=\columnwidth]{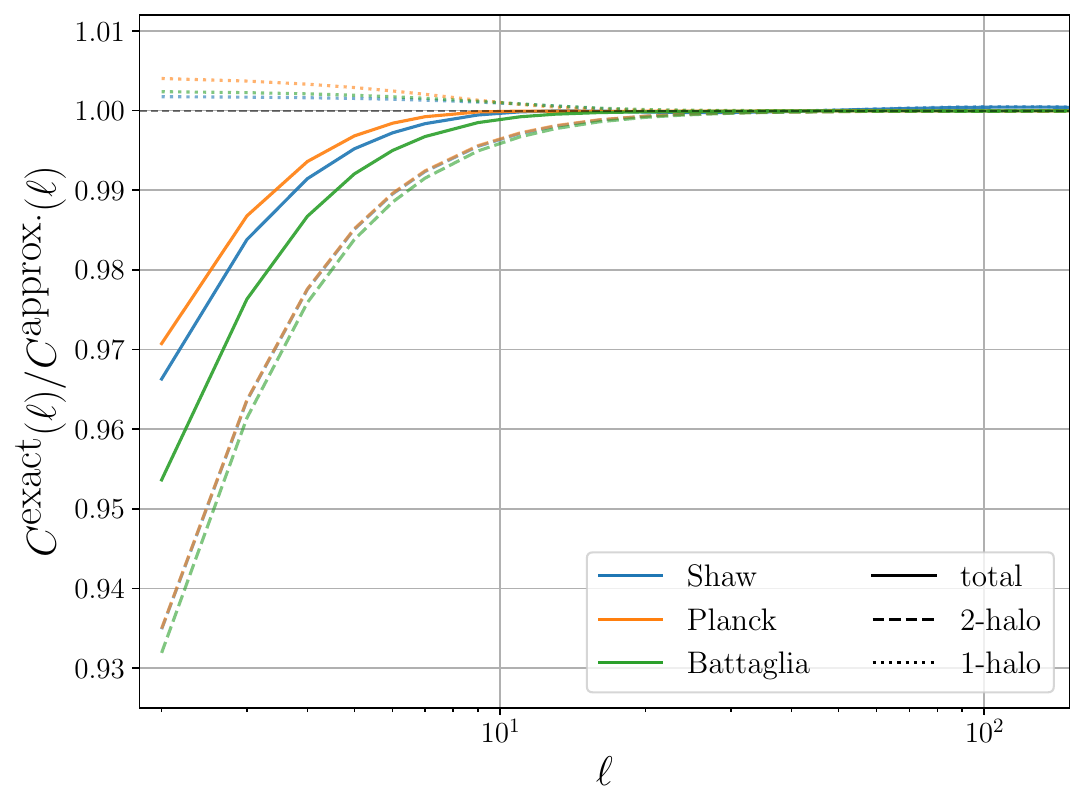}
\caption{Ratio of the exact to the Limber and flat-sky tSZ angular power spectrum $C_{yy}(\ell)$ for the Shaw, Planck, and Battaglia pressure profiles.
The solid, dashed, and dotted lines show the ratios for the total, 2-halo, and 1-halo terms, respectively.
The horizontal dashed line indicates unity.}
\label{fig:nonlimber_ratio_yy}
\end{figure}

\begin{figure}
\includegraphics[width=\columnwidth]{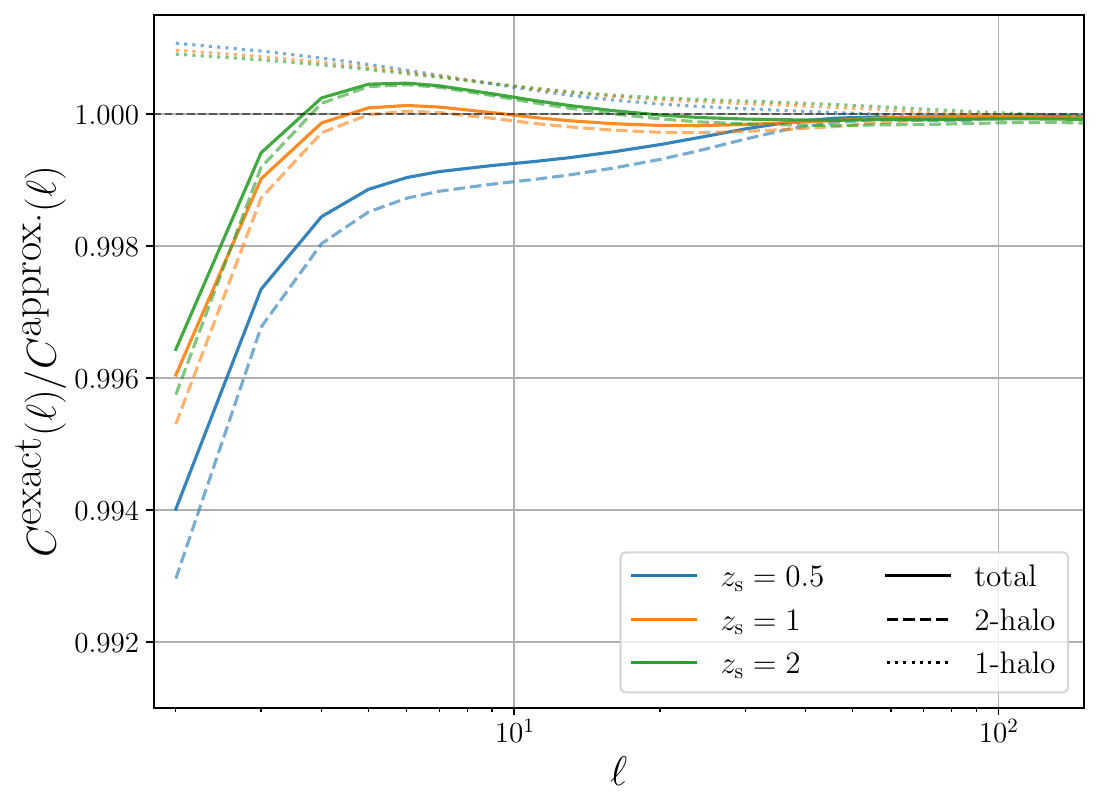}
\caption{Same as Figure~\ref{fig:nonlimber_ratio_yy}, but for the tSZ--convergence cross-power spectrum $C_{y\kappa}(\ell)$ for source redshifts $z_\mathrm{s} = 0.5, 1$, and $2$.}
\label{fig:nonlimber_ratio_yk}
\end{figure}

Figures~\ref{fig:nonlimber_ratio_yy} and \ref{fig:nonlimber_ratio_yk} show the ratios of the exact to the Limber and flat-sky angular power spectra for the tSZ auto-power spectrum with the three pressure profile models and for the tSZ--convergence cross-power spectrum with the source redshifts $z_\mathrm{s} = 0.5, 1$, and $2$, respectively.
For the tSZ auto-power spectrum, the ratio for the 2-halo term is within $0.1\%$ of unity for $\ell \gtrsim 20$.
The curved-sky correction to the 1-halo term is at most $0.4\%$, and thus the deviation of the total power spectrum is dominated by the 2-halo term; the ratio for the total power spectrum is $0.95\text{--}0.97$ at $\ell = 2$, depending on the relative contributions of the 1-halo and 2-halo terms for each pressure profile model.
For the tSZ--convergence cross-power spectrum, the ratio for the 2-halo term is within $0.1\%$ of unity for $\ell \gtrsim 10$.

\begin{acknowledgments}
We would like to thank Kai-Feng Chen, I-Non Chiu, Eiichiro Komatsu, Jeongin Moon, and Ryo Terasawa for their participation in testing an earlier version of our dataset and providing valuable feedback.
This work was supported by MEXT as
``Program for Promoting Researches on the Supercomputer Fugaku''
(Multi-wavelength Cosmological Simulations for Next-generation Surveys, JPMXP1020230407)
and used computational resources of supercomputer Fugaku provided by
the RIKEN Center for Computational Science (Project ID: hp230202, hp240200, hp250219).
This work was supported in part by JSPS KAKENHI Grant Number
JP24H00215 (K.~O. and T.~N.), JP22K21349 (K.~O. and H.~M.), JP25H00662 (K.~O. and T.~I.),
JP25H01513, JP25K17380, JP19KK0076 (K.~O.),
JP23H00108, JP24KK0065 (H.~M.),
JP24H00221 (M.~S. and T.~N.),
JP26H02062 (T.~I.),
JP22K03634, JP26H00402 (T.~N.),
JGC-Saneyoshi Scholarship Foundation (K.~O.),
JST FOREST Program, Grant Number JPMJFR2137 (H.~M.),
JST BOOST, Japan Grant Number JPMJBY24D8 (M.~S.), and
IAAR Research Support Program in Chiba University Japan (T.~I.).
\end{acknowledgments}

\bibliography{main}

\end{document}